\documentclass[hidelinks, 12 pt, a4paper]{article}
\usepackage{amsfonts,amsmath,amssymb,amsthm}
\usepackage{subcaption}
\usepackage[pdftex]{graphicx}
\usepackage{natbib}
\usepackage[hypertexnames=false]{hyperref}
\usepackage{setspace}
\usepackage{pdfsync}
\usepackage{lscape}
\usepackage{verbatim}
\usepackage[hmargin=1.25in,vmargin=1.25in]{geometry}
\usepackage{bibunits}
\usepackage{footmisc}
\usepackage{rotating}
\usepackage{titling}
\usepackage{mathrsfs,dsfont}
\usepackage[dvipsnames,svgnames, x11names]{xcolor}
\usepackage[capposition=top]{floatrow}
\usepackage{booktabs}
\usepackage{multirow}

\usepackage{tikz}
\usepackage{tikz-3dplot}
\usepackage{tkz-euclide}
\usepackage{mathtools}

\newcommand{\reals}{\mathbf{R}}

\def\ArrowheadSize{2mm}
\def\ArrowheadSizeBack{1.5mm}
\def\PlusLength{2pt}
\def\IntersectionProjectionColorTHREE{gray}
\def\NullColor{red}
\def\NullPerpColor{blue}
\def\xstarColor{blue}
\def\NullShiftColor{cyan}
\def\IntersectionColorTWO{cyan}

\def\AxisSizeTWO{3}
\def\AxisColorTWO{lightgray}
\def\PlaneColorTWO{gray}

\def\CameraA{83}
\def\CameraB{195}
\def\AxisSizeTHREE{4}
\def\ScaleX{0.8}
\def\ScaleY{2}
\def\ScaleZ{0.65}
\def\PullbackX{0.7}
\def\PullbackY{2.1}
\def\PullbackZ{0.7}
\def\AxisLabelColorTHREE{darkgray!75}
\def\AxisColorTHREE{darkgray!50}
\def\PlaneColorTHREE{gray!35}
\def\PlaneYZbottom{\AxisSizeTHREE * \PullbackZ}
\def\PlaneYZtop{\AxisSizeTHREE * \ScaleZ * 0.95}
\def\PlaneYZfront{\AxisSizeTHREE * \ScaleY}
\def\PlaneYZback{0}
\def\PlaneXYleft{\AxisSizeTHREE * \ScaleX}
\def\PlaneXYright{0}
\def\PlaneXYfront{\AxisSizeTHREE * \ScaleY}
\def\PlaneXYback{0}
\def\NullPerpFront{2.3}
\def\NullPerpBack{1.5}
\def\NullPerpTop{2.2}
\def\NullPerpBottom{2.2}
\def\NudgeNullPerpLabel{0.2}
\def\NullspaceSize{3.3}
\def\NullspacePullback{0.8}
\def\NudgeNullspaceLabel{0.3}
\def\NullshiftSize{1.2}
\def\NullshiftPullback{0.6}
\def\NudgeNullshiftLabel{0.3}

\tikzset{
    pointstyle/.style={black},
    every circle/.style={radius=2pt},
    2DAxis/.style={-{Latex[length=\ArrowheadSize]}, \AxisColorTWO},
    2DPlane/.style={\PlaneColorTWO, opacity = 0.2},
    2DPlaneLabel/.style={darkgray},
    2DNullspace/.style={
        {Latex[length=\ArrowheadSize]}-{Latex[length=\ArrowheadSize]},
        shorten >= -1.75cm,
        shorten <= -4cm,
        \NullColor
    },
    2DNullspaceLabel/.style={\NullColor},
    2DNullPerp/.style={
        {Latex[length=\ArrowheadSize]}-{Latex[length=\ArrowheadSize]},
        shorten >= -1.5cm,
        shorten <= -4cm,
        \NullPerpColor
    },
    2DNullPerpLabel/.style={\NullPerpColor},
    2DNullShift/.style={
        {Latex[length=\ArrowheadSize]}-{Latex[length=\ArrowheadSize]},
        shorten <= -4cm,
        \NullShiftColor
    },
    2DNullShiftLabel/.style={\NullShiftColor},
    2DIntersection/.style={\IntersectionColorTWO},
    3DAxis/.style={
        -{Latex[length=\ArrowheadSize]},
        \AxisColorTHREE,
        very thick},
    3DAxisBack/.style={
        -{Latex[length=\ArrowheadSizeBack]},
        \AxisColorTHREE,
        densely dotted},
    3DNullPerp/.style={fill=\NullPerpColor, opacity=.05},
    3DNullPerpBasis/.style={
        -{Latex[length=\ArrowheadSize]},
        very thin,
        \NullPerpColor!50
    },
    3DNullPerpBasisBack/.style={
        -{Latex[length=\ArrowheadSizeBack]},
        very thin,
        \NullPerpColor!50
    },
    3DNullPerpLabel/.style={\NullPerpColor},
    3DNullspaceFront/.style={
        -{Latex[length=\ArrowheadSize]},
        very thick,
        \NullColor
    },
    3DNullspaceBack/.style={
        -{Latex[length=\ArrowheadSizeBack]},
        \NullColor!60,
    },
    3DNullspaceLabel/.style={\NullColor},
    3Dxstar/.style={
        -{Latex[length=\ArrowheadSize]},
        very thick,
        \xstarColor!50
    },
    3DNullPerpProjection/.style={dashed, \xstarColor!50},
    3DNullShiftFront/.style={\NullShiftColor},
    3DNullShiftBack/.style={
        -{Latex[length=\ArrowheadSizeBack]},
        \NullShiftColor!60
    },
    3DNullShiftLabel/.style={\NullShiftColor},
    3DPlane/.style={fill=\PlaneColorTHREE,opacity=.5,},
    3DIntersection/.style={
        -{Latex[length=\ArrowheadSize]},
        very thick,
        \NullShiftColor
    },
    3DIntersectionProjection/.style={dashed, \IntersectionProjectionColorTHREE},
    3DAngle/.style={
        color=\NullPerpColor,
        fill=\NullPerpColor!30,
        opacity = 0.50,
        draw,
        ->,
    },
    3DAngleLabel/.style={\NullPerpColor},
    behind/.style={dash pattern={on 2pt off 1pt}},
    plusstyle/.style={ultra thick, gray}
}

\hypersetup{colorlinks, linkcolor = {teal}, citecolor = {blue}, urlcolor ={blue!80!black}}

\makeatletter
\renewcommand\paragraph{\@startsection{paragraph}{4}{\z@}%
            {-2.5ex\@plus -1ex \@minus -.25ex}%
            {1.25ex \@plus .25ex}%
            {\normalfont\normalsize\bfseries}}
\makeatother
\defaultbibliographystyle{ims}
\defaultbibliography{references}

\usetikzlibrary{patterns,shapes, calc, positioning}

\DeclareGraphicsExtensions{.jpg}

\def\qed{\rule{2mm}{2mm}}
\def\indep{\perp \!\!\! \perp}

\newtheorem{theorem}{Theorem}[section]
\newtheorem{lemma}{Lemma}[section]

\newtheorem{example}{Example}[section]

\newtheorem{remark}{Remark}[section]
\newtheorem{assumption}{Assumption}[section]

\let\oldmarginpar\marginpar
\renewcommand{\marginpar}[2][rectangle,draw,fill=black, text=white,text width= 2cm,rounded corners]{
    \oldmarginpar{
    \tiny \tikz \node at (0,0) [#1]{#2};}
    }

\newcommand{\re}{\mathbf{R}}

\begin{document}

\begin{bibunit}

\title{Inference for Large-Scale Linear Systems with Known Coefficients\thanks{We thank Denis Chetverikov, Patrick Kline, and Adriana Lleras-Muney for helpful comments. Omkar Katta and Conroy Lau provided outstanding research assistance. The research of the third author was supported by NSF grant SES-1530661. The research of the fourth author was supported by NSF grant SES-1846832.}}
{\author{
\normalsize Zheng Fang \\ \normalsize Department of Economics \\ \normalsize Texas A$\&$M University \\ \normalsize {\color{blue} zfang@tamu.edu}
\and
\normalsize Andres Santos\\\normalsize  Department of Economics\\\normalsize  UCLA \\\normalsize  {\color{blue} andres@econ.ucla.edu}
\and
\normalsize Azeem M. Shaikh\\ \normalsize Department of Economics \\\normalsize  University of Chicago \\\normalsize  {\color{blue} amshaikh@uchicago.edu}
\and
\normalsize Alexander Torgovitsky\\\normalsize  Department of Economics \\\normalsize  University of Chicago \\\normalsize  {\color{blue} torgovitsky@uchicago.edu}}}

\maketitle

\begin{abstract}
This paper considers the problem of testing whether there exists a non-negative solution to a possibly under-determined system of linear equations with known coefficients.
This hypothesis testing problem arises naturally in a number of settings, including random coefficient, treatment effect, and discrete choice models, as well as a class of linear programming problems.
As a first contribution, we obtain a novel geometric characterization of the null hypothesis in terms of identified parameters satisfying an infinite set of inequality restrictions.
Using this characterization, we devise a test that requires solving only linear programs for its implementation, and thus remains computationally feasible in the high-dimensional applications that motivate our analysis.
The asymptotic size of the proposed test is shown to equal at most the nominal level uniformly over a large class of distributions that permits the number of linear equations to grow with the sample size.
\end{abstract}

\begin{center}
\textsc{Keywords:} linear programming, linear inequalities, moment inequalities, random coefficients, partial identification, exchangeable bootstrap, uniform inference.
\end{center}

\newpage

\onehalfspacing

\section{Introduction}

Given an independent and identically distributed (i.i.d.)\ sample $\{Z_i\}_{i=1}^n$ with $Z$ distributed according to $P \in \mathbf P$, this paper studies the hypothesis testing problem
\begin{equation}\label{eq:null}
H_0: P\in \mathbf P_0 \hspace{0.5 in} H_1:P \in \mathbf P\setminus \mathbf P_0,
\end{equation}
where $\mathbf P$ is a ``large" set of distributions satisfying conditions described below and
$$\mathbf P_0 \equiv \{P \in \mathbf P : \beta(P) = Ax \text{ for some } x\geq 0\}.$$
Here, ``$x\geq 0$" signifies that all coordinates of $x\in \mathbf R^d$ are non-negative, $\beta(P) \in \mathbf R^p$ denotes an unknown but estimable parameter, and the coefficients of the linear system are known in that $A$ is a $p\times d$ known matrix.

As we discuss in Section \ref{sec:examples}, the described hypothesis testing problem plays a central role in a surprisingly varied array of empirical settings. Tests of \eqref{eq:null} can be used for obtaining asymptotically valid confidence regions for counterfactual broadband demand in the analysis of \cite{nevo2016usage}, and for conducting inference on the fraction of employers engaging in discrimination in the audit study of \cite{kline2021reasonable}.
Within the treatment effects literature, tests of \eqref{eq:null} arise naturally when examining the testable implications of the model proposed by \cite{imbens1994identification} and when conducting inference on partially identified parameters, such as in the studies by \cite{kline2016evaluating} and  \cite{kamat2019identification} of the Head Start program, or the analysis of unemployment state dependence by \cite{torgovitsky2019nonparametric}.
The null hypothesis in \eqref{eq:null} has also been shown by \cite{KitamuraStoye2018RUM} to play a central role in testing whether a cross-sectional sample is rationalizable by a random utility model; see \cite{manski2014identification}, \cite{deb2017revealed}, and \cite{lazzati2018nonparametric} for related examples. In addition, we show that for a class of linear programming problems the null hypothesis that the linear program is feasible may be mapped into \eqref{eq:null} -- an observation that enables us to conduct inference in the competing risks model of \cite{honorelleras2006bounds}, the empirical study of the California Affordable Care Act marketplace by \cite{tebaldi2019nonparametric}, and the dynamic discrete choice model of \cite{honore2006bounds}.  

The null hypothesis in \eqref{eq:null} can equivalently be represented as a system of linear inequalities in $\beta(P)$ through, e.g., Fourier-Motzkin elimination.
Such a representation would enable us to test \eqref{eq:null} by relying on approaches devised by the literature on testing for the validity of moment inequalities; see \cite{canay2017practical} for a review.
Unfortunately, in the empirical applications that motivate us the dimensions $p$ and, in particular, $d$ are large, making obtaining such a representation computationally infeasible \citep{KitamuraStoye2018RUM}.
We proceed instead by obtaining a novel geometric characterization of the null hypothesis that forms the cornerstone of our approach to inference.
Specifically, we show that the null hypothesis in \eqref{eq:null} holds if and only if: (i) there is an $x\in \mathbf R^d$ (not necessarily positive) solving $Ax = \beta(P)$; and (ii) the minimum norm solution to $Ax = \beta(P)$, denoted $x^\star(P)$, forms an obtuse angle with any vector in the intersection of the row space of $A$ and the negative orthant in $\mathbf R^d$.
Condition (ii) can be represented as a finite number of linear inequalities in $x^\star(P)$, though enumerating such inequalities can again be computationally prohibitive in applications with large $p$ and/or $d$.
We show that such enumeration is unnecessary: One can instead evaluate whether condition (ii) holds by computing the largest inner product between $x^\star(P)$ and the vectors in the intersection of the row space of $A$ and the negative orthant  -- a task that may be accomplished by solving a linear program.

Our geometric characterization of \eqref{eq:null} can be employed to construct a variety of different tests; see Section \ref{sec:test}.
Guided by a desire for computational and statistical reliability when $p$ and/or $d$ are large, however, we focus on a test that can be computed through linear programming.
Our test statistic employs a linear program to compute the largest violation of the ``inequality" restrictions prescribed by our geometric characterization of the null hypothesis.
While the test statistic is not pivotal, we obtain a critical value by relying on a bootstrap procedure that only requires solving one linear program per bootstrap iteration.
The resulting test is similar in spirit to the generalized moment selection approach of \cite{andrews:soares:2010} in that it aims to learn from the data whether inequalities are ``slack" or ``close" to binding.

Besides delivering computational tractability, the linear programming structure in our test enables us to establish the consistency of our asymptotic approximations under the requirement that $p^2/n$ tends to zero (up to logs).
Leveraging the consistency of such approximations to establish the asymptotic validity of our test further requires us to verify an anti-concentration condition at a particular quantile \citep{chernozhukov2014comparison}.
We show that the required anti-concentration property indeed holds under a condition that relates the allowed rate of growth of $p$ relative to $n$ to the matrix $A$.
This result enables us to derive a sufficient, but more stringent, condition on the rate of growth of $p$ relative to $n$ that delivers anti-concentration universally in $A$.
Furthermore, if, as in much of the related literature, $p$ is fixed with $n$, then our results imply that our test is asymptotically valid under “weak” regularity conditions on $\mathbf P$.

Our paper is related to important work by \cite{KitamuraStoye2018RUM}, who study \eqref{eq:null} in the context of testing the validity of a random utility model.
Their inference procedure, however, relies on conditions on $A$ that can be violated in the broader set of applications that motivate us; see Section \ref{sec:examples}.
\cite{andrews2019inference} and \cite{cox2019simple} propose methods for sub-vector inference in certain conditional moment inequality models that can be related to \eqref{eq:null}.
However, applying their tests, which were designed with a different problem in mind, to \eqref{eq:null} can require non-trivial theoretical extensions or be computationally challenging -- in particular when, as in most of our examples, $\beta(P)$ has non-zero known coordinates and/or $d$ is very large.
On the other hand, we show in Section \ref{sec:cond} that an important insight in \cite{andrews2019inference} allows us to adapt our methodology to conduct subvector inference in a class of conditional moment inequality models.
Our analysis is also conceptually related to work on sub-vector inference in models involving moment inequalities and to a literature on shape restrictions; see, e.g., \cite{romano:shaikh:2008}, \cite{bugni2017inference}, \cite{kaido2019confidence}, \cite{gandhi2017estimating}, \cite{chernozhukov2015constrained}, \cite{zhu2019inference}, and \cite{fang2019general}.
While these procedures are designed for general problems that do not possess the specific structure in \eqref{eq:null}, they are, as a result, less computationally tractable and/or rely on more demanding and high-level conditions than the ones we employ.

The remainder of the paper is organized as follows.
By way of motivation, we first discuss in Section \ref{sec:examples} applications in which the null hypothesis in \eqref{eq:null} arises naturally.
In Sections \ref{sec:geo} and \ref{sec:test}, we establish our geometric characterization of the null hypothesis and the asymptotic validity of our test.
Our simulation studies are contained in Section \ref{sec:simulations}.
The proof of our geometric characterization is contained in the Appendix.
Proofs for all other results and a guide to computation are contained in Supplemental Appendices.
An R package for implementing our test is available at \url{https://github.com/conroylau/lpinfer}.

\section{Applications}\label{sec:examples}

In order to fix ideas, we first discuss a number of empirical settings in which the hypothesis testing problem described in \eqref{eq:null} arises naturally.

\begin{example} \label{ex:dyn} \rm
{\bf (Dynamic Programming).}
Building on \cite{fox2011simple}, \cite{nevo2016usage} estimate a model for residential broadband demand in which there are $h\in \{1,\ldots, d\}$ types of consumers that select among plans $k\in \{1,\ldots, K\}$.
Each plan is characterized by a fee $F_k$, speed $s_k$, usage allowance $\bar C_k$, and overage price $\textsf{p}_k$.
At day $t$, a consumer of type $h$ with plan $k$ has utility over usage $c_t$ and numeraire $y_t$ given by
\begin{equation*}
u_h(c_t,y_t,v_t;k) = v_t(\frac{c_t^{1-\zeta_h}}{1-\zeta_h}) - c_t(\kappa_{1h} + \frac{\kappa_{2h}}{\log(s_k)}) + y_t,
\end{equation*}
where $v_t$ is an i.i.d.\ shock following a truncated log-normal distribution with mean $\mu_h$ and variance $\sigma_h^2$.
The problem faced by a type $h$ consumer with plan $k$ is
\begin{multline}\label{ex:dyn2}
\max_{c_1,\ldots, c_T} \sum_{t=1}^T E[u_h(c_t,y_t,v_t;k)] \\ \text{ s.t. } F_k + \textsf{p}_k \max\{C_T - \bar C_k,0\} + Y_T \leq I, ~ C_T = \sum_{t=1}^T c_t, ~ Y_T = \sum_{t=1}^T y_t,
\end{multline}
where the expectation is over $v_t$ and total wealth $I$ is assumed to be large enough to not restrict usage.
From \eqref{ex:dyn2}, it follows that the distribution of observed plan choice and daily usage, denoted by $Z\in \mathbf R^{T+1}$, for a  consumer of type $h$ is characterized by $\theta_h \equiv (\zeta_h,\kappa_{1h},\kappa_{2h},\mu_h,\sigma_h)$.
Hence, for any function $m$ of $Z$ we obtain the  restriction
\begin{equation*}
E_P[m(Z)] = \sum_{h=1}^d E_{\theta_h}[m(Z)] x_h ,
\end{equation*}
where $E_P$ and $E_{\theta_h}$ denote expectations under the distribution $P$ of $Z$ and under $\theta_h$, respectively, and $x_h$ is the unknown proportion of each type in the population.
After specifying $d = 16807$ different types and $p = 120000$ moments, \cite{nevo2016usage} estimate $x \equiv (x_1,\ldots, x_d)$ by GMM while constraining $x$ to be a probability measure.
The authors then employ the constrained GMM estimator for $x$ and the block bootstrap to conduct inference on counterfactual demand, which equals
\begin{equation*}
\sum_{h=1}^d a(\theta_h) x_h
\end{equation*}
for a known function $a$.
We note, however, that the results in \cite{fang2018inference} imply the bootstrap is \emph{inconsistent} for this problem.
In contrast, the results in this paper enable us to conduct asymptotically valid inference. 
For instance, by setting
\begin{equation}\label{ex:dyn5}
\beta(P) \equiv \left(\begin{array}{c} E_P[m(Z)] \\ 1 \\ \gamma \end{array}\right) \hspace{0.3 in} A \equiv \left(\begin{array}{ccc} E_{\theta_1}[m(Z)] & \cdots & E_{\theta_d}[m(Z)] \\ 1 & \cdots & 1 \\ a(\theta_1) & \cdots & a(\theta_d) \end{array}\right)
\end{equation}
we may obtain a confidence region for counterfactual demand through test inversion (in $\gamma$) of the null hypothesis in \eqref{eq:null} -- here, the final two constraints in \eqref{ex:dyn5} impose that probabilities add up to one and the hypothesized value for counterfactual demand.
Other applications of the approach in  \cite{nevo2016usage} include \cite{blundell2018escalation} and \cite{illanes2019competition}. \qed
\end{example}

\begin{example} \rm \label{ex:TE}
{\bf (Treatment Effects).}
Consider the heterogenous treatment effects model of \cite{imbens1994identification} in which an instrument $W\in \{0,1\}$, potential treatments $(D(0),D(1))$, and potential outcomes $(Y(0),Y(1))$ satisfy
\begin{equation}\label{ex:TE1}
(D(0),D(1),Y(0),Y(1)) \indep W \text{ and } D(1) \geq D(0) \text{ a.s.}
\end{equation}
The requirements in \eqref{ex:TE1} yield testable restrictions on the distributions of observables $(Y,D,W) \equiv (Y(D),D(W),W)$
\citep{balke1994counterfactual, angrist1995two, kitagawa2015test}
that may in fact be mapped into \eqref{eq:null}.
Specifically, assuming for simplicity that $Y$ has discrete support $\mathcal K$, and letting $j^c \equiv 1 - j$, we note that \eqref{ex:TE1} yields
\small
\begin{align}\label{ex:TE2}
P(Y \in B, D = j|W = 0) & = \sum_{l\in\{0,1\}: l \geq j} \sum_{(m,k)\in B\times \mathcal K} P((Y(j),Y(j^c),D(0),D(1)) = (m,k,j,l))  \notag\\
P(Y \in B, D = j|W = 1) & = \sum_{l\in\{0,1\}, l \leq j} \sum_{(m,k)\in B\times \mathcal K} P((Y(j),Y(j^c),D(0),D(1)) = (m,k,l,j)) \notag\\
 1 & = \sum_{l,j\in\{0,1\} : l \geq j} \sum_{m,k\in \mathcal K} P((Y(0),Y(1),D(0),D(1)) = (m,k,j,l))
\end{align}
\normalsize
for any  set $B$.
These restrictions may be written as $\beta(P) = Ax$ for a known $A$ and $x \geq 0$ denoting the joint distribution of $(Y(0),Y(1),D(0),D(1))$.
For $K$ the number of support points of $Y$, in this problem $x\in \mathbf R^d$ with $d = 3K^2$ and $\beta(P)\in \mathbf R^p$ with $p$ as large as $4K+1$. 
For instance, in estimating the distribution of compliers in \cite{angrist1991does}, \cite{imbens1997estimating} let $W$ indicate fourth quarter birth and discretize log weekly earning into 55 bins, yielding $d = 9075$ and $p = 221$.
By proceeding as in Example \ref{ex:dyn}, we may also construct confidence intervals for linear functionals of the distribution of $(Y(0),Y(1),D(0),D(1))$ such as the average treatment effect \citep{balke1997bounds, laffers2019bounding, machado2019instrumental,kamat2019identification,bai:shaikh:vytlacil:2020}. \qed
\end{example}

\begin{example} \label{ex:nonln} \rm
{\bf (Duration Models).}
In studying the efficacy of President Nixon's war on cancer, \cite{honorelleras2006bounds} employ the competing risks model
\begin{equation*}
(T^*,I) = \left\{\begin{array}{cl} (\min\{S_1,S_2\}, \arg\min\{S_1,S_2\}) & \text{ if } W =0 \\ (\min\{\alpha S_1, \beta S_2\}, \arg\min\{\alpha S_1,\beta S_2\}) & \text{ if } W = 1\end{array}\right. ,
\end{equation*}
where $(S_1,S_2)$ are possibly dependent random variables representing duration until death due to cancer and cardio-vascular disease, $W$ is independent of $(S_1,S_2)$ and indicates the implementation of the war on cancer, and $(\alpha,\beta)$ are unknown parameters.
The observed variables are $(T,I,W)$ where $T = t_k$ if $t_k \leq T^* < t_{k+1}$ for $k=1,\ldots,M$ and $t_{M+1} = \infty$, reflecting data sources often contain interval observations of duration.
While $(\alpha,\beta)$ is partially identified, \cite{honorelleras2006bounds} show that there exist known finite sets $\mathcal S(\alpha,\beta)$ and $\mathcal S_{k,i,w}(\alpha,\beta) \subseteq \mathcal S(\alpha,\beta)$ such that $(\alpha,\beta)$ belongs to the identified set if and only if there is a distribution $f(\cdot,\cdot)$ on $\mathcal S(\alpha,\beta)$ satisfying
\begin{multline}\label{ex:nonln2}
\sum_{(s_1,s_2) \in \mathcal S_{k,i,w}(\alpha,\beta)} f(s_1,s_2) = P(T = t_k, I=i| W = w), \\  \sum_{(s_1,s_2) \in \mathcal S(\alpha,\beta)} f(s_1,s_2) = 1, \text{ and } f(s_1,s_2) \geq 0 \text{ for all }(s_1,s_2)\in \mathcal S(\alpha,\beta),
\end{multline}
where the first equality must hold for all $1\leq k\leq M$, $i \in \{1,2\}$, and $w\in \{0,1\}$.
In the context of \cite{honorelleras2006bounds} analysis of the war on cancer, \eqref{ex:nonln2} yields $p = 141$ equality restrictions on $d = 4900$ parameters.
It follows from the representation in \eqref{ex:nonln2} that testing whether a particular $(\alpha,\beta)$ belongs to the identified set is a special case of \eqref{eq:null}.
Through test inversion, the results in this paper therefore allow us to construct a confidence region for the identified set that satisfies the coverage requirement proposed by \cite{imbens:manski:2004}.
We note that, in a similar fashion, our results also apply to the dynamic discrete choice model of \cite{honore2006bounds}.
\qed
\end{example}

\begin{example} \label{ex:discrete} \rm
{\bf (Discrete Choice).}
In their study of demand for health insurance in the California Affordable Care Act marketplace (Covered California), \cite{tebaldi2019nonparametric} model the observed plan choice $Y$ by a consumer according to
\begin{equation*}
Y \equiv \arg\max_{1\leq j\leq J} V_{j} - \textsf{p}_{j},
\end{equation*}
where $J$ denotes the number of available plans, $V = (V_{1},\ldots, V_{J})$ is an unobserved vector of valuations, and $\textsf{p} \equiv (\textsf{p}_{1},\ldots, \textsf{p}_{J})$ denotes post-subsidy prices.
In Covered California, post-subsidy prices satisfy $\textsf{p} = \pi(C)$ for some known function $\pi$ and $C$ a (discrete-valued) vector of individual characteristics that include age and county of residence.
By decomposing $C$ into subvectors $(W,S)$ and assuming $V$ is independent of $S$ conditional on $W$, \cite{tebaldi2019nonparametric} then obtain
\begin{equation*}
P(Y = j|C = c) = \int_{\mathcal V_j(\pi(c))} f_{V|W}(v|w)dv
\end{equation*}
for $f_{V|W}$ the density of $V$ conditional on $W$ and $\mathcal V_j(\textsf{p}) \equiv \{v : v_j - \textsf{p}_j \geq v_k - \textsf{p}_k \text{ for all } k\}$.
The authors further show there is a finite partition $\mathbb V$ of $\re^{J}$ satisfying
\begin{equation}\label{ex:discrete3}
P(Y = j |C= c) = \sum_{\mathcal V \in \mathbb V : \mathcal V \subseteq \mathcal V_j(\pi(x))} \int_{\mathcal V} f_{V|W}(v|w)dv
\end{equation}
and such that counterfactuals, such as the change in consumer surplus due to a change in subsidies, can be written as functionals with the structure
\begin{equation}\label{ex:discrete4}
\sum_{\mathcal V \in \mathbb V} a(\mathcal V) \int_{\mathcal V} f_{V|W}(v|w)dv
\end{equation}
for known function $a:\mathbb V \to \mathbf R$.
Arguing as in Example \ref{ex:dyn}, it then follows from \eqref{ex:discrete3} and \eqref{ex:discrete4} that confidence regions for the desired counterfactuals may be obtained through test inversion of hypotheses as in \eqref{eq:null}.
In \cite{tebaldi2019nonparametric}, the corresponding matrix $A$ has dimensions as high as $253\times 15000$. \qed
\end{example}

\begin{example} \label{ex:reveal} \rm
{\bf (Revealed Preferences).}
Building on \cite{mcfadden1990stochastic}, \cite{KitamuraStoye2018RUM} develop a nonparametric specification test for a random utility model (RUM) by mapping their null hypothesis into \eqref{eq:null}.
In the simplest setting they study, \cite{KitamuraStoye2018RUM} suppose there are $K$ goods and for each individual we observe the prices $\textsf{p}\in \mathbf R^K$ they faced, their budget set $\mathcal B(\textsf{p}) \equiv \{y \in \mathbf R^{K}_+ : \textsf{p}^\prime y = 1\}$, and their chosen consumption bundle $Y\in \mathcal B(\textsf{p})$.
Under the assumption that $\textsf{p}$ has discrete support $\{\textsf{p}_1,\ldots, \textsf{p}_J\}$, the authors build a finite partition $\mathbb V$ of $\bigcup_{j=1}^J \mathcal B(\textsf{p}_j)$ and matrix $A$ such that the distribution $P$ of $(Y,\textsf{p})$ is compatible with RUM if and only if
\begin{equation}\label{ex:reveal1}
\beta(P) = A x \text{ for some } x\geq 0,
\end{equation}
where each coordinate of $\beta(P)$ equals $P(Y \in \mathcal V|\textsf{p} = \textsf{p}_j)$ for some $\mathcal V \in \mathbb V$ and $1\leq j \leq J$.
Each column of $A$ represents a rationalizable non-stochastic demand system and the $x\geq 0$ solving \eqref{ex:reveal1} represents a vector of probabilities over these demand systems.
\cite{KitamuraStoye2018RUM} propose a test for \eqref{ex:reveal1} and implement it using the U.K. Family Expenditure Survey -- an application in which $p$ and $d$ can be as large as 79 and 313440.
We note, however, that the arguments for asymptotic validity of their test rely on a key restriction on $A$: Namely, that $(a_1 - a_0)^\prime (a_2-a_0) \geq 0$ for any distinct column vectors $(a_0,a_1,a_2)$ of $A$.
While this restriction is automatically satisfied in the application that motivates \cite{KitamuraStoye2018RUM} and related work \citep{manski2014identification, deb2017revealed, lazzati2018nonparametric}, it can fail in the previously discussed examples. \qed
\end{example}

\section{Geometry of the Null Hypothesis}\label{sec:geo}

In this section, we obtain a geometric characterization of the condition that a vector $\beta \in \mathbf R^p$ satisfies $\beta = Ax$ for some $x \geq 0$.
This result yields an alternative formulation of the null hypothesis that guides the construction of our test.

In what follows, we let $\mathbf R^k$ be the Euclidean space of dimension $k$ and reserve $p$ and $d$ for denoting the dimensions of the matrix $A$.
For any two column vectors $(v_1,\ldots, v_k)^\prime \equiv v$ and $(u_1,\ldots, u_k)^\prime \equiv u$ in $\mathbf R^k$, we denote their inner product by $\langle v,u\rangle \equiv \sum_{i=1}^k v_i u_i.$
The space $\mathbf R^k$ can be equipped with the norms $\|\cdot\|_q$ given by
\begin{equation*}
\|v\|_q \equiv \{\sum_{i=1}^k |v_i|^q\}^{\frac{1}{q}}
\end{equation*}
for any $1\leq q \leq \infty$, where $\|v\|_\infty$ is understood to equal $\max_{1\leq i \leq k} |v_i|$.
In addition, for any $k\times k$ matrix $M$, the norm $\|\cdot\|_q$ on $\mathbf R^k$ induces the norm
\begin{equation*}
\|M\|_{o,q} \equiv \sup_{\|v\|_q \leq 1} \|M v\|_q
\end{equation*}
on $M$; e.g., $\|M\|_{o,2}$ is the largest singular value of $M$. 
While $\|\cdot\|_1$ and $\|\cdot\|_\infty$ play a crucial role in our statistical analysis, our geometric analysis relies more heavily on the norm $\|\cdot\|_2$.
In particular, for any closed convex set $C\subseteq \mathbf R^k$, we use the properties of the $\|\cdot\|_2$-metric projection operator $\Pi_C : \mathbf R^k \to C$, defined by
\begin{equation*}
\Pi_C(v)\equiv \arg\min_{c\in C} \|v -c\|_2;
\end{equation*}
i.e., $\Pi_C(v)$ denotes the unique closest (under $\|\cdot\|_2$) element in $C$ to $v \in \mathbf R^{k}$.
It will also be helpful to view $A$ as a linear map with range $R$ and null space $N$ given by
\begin{align*}
R \equiv \{b \in \mathbf R^{p} : b = Ax \text{ for some } x\in \mathbf R^d\}
\quad
\text{and}
\quad
N \equiv \{x \in \mathbf R^d : Ax = 0\}.
\end{align*}
The null space $N$ of $A$ induces a decomposition of $\mathbf R^d$ through its orthocomplement
$$N^\perp \equiv \{y \in \mathbf R^d : \langle y,x\rangle =0 \text{ for all } x \in N\};$$
i.e.,\ any vector $x \in \mathbf R^d$ satisfies $x = \Pi_N(x) + \Pi_{N^\perp}(x)$ with $\langle\Pi_N(x),\Pi_{N^\perp}(x)\rangle = 0$.

Our first result is a well known consequence of the orthogonality of $N$ and $N^\perp$.

\begin{lemma}\label{lm:trivial}
For any $\beta \in \mathbf R^p$ there is a unique $x^\star \in N^\perp$ satisfying $\Pi_R(\beta) = Ax^\star$, where $\Pi_R(\beta)$ denotes the projection of $\beta$ onto $R$ (the range of $A$).
\end{lemma}

\noindent If $\beta$ belongs to $R$ then $\Pi_R(\beta) = \beta$ and Lemma \ref{lm:trivial} implies there exists a unique $x^\star\in N^\perp$ satisfying $\beta = A x^\star$.
While $x^\star$ is the unique solution in $N^\perp$, there may exist multiple solutions in $\mathbf R^d$.
In fact, provided $\beta \in R$, Lemma \ref{lm:trivial} implies
$$\{x \in \mathbf R^d : Ax = \beta\} = x^\star + N.$$
Hence, the restriction that $\beta = A x$ for some $x \geq 0$ is equivalent to two conditions:
\begin{equation}\label{eq:geo3}
\text{(i) } \beta \in R \hspace{1 in} \text{(ii) } \{x^\star + N\}\cap \mathbf R^d_+ \neq \emptyset;
\end{equation}
i.e.,\ condition (i) ensures \emph{some} solution to the equation $Ax = \beta$ exists, while condition (ii) ensures a \emph{positive} solution exists.

\begin{figure}[t!]
    \caption{\label{fig:illR2} Illustration of when requirement (ii) in \eqref{eq:geo3} is satisfied}
    \leftskip 8pt
    \resizebox{.8\textwidth}{!}{\input{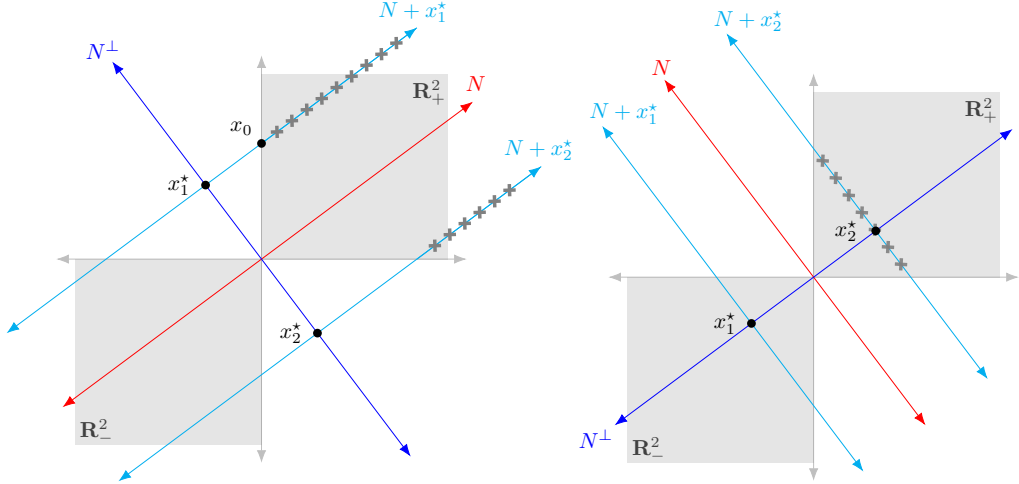}}
    \floatfoot{Left panel: $N$ and $N^\perp$ are such that requirement (ii) in \eqref{eq:geo3} holds regardless of $x^\star$ -- e.g., both $x_1^\star +N$ and $x_2^\star + N$ intersect $\mathbf R^2_+$ with the intersection highlighted with ``$+$" signs. Right panel: $N$ and $N^\perp$ are such that requirement (ii) in \eqref{eq:geo3} holds if and only if $x^\star \in \mathbf R^2_+$.}
\end{figure}

Figure \ref{fig:illR2} illustrates these concepts in the simplest informative setting of $p = 1$ and $d = 2$, in which case $N$ and $N^\perp$ are of dimension one and correspond to a rotation of the coordinate axes.
To develop intuition for requirement (ii) in \eqref{eq:geo3} suppose that $\beta \in R$ so that $Ax^\star = \beta$.
The left panel of Figure \ref{fig:illR2} displays a setting in which condition (ii) holds and an $x \geq 0$ satisfying $Ax = \beta$ may be found even though $x^\star \notin \mathbf R^2_+$ -- e.g., starting from $x_1^\star$ we may move along $N$ until intersecting $\mathbf R^2_+$ at $x_0$. 
In fact, in the left panel of Figure \ref{fig:illR2}, $N$ and $N^\perp$ are such that requirement (ii) in \eqref{eq:geo3} holds regardless of the value of $x^\star$ -- e.g., both $x_1^\star +N$ and $x_2^\star + N$ intersect $\mathbf R^2_+$ with the intersection highlighted with ``$+$" signs.
In contrast, the right panel of Figure \ref{fig:illR2} displays a scenario in which $N$ and $N^\perp$ are such that whether $x^\star +N$ intersects $\mathbf R^2_+$ depends on $x^\star$ -- e.g., $x_2^\star + N$ intersects $\mathbf R^2_+$ while $x^\star_1 + N$ fails to do so.
In fact, in the right panel, condition (ii) in \eqref{eq:geo3}  is satisfied if and only if $x^\star \in \mathbf R^2_+$.

This discussion shows that whether condition (ii) in \eqref{eq:geo3} is satisfied can depend delicately on the orientation of $N$ and $N^\perp$ in $\mathbf R^d$ and the position of $x^\star$ in $N^\perp$.
Our next result provides a tractable geometric characterization of this relationship.

\begin{theorem}\label{th:ineq}
For any $\beta\in \mathbf R^p$ there exists an $x \geq 0$ satisfying $Ax = \beta$ if and only if $\beta \in R$ and $\langle s, x^\star \rangle\leq 0$ for all $s\in N^\perp \cap \mathbf R^d_-$.
\end{theorem}

\begin{figure}[t!]
    \caption{\label{fig:illR3} Illustration of Theorem \ref{th:ineq}}
    \leftskip -5pt
    \resizebox{.8\textwidth}{!}{\centering \input{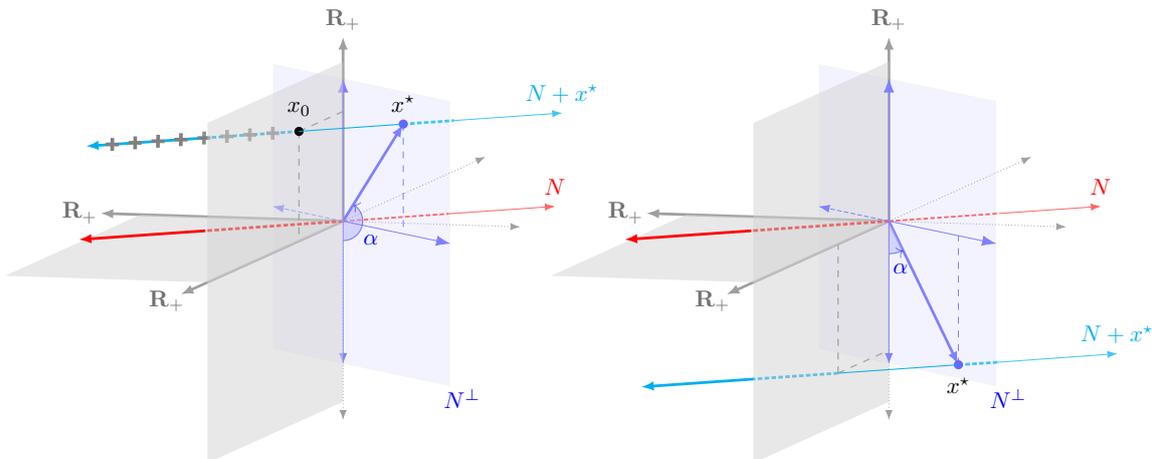}}
    \floatfoot{In this example $N = \{x \in \mathbf R^3 : x = (\lambda,\lambda,0)^\prime \text{ some } \lambda \in \mathbf R\}$ and $N^\perp \cap \mathbf R^3_{-} = \{x \in \mathbf R^3 : x=(0,0,\lambda)^\prime \text{ for some } \lambda \leq 0\}$. $\alpha$ denotes angle between $x^\star$ and $N^\perp \cap \mathbf R^3_{-}$. Left panel: $\alpha$ is obtuse and $x^\star + N$ intersects $\mathbf R^3_+$ with the intersection highlighted with ``$+$" signs. Right panel: $\alpha$ is acute and $x^\star + N$ fails to intersect $\mathbf R^3_+$.}
\end{figure}

Theorem \ref{th:ineq} shows that $\beta = Ax$ for some $x \geq 0$ if and only if $\beta\in R$ and the angle between $x^\star$ and any $s \in N^\perp \cap \mathbf R^d_{-}$ is obtuse.
It is straightforward to verify this relationship in Figure \ref{fig:illR2}.
However, Theorem \ref{th:ineq} is better appreciated in $\mathbf R^3$.
Figure \ref{fig:illR3} illustrates a setting in which $N = \{x\in \mathbf R^3 : x = (\lambda,\lambda,0)^\prime \text{ for some } \lambda \in \mathbf R\}$.
In this case, $x^\star + N$ intersects $\mathbf R^3_+$ if and only if the third coordinate of $x^\star$ is (weakly) positive -- e.g., in the left panel of Figure \ref{fig:illR3} we may move along $N$ until intersecting $\mathbf R^3_+$ at $x_0$.
However, in this illustration $N^\perp \cap \mathbf R^3_{-} = \{x\in \mathbf R^3_{-}:x = (0,0,\lambda)^\prime \text{ for some } \lambda \leq 0\}$ and hence the requirement that the third coordinate of $x^\star$ be positive is also equivalent to the requirement that the angle between $x^\star$ and $N^\perp \cap \mathbf R^3_{-}$ be obtuse.
In Figure \ref{fig:illR3} the angle between $x^\star$ and $N^\perp \cap \mathbf R^3_{-}$ is denoted by $\alpha$ and is obtuse in the left panel (where $(N + x^\star)\cap \mathbf R^3_{+}\neq \emptyset$) and acute in the right panel (where $(N+x^\star)\cap \mathbf R^3_+ = \emptyset$).

\begin{remark}\label{rm:lp} \rm
Any linear program can be written in the standard form
\begin{equation}\label{rm:lp1}
\min_{x\in \mathbf R^d} \langle c, x \rangle \text{ s.t. } Ax = \beta \text{ and } x \geq 0
\end{equation}
for some $c\in \mathbf R^d$, $\beta \in \mathbf R^p$, and $p\times d$ matrix $A$ \citep{luenberger1984linear}.
Representation \eqref{rm:lp1} allows us to conduct inference on the value of a linear program with known $A$ and $c$, and $\beta$ potentially depending on the distribution of the data. This connection was employed in our discussion of some of the examples in Section \ref{sec:examples}, where we mapped the original linear programming formulations employed by the papers cited therein into the hypothesis testing problem in \eqref{eq:null}. \qed
\end{remark}

\begin{remark}\label{rm:slack} \rm
The statement of Theorem \ref{th:ineq} continues to hold if we replace $\mathbf R^{d}_+$ with any closed convex cone $C$ and $\mathbf R^{d}_{-}$ with the polar cone of $C$.
Since Theorem \ref{th:ineq} suffices for addressing all the examples in Section \ref{sec:examples}, however, we do not state such an extension formally.
We further note that restrictions such as $B x \leq 0$ (or $B\beta(P) \leq 0$) for a known matrix $B$ may be incorporated into \eqref{eq:null} by adding the restriction $Bx + y = 0$ (or $BAx + y =0$) for some $y \geq 0$ and appropriately redefining $\beta(P)$, $A$, and $x$ in \eqref{eq:null}. \qed
\end{remark}

\section{The Test} \label{sec:test}

The results in Section \ref{sec:geo} imply that the null hypothesis in \eqref{eq:null} holds if and only if
$$\beta(P) \in R \hspace{0.4 in} \text{ and } \hspace{0.4 in}\langle s,x^\star(P)\rangle \leq0 \text{ for all } s\in N^\perp \cap \mathbf R^d_{-},$$
where  $x^\star(P) \in N^\perp$ satisfies $A(x^\star(P)) = \Pi_R(\beta(P))$.
Based on this characterization, we next develop a test that is computationally feasible in high dimensions.

\subsection{The Test Statistic}

In what follows, we let $A^\dagger $ denote the Moore-Penrose pseudoinverse of $A$, which is a $d\times p$ matrix implicitly defined for any $b\in \mathbf R^p$ through the optimization problem
\begin{equation*}
A^\dagger b \equiv \arg\min_{x \in \mathbf R^d} \|x\|_2^2 \text{ s.t. } x\in \arg\min_{\tilde x \in \mathbf R^d} \|A\tilde x - b\|_2^2;
\end{equation*}
i.e.,\ $A^\dagger b$ is the minimum norm minimizer of $\|Ax - b\|_2$.
Importantly, $A^\dagger b$ is well defined even if there is no $x\in \mathbf R^d$ satisfying $Ax = b$ or the solution is not unique.
It is useful to note that $A^\dagger b$ is the unique element in $N^\perp$ satisfying $A(A^\dagger b) = \Pi_R(b)$, and to thus interpret $A^\dagger$ as a map from $\mathbf R^p$ onto $N^\perp$ \citep{luenberger:1969}.

To build our test statistic, we assume that there is an estimator $\hat \beta_n$ of $\beta(P)$ that is constructed from an i.i.d.\ sample $\{Z_i\}_{i=1}^n$ with $Z_i\in \mathbf Z$ distributed according to $P\in \mathbf P$.
Since $\beta(P) \in R$ under the null hypothesis, Lemma \ref{lm:trivial} implies
\begin{equation}\label{eq:tstat2}
x^\star(P) = A^\dagger \beta(P)
\end{equation}
for any $P\in \mathbf P_0$, which suggests a sample analogue estimator for $x^\star(P)$.
However, when $d < p$, the existence of a solution to the equation $A x = \beta(P)$ locally overidentifies the model in the sense of \cite{chen2018overidentification}, so that a sample analogue estimator may be inefficient.
We therefore instead estimate $x^{\star}(P)$ by
\begin{equation}\label{eq:tstat3}
\hat x^\star_n = A^\dagger \hat C_n \hat \beta_n,
\end{equation}
where $\hat C_n$ is a $p\times p$ matrix satisfying $\hat C_n \beta(P) = \beta(P)$ whenever $P\in \mathbf P_0$.
For example, the sample analogue estimator based on \eqref{eq:tstat2} corresponds to setting $\hat C_n = I_p$ for $I_p$ the $p\times p$ identity matrix.
More generally, the specification in \eqref{eq:tstat3} also accommodates a variety of minimum distance estimators. 

The estimators $\hat \beta_n$ and $\hat x_n^\star$ readily allow us to devise a test based on
Theorem \ref{th:ineq}.
First, note that since the range of $A^\dagger$ equals $N^\perp$, the condition $\langle s, x^\star(P)\rangle \leq 0$ for all $s\in N^\perp \cap \mathbf R^d_{-}$ is equivalent to
\begin{equation}\label{eq:tstat4}
\langle A^\dagger s, x^\star(P)\rangle \leq 0 \text{ for all } s\in \mathbf R^p \text{ s.t. } A^\dagger s \leq 0 \text{ (in $\mathbf R^d$)}.
\end{equation}
To detect violations of condition \eqref{eq:tstat4}, we introduce the statistic
\begin{align}
    &\sup_{s\in \hat {\mathcal V}_n^{\rm i}} \sqrt n \langle A^\dagger s, \hat x_n^\star \rangle  = \sup_{s\in \hat {\mathcal V}_n^{\rm i}} \sqrt n \langle A^\dagger s, A^\dagger \hat C_n \hat \beta_n \rangle \label{eq:tstat5} \\
    \text{where}\qquad &\hat {\mathcal V}_n^{\rm i}  \equiv \{s \in \mathbf R^p : A^\dagger s \leq 0 \text{ and } \|\hat \Omega_n^{\rm i}(AA^\prime)^\dagger s\|_1 \leq 1\}.\label{eq:tstat6}
\end{align}
Here, $\hat \Omega_n^{\rm i}$ is a $p\times p$ symmetric matrix and the ``{\rm i}" superscript alludes to the relation to the ``inequality" condition in Theorem \ref{th:ineq} (i.e.,\ \eqref{eq:tstat4}).
The inclusion of a norm constraint in $\hat {\mathcal V}_n^{\rm i}$ ensures the statistic in \eqref{eq:tstat5} is not infinite with positive probability.
The introduction of $\hat \Omega_n^{\rm i}$ in \eqref{eq:tstat6} provides flexibility in the family of test statistics we examine.
In our simulations, we set $\hat \Omega_n^{\rm i}$ to equal an estimator of the asymptotic standard deviation of $\sqrt n A\hat x_n^\star$, which ensures that \eqref{eq:tstat5} is scale-invariant. 

By Theorem \ref{th:ineq}, any $P\in \mathbf P_0$ must satisfy $\beta(P) \in R$ in addition to \eqref{eq:tstat4}.
To detect violations of this second requirement, we introduce the statistic
\begin{align}
    &\sup_{s\in \hat {\mathcal V}_n^{\rm e}}\sqrt n  \langle s, \hat \beta_n - A\hat x_n^\star \rangle = \sup_{s\in \hat {\mathcal V}_n^{\rm e}} \sqrt n \langle s, (I_p - AA^\dagger \hat C_n)\hat \beta_n \rangle\label{eq:tstat7} \\
    \text{where}
    \qquad
    &\hat {\mathcal V}_n^{\rm e}  \equiv \{s \in \mathbf R^p : \|\hat \Omega_n^{\rm e} s\|_1 \leq 1\}. \notag
\end{align}
Here, $\hat \Omega_n^{\rm e}$ is a $p\times p$ symmetric matrix and the ``{\rm e}" superscript alludes to the relation to the ``equality" condition in Theorem \ref{th:ineq} (i.e.,\ $\beta(P)\in R$).
In particular, note that if $\hat \Omega_n^{\rm e} = I_p$, then \eqref{eq:tstat7} equals $\sqrt n \|\hat \beta_n - A\hat x_n^\star\|_\infty$.
Alternatively, setting $\hat \Omega_n^{\rm e}$ to be an estimate of the asymptotic standard deviation of $\hat \beta_n$ ensures that the statistic in \eqref{eq:tstat7} is scale-invariant.
In applications in which $d\geq p$ and $A$ is full rank, the requirement $\beta(P) \in R$ is automatically satisfied and \eqref{eq:tstat7} is identically zero.

For our test statistic $T_n$, we use the maximum of the statistics in \eqref{eq:tstat5} and \eqref{eq:tstat7}:
\begin{equation}\label{eq:Tndef}
T_n  \equiv \max\{\sup_{s\in \hat{\mathcal V}_n^{\rm e}} \sqrt n \langle s, \hat \beta_n - A\hat x_n^\star\rangle, \sup_{s\in \hat {\mathcal V}_n^{\rm i}}\sqrt n  \langle A^\dagger s, \hat x_n^\star \rangle \} ,
\end{equation}
which can be computed through linear programming.
We do not consider weighting the statistics \eqref{eq:tstat5} and \eqref{eq:tstat7} when taking the maximum because weighting them is numerically equivalent to scaling $\hat \Omega_n^{\rm i}$ and $\hat \Omega_n^{\rm e}$.
A variety of alternative test statistics can of course be motivated by Theorem \ref{th:ineq}. 
A couple of remarks are therefore in order as to why our interest on high-dimensional applications has led us to employing $T_n$.
Focusing on \eqref{eq:tstat5} for conciseness, note that it is a special case of
\begin{equation}\label{eq:stud1}
\sup_{s\in \mathbf R^p} \sqrt n \langle A^\dagger s, \hat x_n^\star\rangle \text{ s.t. } A^\dagger s \leq 0 \text{ and } \hat \omega(s) \leq 1 ,
\end{equation}
where $\hat \omega$ is a convex weight function satisfying $\hat \omega(s) = \hat \omega(-s)$, $\hat \omega(s) \geq 0$, and $\hat \omega(\gamma s) = \gamma \hat \omega(s)$ for any $\gamma \geq 0$ -- e.g., to recover \eqref{eq:tstat5} set $\hat \omega(s) = \|\hat \Omega_n^{\rm i}(AA^\prime)^\dagger s\|_1$.
The linearity of the objective and the homogeneity of $\hat \omega$ imply that \eqref{eq:stud1} in fact equals
\begin{equation}\label{eq:stud2}
\max\{0, \sup_{s\in \mathbf R^p}\frac{\sqrt n\langle A^\dagger s, \hat x_n^\star \rangle}{\hat \omega(s)} \text{ s.t. } A^\dagger s \leq 0 \text{ and } \hat \omega(s) > 0\}.
\end{equation}
Representation \eqref{eq:stud2} shows that \eqref{eq:stud1} implicitly weights each term $\sqrt n \langle A^\dagger s,\hat x_n^\star\rangle$ while remaining computationally tractable -- i.e., \eqref{eq:stud1} can be computed by convex programming, while \eqref{eq:stud2} cannot.
For instance, if we set $\hat \omega(s) = \|\hat \Omega_n^{\rm i}(AA^\prime)^\dagger s\|_2$ with $\hat \Omega_n^{\rm i}$ the sample standard deviation of $\sqrt n A \hat x_n^\star$, then $\hat \omega(s) = (\widehat{\text{Var}}\{\sqrt n \langle A^\dagger s, \hat x_n^\star\rangle\})^{1/2}$ and by \eqref{eq:stud2} the statistic in \eqref{eq:stud1} implicitly studentizes.
In \eqref{eq:tstat5}, we instead use the weighting $\hat \omega(s)= \|\hat \Omega_n^{\rm i}(AA^\prime)^\dagger s\|_1$ because:
(i) It ensures \eqref{eq:tstat5} is a linear program, which scales better than a quadratically-constrained program; and (ii) Using a $\|\cdot\|_1$-constraint allows us to obtain distributional approximations using coupling arguments under $\|\cdot\|_\infty$, which are available under weaker conditions on $p$ than under $\|\cdot\|_2$.
Nonetheless, we emphasize that in certain applications, a researcher may prefer to use weighting functions such as $\hat \omega(s) = \|\hat \Omega_n^{\rm i}(AA^\dagger)s\|_2$ instead.
We expect that, under suitable restrictions, a version of our test that simply replaces $\|\hat\Omega_n^{\rm i}(AA^\prime)^\dagger s\|_1$ with the desired $\hat \omega(s)$ everywhere will be asymptotically valid.

\subsection{The Distribution}

We next introduce assumptions that enable us to approximate the distribution of $T_n$.
Unless otherwise stated, all quantities are allowed to depend on $n$. 

\begin{assumption}\label{ass:weights}
For $\rm j \in \{\rm e,\rm i\}$: (i) $\hat \Omega_n^{\text{\rm j}}$ is symmetric;
(ii) There is a symmetric matrix $\Omega^{\text{\rm j}}(P)$ satisfying $\|(\Omega^{\text{\rm j}}(P))^\dagger(\hat \Omega_n^{\rm j} - \Omega^{\rm j}(P))\|_{o,\infty}  = O_P(a_n/\sqrt{\log(1+p)})$ uniformly in $P\in \mathbf P$;
(iii) $\text{\rm range}\{\hat \Omega_n^{\rm j}\} = \text{\rm range}\{\Omega^{\rm j}(P)\}$ with probability tending to one uniformly in $P\in \mathbf P$.
\end{assumption}

\begin{assumption}\label{ass:beta}
(i) $\{Z_i\}_{i=1}^n$ are i.i.d.\ with $Z_i\in \mathbf Z$ distributed according to $P\in \mathbf P$;
(ii) $\hat x_n^\star = A^\dagger \hat C_n \hat \beta_n$ for some $p\times p$ matrix $\hat C_n$ satisfying $\hat C_n \beta(P) = \beta(P)$ for all $P\in \mathbf P_0$;
(iii) There are $\psi^{\text{\rm i}}(\cdot,P) : \mathbf Z \to \mathbf R^p$ and $\psi^{\text{\rm e}}(\cdot,P) : \mathbf Z \to \mathbf R^p$ satisfying uniformly in $P\in \mathbf P$
\begin{align*}
\|(\Omega^{\text{\rm e}}(P))^\dagger\{(I_p - AA^\dagger\hat C_n)\sqrt n\{\hat \beta_n - \beta(P)\} - \frac{1}{\sqrt n}\sum_{i=1}^n \psi^{\text{\rm e}}(Z_i,P)\}\|_\infty = O_P(a_n)\\
\|(\Omega^{\text{\rm i}}(P))^\dagger \{AA^\dagger \hat C_n \sqrt n\{\hat \beta_n - \beta(P)\} - \frac{1}{\sqrt n}\sum_{i=1}^n \psi^{\text{\rm i}}(Z_i,P)\}\|_\infty = O_P(a_n).
\end{align*}
\end{assumption}

\begin{assumption}\label{ass:moments}
For $\Sigma^{\rm j}(P)\equiv E_P[\psi^{\rm j}(Z,P)\psi^{\rm j}(Z,P)^\prime]$:
(i) $E_P[\psi^{\rm j}(Z,P)] = 0$ for all $P\in \mathbf P$ and $\rm j \in \{\rm e, \rm i\}$;
(ii) The eigenvalues of $(\Omega^{\rm j}(P))^\dagger \Sigma^{\rm j}(P) (\Omega^{\rm j}(P))^\dagger $ are bounded in $\rm j \in \{\rm e, \rm i\}$, $n$, and $P\in \mathbf P$;
(iii) $\Psi(z,P) \equiv \|(\Omega^{\rm e}(P))^\dagger\psi^{\rm e}(z,P)\|_\infty \vee \|(\Omega^{\rm i}(P))^\dagger\psi^{\rm i}(z,P)\|_\infty$ satisfies $\sup_{P\in \mathbf P} \|\Psi(\cdot,P)\|_{P,3} \leq M_{3,\Psi} < \infty$ with $M_{3,\Psi} \geq 1$.
\end{assumption}

\begin{assumption}\label{ass:techrange}
For $\rm j \in \{\rm e,\rm i\}$: (i) $\psi^{\rm j}(Z,P) \in \text{\rm range}\{\Omega^{\rm j}(P)\}$ $P$-almost surely for all $P\in \mathbf P$;
(ii) $(I_p-AA^\dagger \hat C_n )\{\hat \beta_n - \beta(P)\} \in \text{\rm range}\{\Sigma^{\rm e}(P)\}$ and $AA^\dagger \hat C_n \{\hat \beta_n - \beta(P)\} \in \text{\rm range}\{\Sigma^{\rm i}(P)\}$ with probability tending to one uniformly in $P\in \mathbf P$.
\end{assumption}

Because $AA^\dagger \hat C_n $ is often a projection matrix, the relevant asymptotic covariance matrices can be singular.
In order to allow $\hat \Omega_n^{\rm i}$ and $\hat \Omega_n^{\rm e}$ to be sample standard deviation matrices, Assumption \ref{ass:weights} therefore does not assume invertibility.
Instead, Assumption \ref{ass:weights}(ii) requires a suitable form of consistency, and its rate is denoted by $a_n/\sqrt{\log(1+p)}$. Typically $a_n$ will be of order $p/\sqrt n$ (up to logs).
Assumption \ref{ass:weights}(iii) is easily verified when $\hat \Omega^{\rm e}_n$ and $\hat \Omega^{\rm i}_n$ are invertible (e.g., diagonal) or sample standard deviation matrices.
Assumptions \ref{ass:beta}(i)-(ii) formalize previously discussed conditions, while Assumption \ref{ass:beta}(iii) requires our estimators to be asymptotically linear with influence functions whose moments are disciplined by
Assumption \ref{ass:moments}.
Finally, Assumption \ref{ass:techrange}(i), together with Assumption \ref{ass:weights}(iii), restricts the manner in which invertibility of $\hat \Omega_n^{\rm e}$ and $\hat \Omega_n^{\rm i}$ may fail -- this condition is again easily verified if we employ invertible weights or sample standard deviation matrices.
Assumption \ref{ass:techrange}(ii) ensures that the supports of our estimators are contained in the supports of their Gaussian approximations.

Before establishing our distributional approximation to $T_n$, we introduce a final piece of notation.
We denote the population analogues to $\hat {\mathcal V}^{\rm e}_n$ and $\hat {\mathcal V}_n^{\rm i}$ by
\begin{align}\label{eq:Vdef}
\mathcal V^{\rm e}(P) & \equiv \{s \in \mathbf R^p : \|\Omega^{\rm e}(P)s\|_1 \leq 1\} \notag \\
\mathcal V^{\rm i}(P) & \equiv  \{s \in \mathbf R^p : A^\dagger s \leq 0 \text{ and } \| \Omega^{\rm i}(P)(AA^\prime)^\dagger s\|_1 \leq 1\},
\end{align}
and for $\psi^{\rm e}(Z,P)$ and $\psi^{\rm i}(Z,P)$ the influence functions in Assumption \ref{ass:beta}(iii) we set $\psi(Z,P) \equiv (\psi^{\rm e}(Z,P)^\prime, \psi^{\rm i}(Z,P)^\prime)^\prime$ and denote its variance matrix by
\begin{equation}\label{eq:Sigmadef}
\Sigma(P) \equiv E_P[\psi(Z,P)\psi(Z,P)^\prime],
\end{equation}
which has dimension $2p\times 2p$.
For notational simplicity we also define the rate
\begin{equation}\label{eq:cndef}
r_n \equiv M_{3,\Psi}(\frac{ p^2 \log^5(1+p)}{n})^{1/6}  + a_n.
\end{equation}

\noindent Our next theorem gives a distributional approximation for $T_n$ that, under appropriate moment conditions, is valid uniformly in $P\in \mathbf P_0$ provided $p^2\log^5(p)/n = o(1)$.

\begin{theorem}\label{th:dist}
Let Assumptions  \ref{ass:weights}, \ref{ass:beta}, \ref{ass:moments}, \ref{ass:techrange} hold, and $r_n = o(1)$.
Then, there is $(\mathbb G_n^{\text{\rm e}}(P)^\prime,\mathbb G_n^{\text{\rm i}}(P)^\prime)^\prime \equiv \mathbb G_n(P) \sim N(0, \Sigma(P))$ such that uniformly in $P\in \mathbf P_0$
$$ T_n = \max\{\sup_{s\in {\mathcal V}^{\rm e}(P)} \langle s, \mathbb G_n^{\rm e}(P)\rangle , \sup_{s\in \mathcal V^{\rm i}(P)} \langle A^\dagger s, A^\dagger \mathbb G_n^{\text{\rm i}}(P)\rangle + \sqrt n \langle A^\dagger s,A^\dagger \beta(P) \rangle\} + O_P(r_n).$$
\end{theorem}

The asymptotic approximation in Theorem \ref{th:dist} depends on linear programs whose solutions must be attained at one of a finite number of extreme points.
It follows that $T_n$ is asymptotically equivalent to the maximum of a Gaussian vector -- an observation that suggests a connection to the high dimensional central limit theorem of \cite{chernozhukov2019improved}.
The proof of Theorem \ref{th:dist}, however, does not rely on \cite{chernozhukov2019improved} because the number of extreme points depends on $A$ in a non-transparent way and upper bounds, such as that in \cite{mcmullen1970maximum}, are exponential in $p$.
Nonetheless, we note that for certain $A$ and $\hat \beta_n$, \cite{chernozhukov2019improved} may yield better coupling rates than Theorem \ref{th:dist} and allow $p$ to be larger than $n$.
On the other hand, we should not expect such conditions to apply when $\hat \beta_n$ is a vector of empirical probabilities as in Examples \ref{ex:TE}-\ref{ex:reveal} -- a setting we expect to at least require $p/n = o(1)$.

\subsection{The Critical Value}\label{sec:critical-value}

To obtain a critical value, we assume the availability of ``bootstrap" estimates $(\hat {\mathbb G}_n^{\rm e \prime}, \hat {\mathbb G}_n^{\rm i\prime})^\prime$ for the distribution of $(\mathbb G_n^{\rm e}(P)^\prime,\mathbb G_n^{\rm i}(P)^\prime)^\prime$.
Given such estimates, we may follow a number of approaches for obtaining critical values; see, e.g., Section \ref{sec:altcrit}.
Below we focus on an approach that has favorable power properties in simulations.

\noindent {\bf Step 1.}
First, we observe that the main challenge in employing Theorem \ref{th:dist} for inference is the presence of the nuisance function $f(\cdot,P): \mathbf R^p \to \mathbf R$ given by
\begin{equation}\label{step1:eq1}
f(s,P) \equiv \sqrt n \langle A^\dagger s, A^\dagger \beta(P)\rangle .
\end{equation}
While $f(\cdot,P)$ cannot be consistently estimated, we can construct a suitable upper bound for it.
To this end, we note that in applications some coordinates of $\beta(P)$ may equal a known value for all $P\in \mathbf P_0$; see, e.g., Examples \ref{ex:dyn}-\ref{ex:reveal}.
We therefore decompose $\beta(P) = (\beta_{\rm u}(P)^\prime, \beta_{\rm k}^\prime)^\prime$ where $\beta_{\rm k}$ is a known constant for all $P\in \mathbf P_0$, and similarly decompose any $b\in \mathbf R^p$ into subvectors of conformable dimensions $b = (b_{\rm u}^\prime,b_{\rm k}^\prime)^\prime$.
We use definitions to define a restricted estimator for $\beta(P)$ by setting
\begin{equation}\label{step1:eq2}
\hat \beta_n^{\rm r} \in \arg\min_{b=(b_{\rm u}^\prime, b_{\rm k}^\prime)^\prime} \sup_{s\in \hat {\mathcal V}_n^{\rm i}}  |\langle A^\dagger s, \hat x_n^\star - A^\dagger b\rangle| \text{ s.t. } b_{\rm k} = \beta_{\rm k}, ~Ax = b \text{ for some } x\geq 0,
\end{equation}
which may be computed through linear programming; see Appendix \ref{sec:computational-details}. 
Since $f(s,P) \leq 0$ for all $s\in \hat{\mathcal V}_n^{\rm i}$ and $P\in \mathbf P_0$ by Theorem \ref{th:ineq}, it follows that under the null hypothesis $\lambda_n f(s,P) \geq f(s,P)$ for any $\lambda_n \leq 1$ and $s\in \hat {\mathcal V}_n^{\rm i}$.
We therefore set
\begin{equation}\label{step1:eq3}
\hat{\mathbb U}_n(s) \equiv  \lambda_n\sqrt n\langle A^\dagger s,A^\dagger \hat \beta_n^{\rm r}\rangle ,
\end{equation}
which is a consistent estimator for the upper bound $\lambda_n f(s,P)$ provided $\lambda_n \downarrow 0$ at a suitable rate -- we discuss choices of $\lambda_n$ in Section \ref{sec:simulations}.
The upper bound $\hat {\mathbb U}_n$ reflects the structure of the null hypothesis in that: (i) $\hat {\mathbb U}_n(s) \leq 0$ for all $s\in \hat {\mathcal V}_n^{\rm i}$ and (ii) There is a $b \in \mathbf R^p$ satisfying $Ax = b$ for some $x\geq 0$ such that $\hat {\mathbb U}_n(s) =  \langle A^\dagger s, A^\dagger b\rangle$ for all $s\in \mathbf R^p$. \qed 

\noindent {\bf Step 2.} Next, we note that the asymptotic approximation obtained in Theorem \ref{th:dist} is increasing (in a first-order stochastic dominance sense) in the nuisance function $f(\cdot,P)$ (under the pointwise partial order). 
Hence, given the upper bound $\hat {\mathbb U}_n$ defined in Step 1, for a nominal level $\alpha$ test, we may use the bootstrap quantile
\begin{equation*}
\hat c_n(1-\alpha)\equiv \inf\{ u : P(\max\{\sup_{s\in \hat {\mathcal V}_n^{\rm e}} \langle s, \hat {\mathbb G}_n^{\rm e}\rangle, \sup_{s\in \hat {\mathcal V}_n^{\rm i}} \langle A^\dagger s, A^\dagger \hat{\mathbb G}_n^{\rm i}\rangle + \hat{\mathbb U}_n(s)\} \leq u |\{Z_i\}_{i=1}^n) \geq 1-\alpha\}
\end{equation*}
as a critical value for $T_n$.
Computing $\hat c_n(1-\alpha)$ is straightforward as it only requires solving one linear program per bootstrap replication.
We also note that because $0\in \mathcal V^{\rm i}(P)$, any $s \in \mathcal V^{\rm i}(P)$ for which $ \sqrt n \langle A^\dagger s,A^\dagger \beta(P)\rangle$ tends to minus infinity plays an asymptotically negligible role in the distributional approximation of Theorem \ref{th:dist}.
Our critical value reflects this structure because $\hat {\mathbb U}_n(s)$ and $\lambda_n \sqrt n\langle A^\dagger s, A^\dagger \beta(P)\rangle$ are asymptotically equivalent, and thus any $s$ for which $\lambda_n \sqrt n\langle A^\dagger s, A^\dagger \beta(P)\rangle$ tends to minus infinity plays an asymptotic negligible role in determining $\hat c_n(1-\alpha)$. E.g., in an asymptotic setting in which $P$ is fixed and $\lambda_n \sqrt n\to \infty$, any $s$ satisfying $\langle A^\dagger s, A^\dagger \beta(P)\rangle < 0$ plays a negligible role in both the distribution of $T_n$ and our bootstrap approximation. \qed

Given the above definitions, we finally define our test $\phi_n\in \{0,1\}$ to equal
\begin{equation*}
\phi_n \equiv 1\{T_n > \hat c_n(1-\alpha)\};
\end{equation*}
i.e.,\ we reject the null hypothesis whenever $T_n$ exceeds $\hat c_n(1-\alpha)$.
To establish the asymptotic validity of this test, we impose an additional assumption that enables us to derive the asymptotic properties of the bootstrap estimates $(\hat {\mathbb G}_n^{\rm e \prime},\hat{\mathbb G}_n^{\rm i \prime})^\prime$.

\begin{assumption}\label{ass:4bootnew}
(i) There are exchangeable $\{W_{i,n}\}_{i=1}^n$ independent of $\{Z_i\}_{i=1}^n$ with
$$\| (\Omega^{\rm j}(P))^\dagger\{\hat {\mathbb G}^{\rm j}_n - \frac{1}{\sqrt n} \sum_{i=1}^n (W_{i,n} - \bar W_n)\psi^{\rm j}(Z_i,P)\}\|_\infty = O_P(a_n)$$
uniformly in $P\in \mathbf P$ for $\rm j \in \{\rm e, \rm i\}$;
(ii) For some $a,b > 0$, $P(|W_{1,n} - E[W_{1,n}]| > t) \leq 2\exp\{-\frac{t^2}{b + at}\}$ for all $t\in \mathbf R_+$ and $n$;
(iii) $|\sum_{i=1}^n(W_{i,n} - \bar W_n)^2/n -1|= O_P(n^{-1/2})$ and $\sup_n E[|W_{1,n}|^3] < \infty$;
(iv)  $\sup_{P\in \mathbf P} \|\Psi^2(\cdot,P) \|_{P,q} \leq M_{q,\Psi^2} < \infty$ for some $q\in (1,+\infty]$;
(v) For $\rm j\in \{\rm e,\rm i\}$, $\hat {\mathbb G}_n^{\rm j} \in \text{\rm range}\{\Sigma^{\rm j}(P)\}$ with probability tending to one uniformly in $P\in \mathbf P$.
\end{assumption}

Assumption \ref{ass:4bootnew} accommodates a variety of resampling schemes, such as the nonparametric, Bayesian, score, or weighted bootstrap.
In parallel to Assumption \ref{ass:beta}(iii), Assumption \ref{ass:4bootnew}(i) imposes a linearization assumption on our bootstrap estimates that is automatically satisfied whenever $(\hat {\mathbb G}_n^{\rm i\prime},\hat {\mathbb G}_n^{\rm e \prime})^\prime$ is linear in the data.
Assumptions \ref{ass:4bootnew}(ii)(iii) state restrictions on the exchangeable bootstrap weights that are satisfied by commonly used resampling schemes -- e.g., the nonparametric and Bayesian bootstrap, and the score or weighted bootstrap under appropriate choices of weights.
Assumption \ref{ass:4bootnew}(iv) potentially strengthens the moment restrictions in Assumption \ref{ass:moments}(iii) (if $q > 3/2$) and is imposed to sharpen our estimates of the coupling rate for the bootstrap statistics.
Finally, Assumption \ref{ass:4bootnew}(v) is a bootstrap analogue to Assumption \ref{ass:techrange}(ii).

These assumptions suffice for showing that the law of $(\hat{\mathbb G}_n^{\rm i\prime},\hat {\mathbb G}_n^{\rm e \prime})^\prime$ conditional on the data is a suitable estimator of the law of $(\mathbb G_n^{\rm e}(P)^\prime,\mathbb G_n^{\rm i}(P)^\prime)^\prime$.
Formally, we show $(\hat {\mathbb G}_n^{\rm e\prime},\hat {\mathbb G}_n^{\rm i\prime })$ can be coupled (under $\|\cdot\|_\infty$) to a copy of $(\mathbb G_n^{\rm e}(P)^\prime,\mathbb G_n^{\rm i}(P)^\prime)^\prime$ at a rate
\begin{equation*}
b_n \equiv \frac{\sqrt{p\log(1+n)}M_{3,\Psi}}{n^{1/4}} + (\frac{p\log^{5/2}(1+p)M_{3,\Psi}}{\sqrt n})^{1/3} + (\frac{p\log^3(1+p) n^{1/q} M_{q,\Psi^2}}{n})^{1/4}+ a_n;
\end{equation*}
see Lemma \ref{lm:finitebootnew} in the Supplemental Appendix.
In particular, under appropriate moment restrictions, the bootstrap is consistent provided $p^2/n = o(1)$ (up to logs).
The consistency of the exchangeable bootstrap when $p$ grows with $n$ is to our knowledge a novel result that might be of independent interest.

Before establishing the asymptotic validity of our test, we introduce some final pieces of notation.
First, we note that the asymptotic approximation obtained in Theorem \ref{th:dist} contains two linear programs, whose solutions can be shown to belong to the sets
\begin{align*}
\mathcal E^{\rm e}(P) & \equiv \{s \in \mathbf R^p : s \text{ is an extreme point of } \Omega^{\rm e}(P)\mathcal V^{\rm e}(P)\} \\
\mathcal E^{\rm i}(P) & \equiv \{s \in \mathbf R^p : s \text{ is an extreme point of } (AA^\prime)^\dagger \mathcal V^{\rm i}(P)\}.
\end{align*}
For $\rm j \in \{\rm e,\rm i\}$ and $s\in \mathcal E^{\rm j}(P)$, it will also be helpful to define the standard deviations
\begin{align}
\sigma^{\rm e}(s,P) & \equiv \{E_P[(\langle s, (\Omega^{\rm e}(P))^\dagger \mathbb G_n^{\rm e}(P)\rangle)^2]\}^{1/2} \notag \\
\sigma^{\rm i}(s,P) & \equiv \{E_P[(\langle \Omega^{\rm i}(P) s, (\Omega^{\rm i}(P))^\dagger \mathbb G_n^{\rm i}(P)\rangle)^2]\}^{1/2} \notag
\end{align}
and denote their upper and (restricted) lower bounds over $\mathcal E^{\rm e}(P) \cup \mathcal E^{\rm i}(P)$ by
\begin{align*}
\bar \sigma(P) & \equiv \sup_{s \in \mathcal E^{\rm e}(P)}\sigma^{\rm e}(s,P) \vee \sup_{s \in \mathcal E^{\rm i}(P)}\sigma^{\rm i}(s,P)\\
\underline \sigma(P) & \equiv \inf_{s \in \mathcal E^{\rm e}(P) : \sigma^{\rm e}(s,P) > 0} \sigma^{\rm e}(s,P) \wedge \inf_{s \in \mathcal E^{\rm i}(P) : \sigma^{\rm i}(s,P) > 0} \sigma^{\rm i}(s,P),
\end{align*}
where we let $\underline{\sigma}(P) = +\infty$ if $\sigma^{\rm j}(s,P) = 0$ for all $s\in \mathcal E^{\rm j}(P)$, $\rm j \in  \{\rm e,\rm i\}$. For any random variable $V\in \mathbf R$, let $\text{med}\{V\}$ denote its median, and for any $P\in \mathbf P$ define
\begin{equation*}
\text{m}(P) \equiv \text{med}\{\max\{\sup_{s\in \mathcal V^{\rm e}(P)}\langle s, \mathbb G_n^{\rm e}(P)\rangle, \sup_{s\in \mathcal V^{\rm i}(P)} \langle A^\dagger s, A^\dagger \mathbb G_n^{\text{\rm i}}(P)\rangle\}\}.
\end{equation*}
Lastly, we introduce the sequence $\xi_n \equiv r_n \vee b_n\vee \lambda_n\sqrt{\log(1+p)}$.
Our next result establishes the asymptotic validity of the proposed test.

\begin{theorem}\label{th:size}
Let Assumptions
\ref{ass:weights}--\ref{ass:4bootnew}
hold, $\alpha \in (0,0.5)$, and $0\leq \lambda_n\leq 1$. If $\xi_n$ satisfies $\xi_n = o(1)$ and $\sup_{P\in \mathbf P} ({\rm m}(P) +\bar \sigma(P))/\underline{\sigma}^2(P) = o(\xi_n^{-1})$, then
\begin{equation}\label{th:sizedisp}
\limsup_{n\rightarrow \infty}\sup_{P\in \mathbf P_0} E_P[\phi_n] \leq \alpha.
\end{equation}
\end{theorem}

Under additional requirements, it is possible to strengthen the conclusion of Theorem \ref{th:size} to show \eqref{th:sizedisp} holds with equality.
For instance, if in addition $p$ is fixed with $n$ and $\sqrt n\lambda_n \to \infty$, then it is possible to show that $E_P[\phi_n]$ tends to $\alpha$ for any $P$ on the ``boundary" of $\mathbf P_0$ -- a result that, together with Theorem \ref{th:size}, implies the asymptotic size of our test equals $\alpha$.
We also note that Theorem \ref{th:size} imposes a rate condition that constrains how $p$ can grow with $n$.
This rate condition depends on $A$ and the weighting matrices $\Omega^{\rm j}(P)$ for $\rm j\in \{\rm e,\rm i\}$.
As we show in Remark \ref{rm:univbound} below, it is possible to obtain universal (in $A$) bounds for $({\rm m}(P) +\bar \sigma(P))/\underline{\sigma}^2(P)$ when setting $\Omega^{\rm j}(P)$ to be the standard deviation matrix of $\mathbb G_n^{\rm j}(P)$ for $\rm j \in \{\rm e,\rm i\}$.
While such bounds provide sufficient conditions for the rate requirements in Theorem \ref{th:size}, we emphasize that they can be quite conservative for a specific $A$.
Finally, we note that if, as in much of the literature, one considers the case in which $p$ does not grow with $n$, then Remark \ref{rm:univbound} implies that Theorem \ref{th:size} holds under Assumptions \ref{ass:weights}--\ref{ass:4bootnew} and the requirement $\lambda_n = o(1)$.

\begin{remark}\label{rm:univbound} \rm
Whenever $\Omega^{\rm j}(P)$ equals the standard deviation matrix of $\mathbb G_n^{\rm j}(P)$ for $\rm j \in \{\rm e,\rm i\}$, it is possible to obtain universal (in $A$) bounds on $\bar \sigma(P)$, $\underline{\sigma}(P)$, and $\text{m}(P)$.
Under such choice of $\Omega^{\rm j}(P)$, it is straightforward to show $\bar \sigma(P) \leq 1$ by employing the eigen-decomposition of $\Omega^{\rm i}(P)$.
Similar arguments imply
\begin{multline}\label{rm:univbound2}
\min_{s \in \mathcal E^{\rm i}(P) : \sigma^{\rm i}(s,P) > 0} \sigma^{\rm i}(s,P) \\
\geq \inf_{s : \|\Omega^{\rm i}(P)s\|_1 = 1} \{E_P[(\langle \Omega^{\rm i}(P) s, (\Omega^{\rm i}(P))^\dagger \mathbb G_n^{\rm i}(P)\rangle)^2]\}^{1/2} \geq \inf_{s: \|s\|_1 = 1} \|s\|_2 = \frac{1}{\sqrt p}
\end{multline}
and $\underline \sigma(P) \geq 1/\sqrt p$, while a maximal inequality yields ${\rm m}(P) \lesssim \sqrt{\log(1+p)}$.
The universal (in $A$) bound in \eqref{rm:univbound2} can, however, be quite conservative for  specific $A$.\qed
\end{remark}

\subsection{Extensions}

We next discuss extensions to our results.
For conciseness, we omit a formal analysis, but they follow by similar arguments to those employed in Theorem \ref{th:size}.

\subsubsection{Two Stage Critical Value} \label{sec:altcrit}

We have focused on a particular choice of critical value due to its favorable power properties in our simulations.
It is important to note, however, that other approaches are also available.
For instance, an alternative critical value may be obtained by proceeding in a manner that is similar in spirit to the procedure proposed by \cite{romano2014practical} and \cite{bai2019practical} for testing whether a finite-dimensional vector of populations means is nonnegative.
Specifically, for some pre-specified $\gamma \in (0,\alpha)$, define
\begin{equation*}
\hat c^{(1)}_n(1-\gamma) \equiv \inf \{u : P( \sup_{s\in \hat {\mathcal V}_n^{\rm i}} \langle A^\dagger s, - A^\dagger  \hat {\mathbb G}_n^{\text{i}}\rangle \leq u | \{Z_i\}_{i=1}^n) \geq 1- \gamma \},
\end{equation*}
and, in place of $\hat {\mathbb U}_n$ as introduced in \eqref{step1:eq3}, define the upper bound $\tilde {\mathbb U}_n$ to be
\begin{equation*}
\tilde {\mathbb U}_n(s) \equiv  \min\{\sqrt n\langle A^\dagger s,\hat x_n^\star \rangle + \hat c^{(1)}_n(1-\gamma),0\}.
\end{equation*}
The function $\tilde {\mathbb U}_n : \mathbf R^p\to \mathbf R$ may be interpreted as an upper confidence region for $f(\cdot,P)$ (as in \eqref{step1:eq1}) with uniform (in $P\in \mathbf P_0$) asymptotic coverage probability $1-\gamma$.
For a nominal level $\alpha$ test, we may then compare $T_n$ to the critical value
\begin{multline*}
\hat c_n^{(2)}(1-\alpha + \gamma) \equiv \\ \inf \{ u : P( \max\{\sup_{s\in \hat {\mathcal V}^{\rm e}_n} \langle s, \hat{\mathbb G}_n^{\rm e}\rangle , \sup_{s\in \hat{\mathcal V}_n^{\rm i}} \langle A^\dagger s, A^\dagger \hat {\mathbb G}_n^{\text{\rm i}}\rangle + \hat{\mathbb U}_n(s)\} \leq u|\{Z_i\}_{i=1}^n) \geq 1-\alpha + \gamma\}.
\end{multline*}
The $1-\alpha+\gamma$ quantile is used instead of the $1-\alpha$ quantile to account for the possibility that $f(s,P) > \tilde {\mathbb U}_n(s)$ for some $s\in \hat{\mathcal V}_n^{\rm i}$.
The resulting test can be shown to be asymptotically valid under the same conditions imposed in Theorem \ref{th:size}.
An appealing feature of the described approach is that it does not require selecting a ``bandwidth" $\lambda_n$.
However, we find in simulations that the power of the resulting test is lower than that of the test $\phi_n$.
Intuitively, this is due to $\tilde {\mathbb U}_n$ not satisfying $\tilde {\mathbb U}_n(s) = \langle A^\dagger s, A^\dagger b\rangle$ for some $b\in \mathbf R^p$ such that $A x = b$ with $x\geq 0$.
As a result, the upper bound $\tilde {\mathbb U}_n$ does not fully reflect the structure of the null hypothesis.

\subsubsection{Alternative Sampling Frameworks}\label{sec:cond}

While we have focused on i.i.d.\ settings for simplicity, we note that extensions to other asymptotic frameworks are conceptually straightforward.
One interesting such extension is to combine our analysis with the insights in \cite{andrews2019inference} concerning the problem sub-vector inference in a class of models defined by conditional moment inequalities.
In particular, \cite{andrews2019inference} note that in an empirically relevant class of models the parameter of interest $\pi$ satisfies
\begin{equation}\label{eq:alt1}
E_P[G(D,\pi) - M(W,\pi)\delta |W] \leq 0 \text{ for some } \delta \in \mathbf R^{d_\delta}
\end{equation}
where $G(D,\pi) \in \mathbf R^p$, $M(W,\pi)$ is a $p\times d_\delta$ matrix, and both are known functions of $(D,W,\pi)$.
\cite{andrews2019inference} observe that the structure of these models is such that testing whether a specified value $\pi_0$ satisfies \eqref{eq:alt1} is facilitated by conditioning on $\{W_i\}_{i=1}^n$.
As we next argue, their important insight carries over to our framework.

For any $\delta \in \mathbf R^{d_\delta}$, let $\delta^+ \equiv \delta \vee 0$ and $\delta_{-} \equiv -(\delta \wedge 0)$, where $\vee$ and $\wedge$ denote coordinate-wise maximums and minimums. 
We then observe that if $\pi_0$ satisfies \eqref{eq:alt1}, then
\begin{equation*}
\frac{1}{n}\sum_{i=1}^n E_P[G(D,\pi_0)|W_i] = \frac{1}{n}\sum_{i=1}^n M(W_i,\pi_0)(\delta^+ - \delta^{-}) - \Delta \text{ for some } \Delta \in  \mathbf R^p_{+}, ~ \delta \in \mathbf R^{d_\delta}.
\end{equation*}
Hence, by setting $P$ to denote the distribution of $\{D_i\}_{i=1}^n$ conditional on $\{W_i\}_{i=1}^n$, we may test the null hypothesis that $\pi_0$ satisfies \eqref{eq:alt1} by letting
\begin{equation*}
\beta(P) \equiv \frac{1}{n}\sum_{i=1}^n E_P[G(D,\pi_0)|W_i] \hspace{0.3 in} A \equiv [\frac{1}{n}\sum_{i=1}^n M(W_i,\pi_0), ~ -\frac{1}{n}\sum_{i=1}^n M(W_i,\pi_0), ~ -I_p]
\end{equation*}
and testing whether $\beta(P) = Ax$ for some $x \geq 0$ -- note $A$ does not depend on $P$ due to the conditioning on $\{W_i\}_{i=1}^n$.
By letting $\hat \beta_n\equiv \frac{1}{n}\sum_{i=1}^n G(D_i,\pi_0)$,
our test remains largely the same, with the exception that $(\hat {\mathbb G}_n^{\rm e \prime}, \hat {\mathbb G}_n^{\rm i\prime})^\prime$ must be consistent for the law of
\begin{equation*}
((\Omega^{\rm e}(P))^\dagger (I_p - AA^\dagger \hat C_n)\sqrt n\{\hat \beta_n - \beta(P)\}^\prime, (\Omega^{\rm i}(P))^\dagger AA^\dagger \hat C_n\sqrt n\{\hat \beta_n - \beta(P)\}^\prime)^\prime
\end{equation*}
conditional on $\{W_i\}_{i=1}^n$ (instead of unconditionally, as in Theorem \ref{th:size}).

\section{Simulations with a Mixed Logit Model}\label{sec:simulations}

\subsection{The Model}\label{sec:model}

Example \ref{ex:dyn} is an example of a class of mixture models considered by \cite{fox2011simple}.
A simpler example with the same structure is a static, binary choice logit with random coefficients.
In this model, a consumer chooses $Y \in \{0,1\}$ by
\begin{align*}
    Y = 1\left\{C_{0} + C_{1}W - U \geq 0\right\},
\end{align*}
where $W$ is an observed variable which we will think of as the price of buying a good ($Y = 1$), and $V\equiv (C_{0}, C_{1})$ and $U$ are latent variables. 
The unobservable $U$ is assumed to follow a standard logistic distribution, independently of $(V,W)$.

A consumer of type $v = (c_{0}, c_{1})$ facing price $w$ buys the good with probability
\begin{align}
    \label{eq:decision-per-type}
    P(Y = 1 \vert W = w, V = v) = \frac{1}{1 + \exp(-c_{0} - c_{1}w)} \equiv \ell(w, v).
\end{align}
\cite{bajarifoxryan2007aer} and \cite{fox2011simple} assume $V$ is independent of $W$ and approximate the distribution of $V$ using a discrete distribution with known support points $(v_{1},\ldots,v_{d})$ and unknown respective probabilities $x \equiv (x_{1},\ldots,x_{d})$.
Under these assumptions, \eqref{eq:decision-per-type} can be aggregated into a conditional moment equality:
\begin{align}
    \label{eq:conditional-moment}
    P(Y = 1 \vert W = w) = \sum_{j=1}^{d} x_{j}\ell(w, v_{j}).
\end{align}

A natural quantity of interest in this model is the price elasticity of purchase probability.
For a consumer of type $v =(c_0,c_1)$ facing price $\bar{w}$, this is
\begin{align*}
    \epsilon(v, \bar{w})
    \equiv
    \left(
        \frac{\partial}{\partial w}
        \ell(v, w)
        \Big\vert_{w = \bar{w}}
    \right)
    \frac{\bar{w}}{\ell(v, \bar{w})}
    =
    c_{1}\bar{w}(1 - \ell(v, \bar{w})).
\end{align*}
The cumulative distribution function (c.d.f.) of this elasticity is
\begin{align}
    F_{\epsilon}(t \vert \bar{w})
    \equiv
    P(\epsilon(V, \bar{w}) \leq t)
    =
    \sum_{j=1}^{d}
    1\{\epsilon(v_{j},\bar{w}) \leq t\}x_{j}
    \equiv
    a(t, \bar{w})'x,
    \label{eq:target-parameter}
\end{align}
where $a(t, \bar{w}) \equiv (a_{1}(t, \bar{w}),\ldots,a_{d}(t, \bar{w}))'$ with $a_{j}(t, \bar{w}) \equiv 1\{\epsilon(v_{j}, \bar{w}) \leq t\}$.
We take the c.d.f.\ $F_{\epsilon}(\cdot \vert \bar{w})$ as our parameter of interest in the discussion ahead.

\subsection{Data Generating Processes}
\label{sec:dgps}

In our simulations we generate data from a class of mixed logit models parameterized as follows.
The distribution of $W$ is uniform over $p-2$ evenly spaced points between $0$ and $2$, inclusive.
The known support of $C_{0}$ is generated by taking a Sobol sequence of length $\sqrt{d}$ and rescaling it to lie in $[.5, 1.0]$.
Similarly, the support of $C_{1}$ is a Sobol sequence of length $\sqrt{d}$ rescaled to $[-3, 0]$.
The distribution of $V\equiv (C_0,C_1)$ is taken to be uniform over the product of the two marginal supports, so that it has $d$ support points.


\begin{figure}[t!]
    \centering
    \caption{\label{fig:idsets} Bounds on the distribution of price elasticity $F_{\epsilon}(t \vert 1)$}
    \includegraphics[scale = .75]{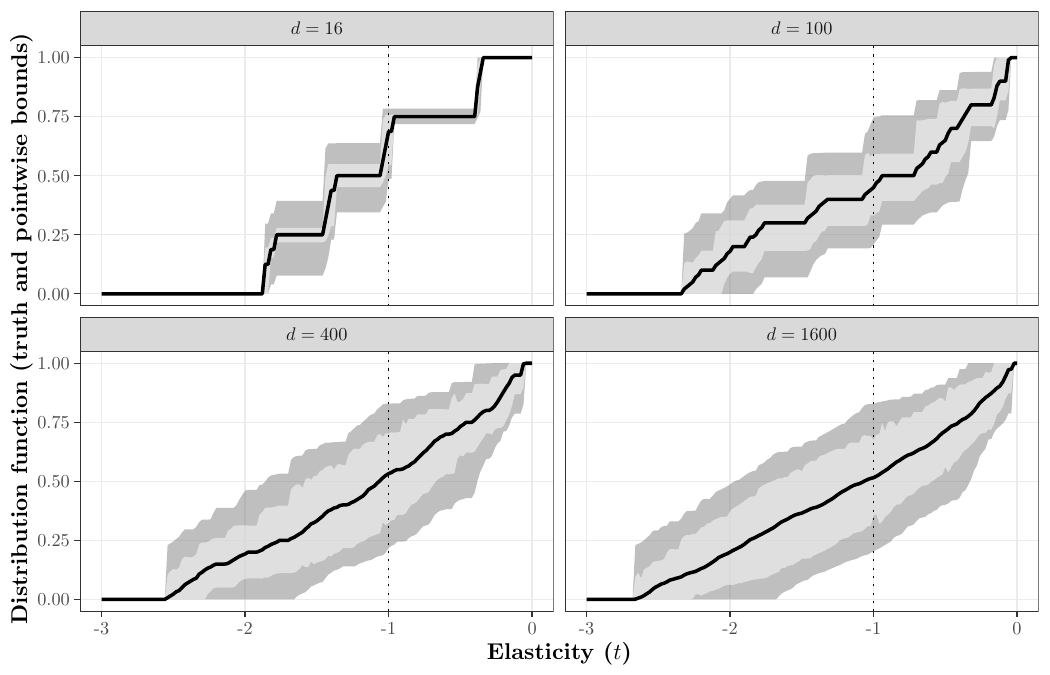}
    \floatfoot{These plots are based on the data generating processes described in Section \ref{sec:dgps}.
        The solid black line is the actual value of $F_{\varepsilon}(t \vert 1)$.
        The lighter color is $\mathbb{A}^{\star}(t,1|P)$ when the support of $W$ has sixteen points.
        The darker color is the same set when the support of $W$ has only four points.
        The dotted vertical is the value $t = -1$ used in the Monte Carlo simulations in Section \ref{sec:monte-carlo}.
    }
\end{figure}

\cite{foxkimryanetal2012joe} provide identification results that apply to the binary mixed logit model. 
However, their conditions require $W$ to be continuously distributed.
When $W$ is discretely distributed, one might expect the distributions of $V$ and thus of $\epsilon(V,\bar{w})$ are only partially identified.
We explore this conjecture computationally.
We denote the identified set for the distribution of $V$ as
\begin{align*}
    \mathbb{X}^{\star}(P)
    \equiv
    \{
        x \in \re^{d}_{+} :
        \sum_{j=1}^{d} x_{j} = 1, ~
        \sum_{j=1}^{d}
        x_{j}\ell(w, v_{j})
        =
        P(Y=1|W=w)
        \text{ for all $w \in \mathcal W$}
    \},
\end{align*}
where $\mathcal W$ is the support of $W$.
In addition, for any $t\in \mathbf R$, we denote the identified set for $F_\epsilon(t |\bar w)$ by $\mathbb A^\star(t,\bar w|P)$, which simply equals the projection of $\mathbb{X}^{\star}(P)$ under the linear map introduced in \eqref{eq:target-parameter}:
\begin{align*}
    \mathbb{A}^{\star}(t,\bar w|P)
    \equiv
    \left\{
        a(t, \bar{w})'x
        :
        x \in \mathbb{X}^{\star}(P)
    \right\}.
\end{align*}
Since $\mathbb{X}^{\star}(P)$ is a system of linear equalities and inequalities, and $x\mapsto a(t, \bar{w})^\prime x$ is scalar-valued and linear, $\mathbb{A}^{\star}(t,\bar w|P)$ is a closed interval \citep[see, e.g.][for a similar argument]{mogstadsantostorgovitsky2018e}.
The left endpoint of this interval is given by
\begin{equation}
    \min_{x \in \re^{d}_{+}}
    a(t, \bar{w})'x
    ~
    \text{s.t.}
    ~
    \sum_{j=1}^{d}x_{j} = 1,
    ~
    \sum_{j=1}^{d}x_{j}\ell(w, v_{j})
    =
   P(Y=1|W=w)
    \text{ for all $w \in \mathcal W$},
    \label{eq:lp-solve-for-bounds-mixedlogit}
\end{equation}
and the right endpoint is equal to its maximization counterpart.

Figure \ref{fig:idsets} depicts $ \mathbb{A}^{\star}(t,\bar w|P)$ as a function of $t$ for $\bar{w} = 1$.
The outer and inner bands depict the identified set when the support of $W$ has four and sixteen points, respectively, while the solid line indicates the distribution under the actual data generating process.
The identified sets are non-trivial and widen with the number of support points $d$ for the unobservable $V$.
For $d = 16$, the bounds when $W$ has sixteen support points are narrow, but numerically distinct from a point.
This is because the system of moment equations that defines $\mathbb{X}^{\star}(P)$, while known to be nonsingular in principle, is sufficiently close to singular to matter numerically.

\subsection{Test Implementation}
\label{sec:test-implementation}


As in Example \ref{ex:dyn}, we may use our results to test whether a hypothesized $\gamma \in \mathbf R$ belongs to the identified set for $F_\epsilon(t|\bar w)$.
Using \eqref{eq:conditional-moment} and recalling $W$ was set to have $p-2$ support points, we may then map such hypothesis into \eqref{eq:null} by setting
\begin{align*}
    \beta(P)
    =
    \begin{pmatrix}
        P(Y=1|W = w_1) \\
        \vdots \\
        P(Y=1|W=w_{p-2}) \\
        1 \\
        \gamma
    \end{pmatrix}
    \quad
    A
    =
    \begin{pmatrix}
        \ell(w_{1},v_{1}) & \cdots & \ell(w_{1}, v_{d}) \\
        \vdots & \vdots & \vdots \\
        \ell(w_{p-2}, v_{1}) & \cdots & \ell(w_{p-2}, v_{d}) \\
        1 & \cdots & 1 \\
        a_{1}(t, \bar w) & \cdots & a_{d}(t,\bar w)
    \end{pmatrix}.
\end{align*}
We take $\hat \beta_n \equiv (\hat \beta_{{\rm u},n}, 1, \gamma)^\prime \in \mathbf R^p$, where $\hat{\beta}_{{\rm u},n}$ is the sample analogue to the first $p-2$ components of $\beta(P)$.
We set $\hat x_n^\star = A^\dagger \hat \beta_n$ for designs with $d\geq p$, and let
\begin{align*}
    \hat x_n^\star \equiv  \arg\min_{x \in \re^{d}}
    \left(
        \hat \beta_{{\rm u},n}
        -
        A_{\rm u} x
    \right)'
    \hat{\Xi}_{n}^{-1}
    \left(
        \hat \beta_{{\rm u},n}
        -
        A_{\rm u} x
    \right)
    \quad
    \text{s.t.}
    \quad
    \sum_{j=1}^{d} x = 1
    \quad
    \text{and}
    \quad
    a(t, \bar{w})' x = \gamma,
\end{align*}
when $d  < p$, where $A_{\rm u}$ corresponds to the first $p-2$ rows of $A$ and $\hat{\Xi}_{n}$ is the sample analogue estimator of asymptotic variance matrix of $\hat{\beta}_{{\rm u},n}$.
We let $\hat \Omega_n^{\rm e}$ be the sample standard deviation matrix of $\hat \beta_n$, and $\hat \Omega_n^{\rm i}$ be the sample standard deviation of $A\hat x_n^\star$ computed from 250 draws of the nonparametric bootstrap.


\begin{figure}[t!]
    \centering
    \caption{\label{fig:size} Null rejection probabilities for (nearly) point-identified designs}
    \includegraphics[scale = .75]{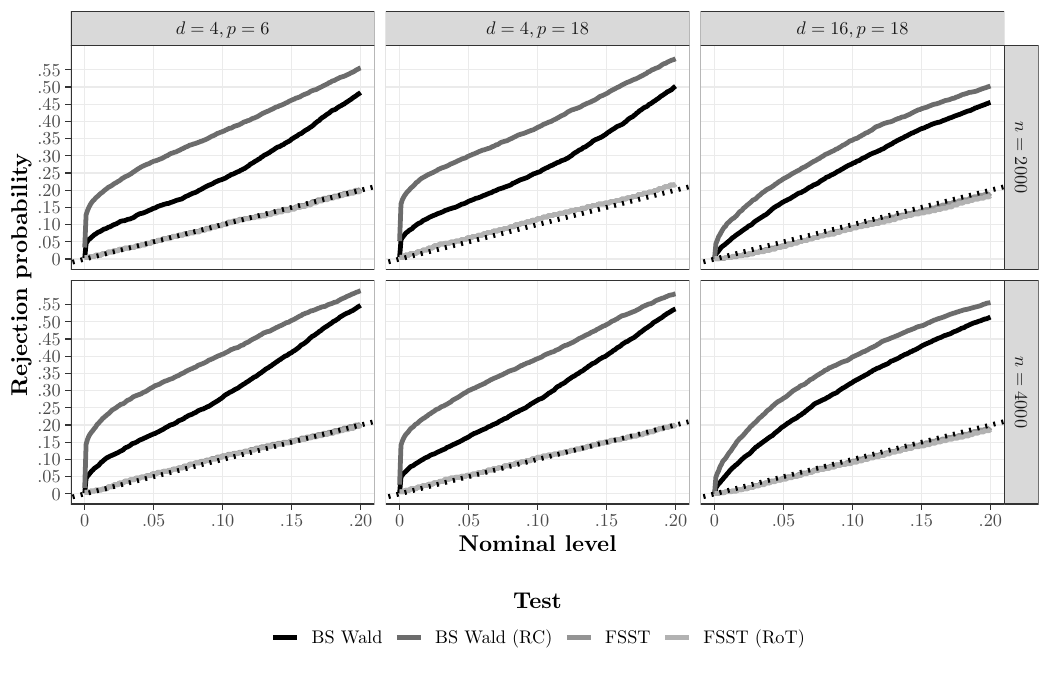}
    \floatfoot{The dotted line is the 45 degree line.
        ``FSST'' refers to the test developed in this paper with $\lambda_{n}^{\rm b}$, whereas ``FSST (RoT)'' uses the rule of thumb choice $\lambda_{n}^{\rm r}$.
        ``BS Wald'' corresponds to a Wald test using bootstrap estimates of the standard errors. 
        ``BS Wald (RC)'' is the same procedure but with standard errors based on bootstrapping with a re-centered GMM criterion.
        The null hypothesis is that $F_{\epsilon}(-1 \vert 1)$ is equal to its true value.
        In the case of $d = 16, p = 18$, which is set identified but with a very narrow identified set, we test the null hypothesis that $F_{\epsilon}(-1 \vert 1)$ is equal to the lower bound of the identified set.
    }
\end{figure}

We explore two rules for selecting $\lambda_n$.
To motivate them, we note that an important theoretical restriction on $\lambda_n$ is that, uniformly in $P\in \mathbf P_0$, it satisfy
\begin{equation}\label{eq:lambda1}
\lambda_n \sqrt n \sup_{s\in \hat{\mathcal V}_n^{\rm i}} \langle A^\dagger s, A^\dagger A( \hat x_n^\star - x^\star(P))\rangle = o_P(1);
\end{equation}
see Lemma \ref{lm:first}.
Employing our coupling $\sqrt n A(\hat x_n^\star - x^\star(P)) \approx \mathbb G^{\rm i}_n(P)$ and $\Omega_n^{\rm i}(P)$ being the standard deviation matrix of $\mathbb G_n^{\rm i}(P)$ suggests selecting $\lambda_n$ to satisfy $\lambda_n \sqrt{\log(e\vee p)} = o(1)$ -- here $a\vee b \equiv \max\{a,b\}$.
For a concrete choice of $\lambda_n$, we rely on the law of iterated logarithm and let $\lambda_n^{\rm r} = 1/\sqrt{\log(e\vee p)\log(e\vee \log(e\vee n))}$.
As an alternative to $\lambda_n^{\rm r}$, we employ the bootstrap to approximate the law of \eqref{eq:lambda1}.
In particular, for some $\delta_n \downarrow 0$ we let $\lambda_n^{\rm b} \equiv \min\{1,\hat \tau_n(1-\delta_n)\}$ where $\hat\tau_n(1-\delta_n)$ denotes the $1-\delta_n$ quantile of
\begin{equation}
\label{eq:lambda2}
\sup_{s\in \hat{\mathcal V}_n^{\rm i}} \langle A^\dagger s, A^\dagger \hat {\mathbb G}_n^{\rm i}\rangle
\end{equation}
conditional on the data. For a concrete choice of $\delta_n$ we let $\delta_n = 1/\sqrt{\log(e\vee \log(e\vee n))}$.

In Appendix \ref{sec:computational-details}, we describe the computation of our test in more detail.
In particular, we show how to reformulate all optimization problems into linear programming problems that do not require explicitly computing $A^\dagger$.
An R package for implementing our test is available at \url{https://github.com/conroylau/lpinfer}.

\begin{table}[t!]
        \scriptsize
    \centering
        \begin{tabular}{@{}c c cc c cc c cc c cc c cc c cc c c    c c c c c c c c @{}}
	\toprule
    &  & \multicolumn{7}{c}{(a) Results for $\lambda_n^{\rm b}$} & &  \multicolumn{7}{c}{(b) Results for $\lambda_n^{\rm r}$}\\
	\cmidrule(lr){3-9} \cmidrule(lr){11-17}
    &  & \multicolumn{7}{c}{$d$} & & \multicolumn{7}{c}{$d$}\\
	\cmidrule(lr){3-9} \cmidrule(lr){11-17}
	$n$ & $p$ & 100 & 400 & 1600 & 4900 & $100^{2}$ & $225^{2}$ & $317^{2}$ & & 100 & 400 & 1600 & 4900 & $100^{2}$ & $225^{2}$ & $317^{2}$ \\
	\midrule
	\multirow{2}{*}{1000} & 6 & .036 & .034 & .034 & .037 & .038 & .036 & .036 & & .020 & .019 & .021 & .021 & .022 & .019 & .021\\
	 & 18 & .040 & .035 & .036 & .041 & .039 & .038 & .036 & & .037 & .029 & .029 & .033 & .033 & .031 & .030\\
	\cmidrule(lr){1-17}
	\multirow{3}{*}{2000} & 6 & .042 & .042 & .049 & .046 & .047 & .052 & .061 & & .030 & .025 & .033 & .032 & .033 & .027 & .039\\
	 & 18 & .031 & .028 & .032 & .032 & .030 & .030 & .028 & & .023 & .021 & .028 & .027 & .025 & .027 & .020 \\
	 & 38 & .053 & .046 & .051 & .052 & .052 & .067 & .053 & & .048 & .039 & .043 & .045 & .047 & .062 & .046 \\
	\cmidrule(lr){1-17}
	\multirow{4}{*}{4000} & 6 & .045 & .048 & .049 & .054 & .058 & .051 & .065 & & .034 & .034 & .038 & .042 & .046 & .035 & .058 \\
	 & 18 & .028 & .031 & .029 & .028 & .030 & .038 & .035 & & .023 & .026 & .024 & .022 & .025 & .032 & .028 \\
	 & 38 & .031 & .034 & .039 & .036 & .040 & .035 & .037 & & .026 & .029 & .033 & .032 & .035 & .032 & .033\\
	 & 51 & .042 & .051 & .051 & .040 & .047 & .047 & .030 & & .038 & .044 & .045 & .034 & .042 & .041 & .027\\
	\cmidrule(lr){1-17}
	\multirow{5}{*}{8000} & 6 & .049 & .055 & .056 & .048 & .054 & .055 & .073 & & .040 & .046 & .048 & .040 & .046 & .050 & .061 \\
	 & 18 & .034 & .035 & .036 & .030 & .032 & .040 & .041 & & .028 & .028 & .032 & .025 & .027 & .032 & .034 \\
	 & 38 & .033 & .035 & .035 & .037 & .037 & .025 & .047 & & .027 & .029 & .030 & .032 & .032 & .021 & .043 \\
	 & 51 & .034 & .043 & .035 & .040 & .037 & .035 & .038 & & .029 & .036 & .028 & .034 & .033 & .030 & .031 \\
	 & 83 & .043 & .042 & .050 & .048 & .042 & .054 & .046 & & .038 & .035 & .046 & .041 & .034 & .048 & .042  \\
	\bottomrule
\end{tabular}

    \caption{\label{tab:size} Null rejection probabilities for a nominal 0.05 test}
    \floatfoot{The null hypothesis is that $F_{\epsilon}(-1 \vert 1)$ is equal to the lower bound of the population identified set.
    }
\end{table}

\subsection{Monte Carlo Simulations}
\label{sec:monte-carlo}

We start by examining the null rejection probabilities of our testing procedure by setting $\gamma$ to be the lower bound of the population identified set computed via \eqref{eq:lp-solve-for-bounds-mixedlogit} with $t = -1$ and $\bar{w} = 1$.
In unreported simulations we found setting $\gamma$ to be the upper bound of the identified set yielded similar results.
We consider sample sizes of $n = 1000$, $2000$, $4000$, and $8000$ for each of the data generating processes discussed in Section \ref{sec:dgps}.
Results with $d \leq 10000$ are based on $5000$ Monte Carlo replications and 250 nonparametric bootstrap draws.
When $d > 10000$, we use $1000$ Monte Carlo replications.

\begin{figure}[t!]
    \centering
    \caption{\label{fig:power} Power curves for FSST nominal $0.10$ test}
    \includegraphics[scale = .75]{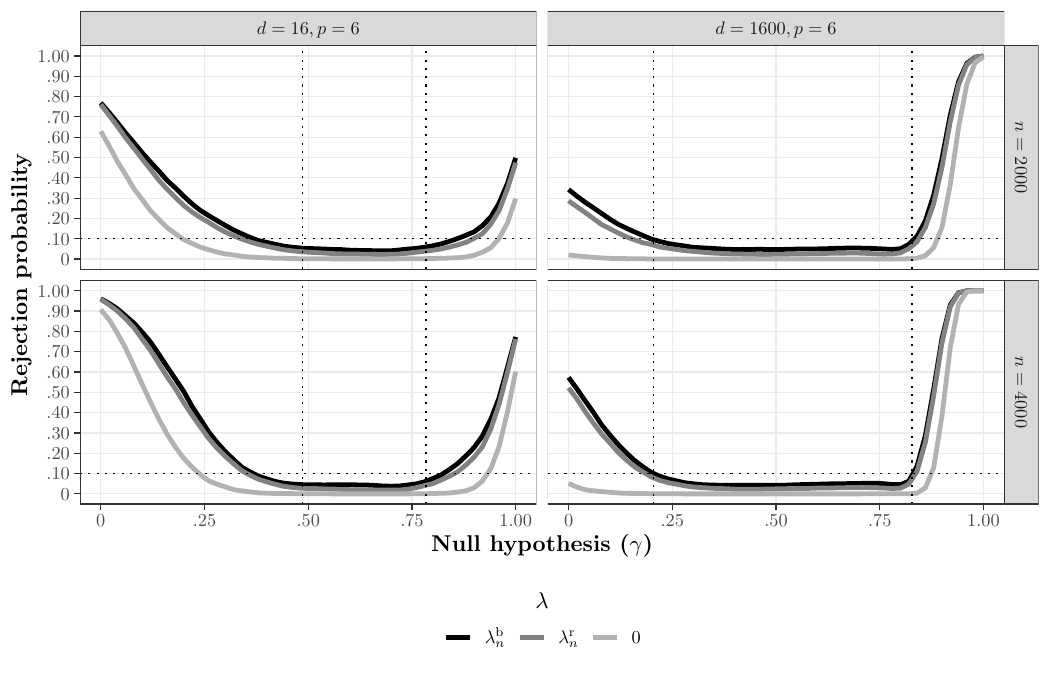}
    \floatfoot{The vertical dotted lines indicate the lower and upper bounds of the population identified set. The horizontal dotted line indicates the nominal level ($0.10$).
    }
\end{figure}

We first consider the designs in which $p-2 \geq d$ so that $F_{\epsilon}(-1 \vert 1)$ is (nearly) point identified.
In this case, one might alternatively consider estimating probability weights $x_0$ satisfying the moment restrictions in \eqref{eq:conditional-moment} by constrained GMM, and then conducting inference on $F_{\epsilon}(-1\vert 1)$ using a bootstrapped Wald test.
For example, this is the approach that appears to have been taken by \cite{nevo2016usage} in the related setting discussed in Example \ref{ex:dyn}.
However, the non-negativity constraints on $x_0$ imply that the bootstrap will generally not be consistent in this case \citep{fang2018inference}.

We demonstrate this point in Figure \ref{fig:size} with plots of the actual and nominal level for both our (FSST) and for the bootstrapped Wald test based on constrained GMM.
The latter exhibits large size distortions.
For example the GMM test with nominal level $5\%$ rejects in over 15\% of draws $d = 16, p = 18$ and $n = $ 2,000, and a nominal level $10\%$ test rejects in over 25\% of draws when $d = 4, p = 18$, and $n = $ 4,000.
Re-centering the GMM criterion before conducting this test \citep[e.g.][]{hallhorowitz1996e} leads to even greater over-rejection.
In contrast, FSST has nearly equal nominal and actual levels across the examined designs.

In Table \ref{tab:size}, we report empirical rejection rates for our procedure using partially identified designs that range in size from relatively small ($d = 100, p = 6$) to enormous ($p = 83, d = 317^{2} \approx 10^{5}$).
We note that in this application, $p/n$ should be small because otherwise we will draw samples (or bootstrap samples) that do not contain all the support points of $W$.
Reflecting this constraint, in Table \ref{tab:size} we let $p$ grow with $n$ but keep the largest values of $p/n$ at approximately $.01$.
No such restriction is imposed on $d$ and we consider designs in which $d$ far exceeds $n$ (e.g., with $d/n$ as large as 100).
Across all different data generating processes and sample sizes, even in the largest models, we find the null rejection probabilities remain approximately no greater than the nominal level.

Comparing panels (a) and (b) of Table \ref{tab:size}, we see that the occasional (and mild) over-rejections can be controlled by using $\lambda_{n}^{\rm r}$ instead of $\lambda_{n}^{\rm b}$.
Figure \ref{fig:power} illustrates the impact that the choice of $\lambda_{n}$ has on power for two of the smaller designs.
Both $\lambda_{n}^{\rm b}$ and $\lambda_{n}^{\rm r}$ provide considerable power gains over the conservative choice of $\lambda_{n} = 0$.


\begin{center}
    {\Large  \sc {Appendix}}
\end{center}

\renewcommand{\thesection}{A.\arabic{section}}
\renewcommand{\theequation}{A.\arabic{equation}}
\renewcommand{\thelemma}{A.\arabic{section}.\arabic{lemma}}
\renewcommand{\thecorollary}{A.\arabic{section}.\arabic{corollary}}
\renewcommand{\thetheorem}{A.\arabic{section}.\arabic{theorem}}
\renewcommand{\theassumption}{A.\arabic{section}.\arabic{assumption}}
\setcounter{lemma}{0}
\setcounter{theorem}{0}
\setcounter{corollary}{0}
\setcounter{equation}{0}
\setcounter{remark}{0}
\setcounter{section}{0}
\setcounter{assumption}{0}

This Appendix contains the proofs of all the results in Section \ref{sec:geo}.
The proofs for all other results in the paper are included in the Supplemental Appendices.

\noindent \emph{Proof of Lemma \ref{lm:trivial}:} First note that by definition of $R$, there exists a $x \in \mathbf R^d$ such that $\Pi_R(\beta) = Ax$.
Moreover, by Theorem 3.4.1 in \cite{luenberger:1969} we may decompose $x$ as $x = \Pi_N(x) + \Pi_{N^\perp}(x)$.
Hence, defining $x^\star$ to equal $x^\star \equiv \Pi_{N^\perp}(x)$ and using that $A(\Pi_N(x)) = 0$ by definition of $N$, we obtain that
$$\Pi_R(\beta) = Ax = A(\Pi_{N^\perp }x + \Pi_N x) = Ax^\star.$$
To see $x^\star$ is the unique element in $N^\perp$ satisfying $\Pi_R(\beta) = Ax^\star$, let $\tilde x\in N^\perp$ be any element satisfying $A\tilde x  = \Pi_R(\beta) = Ax^\star$.
Since $A(\tilde x - x^\star) = 0$, it then follows that $\tilde x - x^\star \in N$.
However, we also have $\tilde x - x^\star \in N^\perp$ since $\tilde x,x^\star \in N^\perp$ and $N^\perp$ is a vector subspace of $\mathbf R^d$.
Thus, we obtain $x^\star -\tilde x\in N \cap N^\perp$, and since $N \cap N^\perp = \{0\}$ we can conclude $\tilde x = x^\star$, which establishes $x^\star$ is indeed unique. \qed

\noindent \emph{Proof of Theorem \ref{th:ineq}:}
Fix any $\beta \in \mathbf R^{p}$ and recall $\Pi_R(\beta)$ denotes its projection under $\|\cdot\|_2$ onto $R$ (the range of $A$).
Next note that by Farkas' Lemma (see, e.g., Corollary 5.85 in \cite{aliprantis:border:2006}) it follows that the statement
\begin{equation}\label{th:ineq1}
\Pi_R(\beta) = A \tilde x \text{ for some } \tilde x \geq 0
\end{equation}
holds if and only if there \emph{does not} exist a $y\in \mathbf R^p$ satisfying the inequalities:
\begin{equation}\label{th:ineq2}
A^\prime y \leq 0 ~(\text{in } \mathbf R^d) \text{ and } \langle y, \Pi_R(\beta)\rangle > 0.
\end{equation}
In particular, there being no $y\in \mathbf R^p$ satisfying \eqref{th:ineq2} is equivalent to the statement
\begin{equation}\label{th:ineq3}
\langle y,\Pi_R(\beta) \rangle \leq 0 \text{ for all } y \in \mathbf R^p \text{ such that }A^\prime y \leq 0 ~(\text{in } \mathbf R^d).
\end{equation}
Next note Lemma \ref{lm:trivial} implies that there is a unique $x^{\star} \in N^{\perp}$ such that $\Pi_R(\beta) = Ax^{\star}$.
Therefore, $\langle y,Ax^\star\rangle = \langle A^\prime y,x^\star\rangle$ implies \eqref{th:ineq3} is equivalent to
\begin{equation}\label{th:ineq4}
\langle A^\prime y, x^\star\rangle \leq 0 \text{ for all } y\in \mathbf R^p \text{ such that } A^\prime y \leq 0 ~(\text{in } \mathbf R^d).
\end{equation}
Since $\{ A^\prime y : y \in \mathbf R^p \text{ and } A^\prime y \leq 0\} = \text{range}\{A^\prime\}\cap \mathbf R^d_{-}$, \eqref{th:ineq4} is equivalent to
\begin{equation}\label{th:ineq5}
\langle s, x^\star \rangle \leq 0 \text{ for all } s \in \text{range}\{A'\}\cap \mathbf R_-^d.
\end{equation}
However, since $\text{range}\{A^\prime\}$ is closed, Theorem 6.6.3 in \cite{luenberger:1969} further implies that $\text{range}\{A^\prime\} = N^\perp$.
Therefore, condition \eqref{th:ineq5} is satisfied if and only if
\begin{equation}\label{th:ineq6}
\langle s, x^\star \rangle \leq 0 \text{ for all } s \in N^\perp \cap \mathbf R_-^d.
\end{equation}
In summary, we have shown that \eqref{th:ineq1} is satisfied if and only if  \eqref{th:ineq6} holds.
Since in addition $\beta \in R$ if and only if $\beta = \Pi_R(\beta)$, the claim of the theorem follows. \qed

\phantomsection
\addcontentsline{toc}{section}{References}

{\small
\singlespace
\setstretch{0.99}
\putbib
}

\end{bibunit}

\newpage


\begin{bibunit}

\renewcommand{\thesection}{S.\arabic{section}}
\renewcommand{\theequation}{S.\arabic{equation}}
\renewcommand{\thelemma}{S.\arabic{lemma}}
\renewcommand{\thecorollary}{S.\arabic{corollary}}
\renewcommand{\thetheorem}{S.\arabic{theorem}}
\renewcommand{\theassumption}{S.\arabic{assumption}}
\setcounter{lemma}{0}
\setcounter{theorem}{0}
\setcounter{corollary}{0}
\setcounter{equation}{0}
\setcounter{remark}{0}
\setcounter{section}{0}
\setcounter{assumption}{0}

\setcounter{page}{1}


\begin{center}
    {\Large {\sc Supplemental Appendix I}}
\end{center}

\onehalfspacing

This Supplemental Appendix contains the proofs of Theorems \ref{th:dist}, \ref{th:size} and some auxiliary results.
The auxiliary Lemmas \ref{lm:deg}-\ref{lm:Vpoly} are stated, but their proofs, together with computational details on the implementation of our procedure, can be found in Supplemental Appendix II (available at the authors' websites).



\noindent \emph{Proof of Theorem \ref{th:dist}:} First note that by Lemma \ref{lm:betacoup} there exists a Gaussian vector $(\mathbb G_n^{\text{e}}(P)^\prime,\mathbb G_n^{\text{i}}(P)^\prime)^\prime \equiv \mathbb G_n(P) \in \mathbf R^{2p}$ with $\mathbb G_n(P) \sim N(0,\Sigma(P))$ satisfying
\begin{align}
\|(\Omega^{\text{\rm e}}(P))^\dagger \{(I_p - AA^\dagger\hat C_n)\sqrt n\{\hat \beta_n - \beta(P)\} - \mathbb G_n^{\text{\rm e}}(P)\}\|_\infty = O_P(r_n) \notag \\
\|(\Omega^{\text{\rm i}}(P))^\dagger \{AA^\dagger\hat C_n \sqrt n\{\hat \beta_n - \beta(P)\} - \mathbb G_n^{\text{\rm i}}(P)\}\|_\infty = O_P(r_n)  \label{th:dist2}
\end{align}
uniformly in $P\in \mathbf P$.
Further note that Assumption \ref{ass:techrange}(i) implies $\text{range}\{\Sigma^{\rm j}(P)\} \subseteq \text{range}\{\Omega^{\rm j}(P)\}$ for $\rm j \in \{\rm e,\rm i\}$ and $P\in \mathbf P$.
Therefore, Assumption \ref{ass:techrange}(ii) yields
\begin{align}
(I_p - AA^\dagger\hat C_n)\sqrt n\{\hat \beta_n - \beta(P)\} \in \text{range}\{\Omega^{\rm e}(P)\} \notag \\ 
AA^\dagger\hat C_n \sqrt n\{\hat \beta_n - \beta(P)\} \in \text{range}\{\Omega^{\rm i}(P)\} \label{th:dist4}
\end{align}
with probability tending to one uniformly in $P\in \mathbf P$.
Next, note $AA^\dagger s = s$ for all $s \in R$ and Theorem \ref{th:ineq} imply $(I_p - AA^\dagger)\beta(P) = 0$ for all $P\in \mathbf P_0$.
Hence, $\hat x_n^\star = A^\dagger \hat C_n \hat \beta_n$ and $\hat C_n \beta(P) = \beta(P)$ for all $P\in \mathbf P_0$ by Assumption \ref{ass:beta}(ii) yield
\begin{multline}\label{th:dist5}
\sup_{s\in \hat{\mathcal V}_n^{\rm e}} \sqrt n \langle s, \hat \beta_n - A \hat x_n^\star\rangle \\
= \sup_{s\in \hat{\mathcal V}_n^{\rm e}} \langle s, (I_p - AA^\dagger \hat C_n)\sqrt n \hat \beta_n\rangle = \sup_{s\in \hat{\mathcal V}_n^{\rm e}} \langle s, (I_p - AA^\dagger \hat C_n)\sqrt n \{\hat \beta_n- \beta(P)\}\rangle
\end{multline}
for all $P\in \mathbf P_0$.
Similarly, employing that $A^\dagger AA^\dagger = A^\dagger$ (see Proposition 6.11.1(5) in \cite{luenberger:1969}) together with $A\hat x_n^\star = AA^\dagger \hat C_n \hat \beta_n$ and $\hat C_n \beta(P) =\beta(P)$ for all $P\in \mathbf P_0$ by Assumption \ref{ass:beta}(ii), implies that for all $P\in \mathbf P_0$ we have
\begin{multline}\label{th:dist6}
\sup_{s\in \hat{\mathcal V}_n^{\rm i}} \sqrt n  \langle A^\dagger s, \hat x_n^\star \rangle = \sup_{s\in \hat{\mathcal V}_n^{\rm i}} \sqrt n  \langle A^\dagger s, A^\dagger \hat C_n \hat \beta_n\rangle \\
=\sup_{s\in \hat{\mathcal V}_n^{\rm i}}  \langle A^\dagger s, A^\dagger AA^\dagger \hat C_n \sqrt n \{\hat \beta_n - \beta(P)\} \rangle + \sqrt n  \langle A^\dagger s, A^\dagger \beta(P)\rangle .
\end{multline}
Moreover, if $P\in \mathbf P_0$, then $\sqrt n \langle A^\dagger s, A^\dagger \beta(P)\rangle \leq 0$ for all $s$ satisfying $A^\dagger s \leq 0$ by Theorem \ref{th:ineq}, $A^\dagger s \in N^\perp \cap \mathbf R^d_{-}$ whenever $A^\dagger s \leq 0$, and $x^\star(P) = A^\dagger \beta(P)$.
Hence, $r_n = o(1)$, \eqref{th:dist2}, \eqref{th:dist4}, \eqref{th:dist5}, \eqref{th:dist6}, and Theorem \ref{th:master} applied with $\hat {\mathbb W}_n^{\rm e}(P) = (I_p - AA^\dagger\hat C_n)\sqrt n\{\hat \beta_n - \beta(P)\}$, $\hat {\mathbb W}_n^{\rm i}(P) =  AA^\dagger\hat C_n \sqrt n\{\hat \beta_n - \beta(P)\}$, $\hat f_n(s,P) = \sqrt n \langle A^\dagger s, A^\dagger \beta(P)\rangle $, $\mathbf Q = \mathbf P_0$, and $\omega_n = r_n$ together with $a_n + r_n = O(r_n)$ imply
\begin{align*}
\sup_{s\in \hat{\mathcal V}_n^{\rm e}} \sqrt n \langle s, \hat \beta_n - A \hat x_n^\star\rangle  & = \sup_{s\in {\mathcal V}^{\rm e}(P)} \langle s, \mathbb G_n^{\rm e}(P)\rangle + O_P(r_n)  \\
\sup_{s\in \hat{\mathcal V}_n^{\rm i}} \sqrt n  \langle A^\dagger s, \hat x_n^\star \rangle  & = \sup_{s\in \mathcal V^{\rm i}(P)} \langle A^\dagger s, A^\dagger \mathbb G_n^{\rm i}(P)\rangle + \sqrt n\langle A^\dagger s, A^\dagger \beta(P)\rangle + O_P(r_n),
\end{align*}
uniformly in $P\in \mathbf P_0$, from which the claim of the theorem follows. \qed

\noindent \emph{Proof of Theorem \ref{th:size}:} For notational simplicity we first set $\eta \equiv 1-\alpha$ and define
\begin{align}
\mathbb M_n(s,P) & \equiv \langle A^\dagger s, A^\dagger \mathbb G_n^{\text{i}}(P)\rangle \hspace{0.3 in} &\mathbb U_n(s,P)& \equiv \sqrt n \langle A^\dagger s, A^\dagger \beta(P)\rangle  \label{th:sizedef1} \\
\mathbb A_n^{\rm e}(s,P) & \equiv \langle s, (\Omega^{\rm e}(P))^\dagger \mathbb G_n^{\rm e}(P)\rangle \hspace{0.3 in} & \mathbb A_n^{\rm i}(s,P) &\equiv \langle s, \mathbb G_n^{\rm i}(P) + \sqrt n\beta(P)\rangle \label{th:sizedef2}.
\end{align}
Also set sequences $\ell_n \downarrow 0$ and $\tau_n \uparrow 1$ to satisfy $r_n \vee b_n \vee \lambda_n \sqrt{\log(1+p)} = o(\ell_n)$ and
\begin{equation}\label{th:size8}
\sup_{P\in \mathbf P} \frac{\text{m}(P) + \bar \sigma(P)z_{\tau_n}}{\underline{\sigma}^2(P)} = o(\ell_n^{-1}),
\end{equation}
which  is feasible by hypothesis.
Further note that since $\eta > 0.5$, there is $\epsilon>0$ such that $\eta - \epsilon > 0.5$ and for $z_{\eta - \epsilon}$ the $\eta - \epsilon$ quantile of a standard normal, let
\begin{align}
E_{1n}(P) & \equiv \{ \hat c_n(\eta) \geq (\underline{\sigma}(P)z_{\eta-\epsilon})/2\} \\ 
E_{2n}(P) & \equiv \{ \mathbb U_n(s,P) \leq \hat {\mathbb U}_{n}(s) + \ell_n \text{ for all } s \in \hat {\mathcal V}_n^{\rm i}\} \label{th:sizedef5}.
\end{align}

Next, note that $0\in \hat {\mathcal V}_n^{\rm e}$ and $0\in \hat{\mathcal V}_n^{\rm i}$ together yield that $\hat c_n(\eta) \geq 0$.
Therefore, $\phi_n = 1$ implies $T_n > 0$, which together with Lemma \ref{lm:deg} implies that the conclusion of the theorem is immediate on the set $\mathbf D_0 \equiv \{P\in \mathbf P_0 : \sigma^{\rm j}(s,P) = 0 \text{ for all } s\in \mathcal E^{\rm j}(P) \text{ and all } \rm j \in \{\rm e,\rm i\}\}$.
We therefore assume without loss of generality that for all $P\in \mathbf P_0$, $\sigma^{\rm j}(s,P) > 0$ for some $s\in \mathcal E^{\rm j}(P)$ and some $\rm j \in \{\rm e,\rm i\}$.
Next, we also observe that since $\phi_n = 1$ implies $T_n > 0$, Lemma \ref{aux:critbound} yields
\begin{equation}\label{th:size1}
\limsup_{n\rightarrow \infty} \sup_{P\in \mathbf P_0} P(\phi_n = 1) \\ = \limsup_{n\rightarrow \infty} \sup_{P\in \mathbf P_0} P(T_n > \hat c_n(\eta); ~ E_{1n}(P)).
\end{equation}
Moreover, for $\rm j \in \{\rm e,\rm i\}$,  $\mathbb G_n^{\rm j}(P) \in \text{range}\{\Sigma^{\rm j}(P)\} \subseteq \text{range}\{\Omega^{\rm j}(P)\}$ almost surely by Theorem 3.6.1 in \cite{bogachev1998gaussian} and Assumption \ref{ass:techrange}(i).
Hence, it follows that $\Omega^{\rm j}(P)(\Omega^{\rm j}(P))^\dagger \mathbb G_n^{\rm j}(P) = \mathbb G_n^{\rm j}(P)$ almost surely for $\rm j \in \{\rm e,\rm i\}$, which together with H\"older's inequality, Assumption \ref{ass:weights}(ii), the definitions of $\mathcal V^{\rm e}(P)$ and $\mathcal V^{\rm i}(P)$, and $\mathbb U_n(s,P) \leq 0$ for $s \in \mathcal V^{\rm i}(P)$ and $P\in \mathbf P_0$ by Theorem \ref{th:ineq} imply that almost surely
\begin{align*}
\sup_{s\in \mathcal V^{\rm e}(P)} \langle s,\mathbb G_n^{\rm e}(P)\rangle & = \sup_{s\in \mathcal V^{\rm e}(P)} \langle \Omega^{\rm e}(P) s,(\Omega^{\rm e}(P))^\dagger \mathbb G_n^{\rm e}(P)\rangle < \infty \\ 
\sup_{s\in \mathcal V^{\rm i}(P)} \mathbb M_n(s,P) +\mathbb U_n(s,P)& = \sup_{s\in \mathcal V^{\rm i}(P)} \langle \Omega^{\rm i}(P)(AA^\prime)^\dagger s, (\Omega^{\rm i}(P))^\dagger\mathbb G_n^{\rm i}(P)\rangle + \mathbb U_n(s,P) < \infty. 
\end{align*}
Thus, by Theorem \ref{th:dist} and Lemmas \ref{lm:auxlp}, \ref{lm:Vpoly} we obtain uniformly in $P\in \mathbf P_0$
\begin{equation}\label{th:size4}
T_n  = \max_{s\in \mathcal E^{\rm e}(P)} \mathbb A^{\rm e}_n(s,P) \vee \max_{s\in \mathcal E^{\rm i}(P)} \mathbb A_n^{\rm i}(s,P)  + O_P(r_n).
\end{equation}
For any $\tau \in (0,1)$ and $\mathbb M_n(s,P)$ as in \eqref{th:sizedef1}, we next let $c_n^{(1)}(\tau,P)$ be given by
\begin{equation}\label{th:size5}
c_n^{(1)}(\tau,P) \equiv \inf\{ u : P(\sup_{s \in \mathcal V^{\rm i}(P)} \mathbb M_n(s,P) \leq u) \geq \tau \}.
\end{equation}
Employing $c_n^{(1)}(\tau,P)$ we further define a ``truncated" subset $\mathcal E^{{\rm i},\tau}(P) \subseteq \mathcal E^{\rm i}(P)$ by
\begin{equation}\label{th:size6}
\mathcal E^{\rm i,\tau}(P) \equiv \{s \in \mathcal E^{\rm i}(P) : - \langle s, \sqrt n\beta(P)\rangle \leq c_n^{(1)}(\tau,P)\}.
\end{equation}
Next note that $0 \in \mathcal V^{\rm i}(P)$ satisfying $\mathbb M_n(0,P) = 0$ implies $\sup_{s \in \mathcal V^{\rm i}(P)} \mathbb M_n(s,P)$ is nonnegative almost surely and therefore $c_n^{(1)}(\tau,P) \geq 0$.
Since in addition $0 \in \mathcal E^{\rm i}(P)$ by Lemma \ref{lm:Vpoly}, it follows $0 \in \mathcal E^{\rm i,\tau}(P)$ and therefore we obtain that
\begin{multline*}
P(\max_{s\in \mathcal E^{\rm i}(P)} \mathbb A_n^{\rm i}(s,P) = \max_{s\in \mathcal E^{\rm i,\tau}(P)} \mathbb A_n^{\rm i}(s,P))\\ \geq P(\max_{s\in \mathcal E^{\rm i}(P)\setminus \mathcal E^{\rm i,\tau}(P)} \mathbb A_n^{\rm i}(s,P) \leq 0) \geq P(\sup_{s\in \mathcal V^{\rm i}(P)} \mathbb M_n(s,P)  \leq c_n^{(1)}(\tau,P)) \geq \tau,
\end{multline*}
where the second and final inequalities hold by definitions \eqref{th:sizedef1} and \eqref{th:size5}, and $ \mathcal E^{\rm i}(P) \subseteq (AA^\prime)^\dagger \mathcal V^{\rm i}(P)$.
Next define the sets $\mathcal C_n({\rm j},P)$ according to the relation
\begin{equation*}
\mathcal C_n({\rm j},P) \equiv \left\{\begin{array}{cl} \mathcal E^{\rm e}(P) & \text{ if } \rm j = \rm e \\ \mathcal E^{{\rm i},\tau_n}(P) & \text{ if } \rm j = \rm i\end{array}\right. .
\end{equation*}
Given these definitions, we then obtain from results \eqref{th:size8}, \eqref{th:size1}, and \eqref{th:size4} that
\begin{align}\label{th:size10}
\limsup_{n\rightarrow \infty} &\sup_{P\in \mathbf P_0} P(\phi_n = 1) \notag \\
&\leq \limsup_{n\rightarrow \infty} \sup_{P\in \mathbf P_0} P(\max_{s\in \mathcal E^{\rm e}(P)} \mathbb A_n^{\rm e}(s,P) \vee \max_{s\in \mathcal E^{{\rm i},\tau_n}(P)}\mathbb A_n^{\rm i}(s,P)  > \hat c_n(\eta) - \ell_n; ~ E_{1n}(P)) \notag \\
& = \limsup_{n\rightarrow \infty} \sup_{P\in \mathbf P_0} P(\max_{\rm j\in \{\rm e,\rm i\}} \max_{s\in \mathcal C_n({\rm j},P)} \mathbb A_n^{\rm j}(s,P) > \hat c_n(\eta) - \ell_n; ~ E_{1n}(P))
\end{align}
due to $\tau_n \uparrow 1$ and $r_n = o(\ell_n)$ by construction.
Further  define the set $\mathcal A_n(P)$ by
\begin{equation*}
\mathcal A_n(P) \equiv \{({\rm j},s) : {\rm j \in \{ e,i\}}, ~ s \in \mathcal C_n({\rm j},P), ~ \sigma^{\rm j}(s,P) > 0\},
\end{equation*}
and note that, for $n$ sufficiently large, $\inf_{P\in \mathbf P} (\underline{\sigma}(P) z_{\eta -\epsilon}) - 2\ell_n >0$ by \eqref{th:size8}, in which case $E_{1n}(P)$ implies $\hat c_n(\eta) -\ell_n> 0$.
Since for all $P\in \mathbf P_0$ we have $E[\mathbb A_n^{\rm e}(s,P)] = 0$ for all $s\in \mathcal E^{\rm e}(P)$ and $E[\mathbb A_n^{\rm i}(s,P)] \leq 0$ for all $s\in \mathcal E^{{\rm i},\tau_n}(P)$ due to $\langle (AA^\prime)^\dagger s, \beta(P)\rangle \leq 0$ for all $s\in \mathcal V^{\rm i}(P)$ by Theorem \ref{th:ineq}, we can conclude from \eqref{th:size10} that the claim of the theorem holds if $\mathcal A_n(P) = \emptyset$.
Hence, assuming without loss of generality that $\mathcal A_n(P) \neq \emptyset$ we obtain from the same observations that
\begin{multline}\label{th:size12}
\limsup_{n\rightarrow \infty} \sup_{P\in \mathbf P_0} P(\phi_n = 1)
\leq \limsup_{n\rightarrow \infty} \sup_{P\in \mathbf P_0} P(\max_{({\rm j},s) \in \mathcal A_n(P)} \mathbb A_n^{\rm j}(s,P) > \hat c_n(\eta) - \ell_n) \\
 =\limsup_{n\rightarrow \infty} \sup_{P\in \mathbf P_0} P(\max_{({\rm j},s) \in \mathcal A_n(P)} \mathbb A_n^{\rm j}(s,P) > \hat c_n(\eta) - \ell_n; ~ E_{2n}(P)),
\end{multline}
where the final inequality holds for $E_{2n}(P)$ as defined in \eqref{th:sizedef5} by Lemma \ref{lm:first}.

For any $P\in \mathbf P_0$, it follows that under $E_{2n}(P)$, $\hat c_n(\eta)$ is $P$-almost surely bounded from below by the conditional on $\{Z_i\}_{i=1}^n$ $\eta$ quantile of the random variable
\begin{equation}\label{th:size13}
\max\{\sup_{s\in \hat{\mathcal V}_n^{\rm e}} \langle s, \hat{\mathbb G}_n^{\rm e}\rangle, \sup_{s\in \hat{\mathcal V}_n^{\rm i}} \langle A^\dagger s, A^\dagger \hat{\mathbb G}_n^{\rm i}\rangle + \mathbb U_n(s,P)\} - \ell_n.
\end{equation}
Moreover, by Theorem \ref{lm:finitebootnew} there is a Gaussian vector $(\mathbb G_{n}^{\text{e}\star}(P)^\prime,\mathbb G_{n}^{\text{i}\star}(P)^\prime)^\prime \equiv \mathbb G_n^\star(P)$ with $\mathbb G_n^\star(P) \sim N(0,\Sigma(P))$, independent of $\{Z_i\}_{i=1}^n$, and satisfying
\begin{equation*}
\|(\Omega^{\rm e}(P))^\dagger\{\hat {\mathbb G}^{\rm e}_n - \mathbb G^{\rm e\star}_n(P)\}\|_\infty \vee \|(\Omega^{\rm i}(P))^\dagger\{\hat {\mathbb G}^{\rm i}_n - \mathbb G^{\rm i\star }_n(P)\}\|_\infty= O_P(b_n)
\end{equation*}
uniformly in $P\in \mathbf P$.
Since $r_n = o(1)$ implies $a_n = o(1)$, we may apply Theorem \ref{th:master} with $\hat {\mathbb W}_n = \hat {\mathbb G}_n$, $\mathbb W_n(P) = \mathbb G_n^\star(P)$, and $\hat f_n(s,P) = \mathbb U_n(s,P)$ to obtain
\begin{align}\label{th:size15}
\max&\{\sup_{s\in \hat{\mathcal V}_n^{\rm e}} \langle s, \hat{\mathbb G}_n^{\rm e}\rangle, \sup_{s\in \hat{\mathcal V}_n^{\rm i}} \langle A^\dagger s, A^\dagger \hat{\mathbb G}_n^{\rm i}\rangle + \mathbb U_n(s,P)\} \notag \\
 = & \max\{\sup_{s\in \mathcal V^{\rm e}(P)} \langle s, {\mathbb G}_n^{\rm e\star }(P)\rangle, \sup_{s\in \mathcal V^{\rm i}(P)} \langle A^\dagger s, A^\dagger \mathbb G_n^{ \rm i\star}(P)\rangle + \mathbb U_n(s,P)\} + O_P(b_n)\notag \\
 = &\max\{\max_{s\in \mathcal E^{\rm e}(P)} \langle s, (\Omega^{\rm e}(P))^\dagger {\mathbb G}_n^{\rm e\star }(P)\rangle, \max_{s\in \mathcal E^{\rm i}(P)} \langle s, \mathbb G_n^{\rm i \star}(P) + \sqrt n \beta(P)\rangle\} + O_P(b_n)
\end{align}
uniformly in $P \in \mathbf P_0$, and where the second equality follows by arguing as in \eqref{th:size4}.
Therefore, defining $c^{(2)}_n(\eta,P)$ to be the following $\eta$ quantile
\begin{equation*}
c^{(2)}_n(\eta,P) \equiv \inf\{ u : P(\max_{({\rm j},s) \in \mathcal A_n(P)} \mathbb A_n^{\rm j}(s,P) \leq u) \geq \eta\},
\end{equation*}
we obtain from $E_{2n}(P)$ implying that $\hat c_n(\eta)$ is $P$-almost surely bounded from below by the conditional on $\{Z_i\}_{i=1}^n$ $\eta$ quantile of \eqref{th:size13} for any $P\in \mathbf P_0$, \eqref{th:size12} and \eqref{th:size15}, $\mathbb G_n(P)$ and $\mathbb G_n^\star(P)$ sharing the same distribution, $\mathbb G_n^\star(P)$ being independent of $\{Z_i\}_{i=1}^n$, Lemma 11 in \cite{chernozhukov:lee:rosen:2013}, and $b_n = o(\ell_n)$ that
\begin{equation}\label{th:size17}
 \limsup_{n\rightarrow \infty} \sup_{P\in \mathbf P_0} P(\phi_n = 1)
 \leq \limsup_{n\rightarrow \infty} \sup_{P\in \mathbf P_0} P(\max_{({\rm j},s) \in \mathcal A_n(P)} \mathbb A_n^{\rm j}(s,P)> c^{(2)}_n(\eta_n,P) - 3\ell_n)
\end{equation}
for some sequence $\eta_n$ satisfying $\eta_n \uparrow \eta$.

To conclude, for any $({\rm j},s) \in \mathcal A_n(P)$ we define the random variable
\begin{equation*}
\mathbb N(({\rm j},s),P) \equiv \frac{\mathbb A^{\rm j}_n(s,P) - c_n^{(2)}(\eta_n,P)}{\sigma^{\rm j}(s,P)} + \frac{c^{(1)}_n(\tau_n,P) + 0 \vee c^{(2)}_n(\eta_n,P)}{\underline{\sigma}(P)}.
\end{equation*}
Then note that $E[\mathbb N(({\rm j},s),P)] \geq 0$ for any $({\rm j},s) \in \mathcal A_n(P)$, by definition of $\mathcal E^{{\rm i},\tau_n}(P)$, $c_n^{(1)}(\eta_n,P) \geq 0$, and $\sigma^{\rm j}(s,P) \geq \underline{\sigma}(P)$ for all $({\rm j},s)\in \mathcal A_n(P)$.
Thus, since in addition $\text{Var}\{\mathbb N(({\rm j},s),P)\} = 1$ for any $({\rm j},s) \in \mathcal A_n(P)$ and $\mathcal A_n(P)$ is finite due to $\mathcal E^{\rm e}(P)$ and $\mathcal E^{\rm i}(P)$ being finite by Corollary 19.1.1 in \cite{rockafellar1970convex}, Lemma \ref{lm:auxdensity} implies
\begin{align}\label{th:size19}
P(|\max_{({\rm j},s) \in \mathcal A_n(P)} & \mathbb A_n^{\rm j}(s,P) - c^{(2)}_n(\eta_n,P)|\leq 3\ell_n) \notag \\
& \leq  P(|\max_{({\rm j},s) \in \mathcal A_n(P)} \mathbb N(({\rm j},s),P) - \frac{ c^{(1)}_n(\tau_n,P) + 0 \vee c^{(2)}_n(\eta_n,P)}{\underline{\sigma}(P)} |\leq \frac{3\ell_n}{\underline{\sigma}(P)}) \notag \\
& \leq \frac{12\ell_n}{\underline{\sigma}(P)} \max\{\text{med}\{\max_{({\rm j},s)\in \mathcal A_n(P)} \mathbb N(({\rm j},s),P)\},1\}
\end{align}
for any $P\in \mathbf P_0$.
Next note the definition of $\mathbb N(({\rm j},s),P)$, $\Omega^{\rm j}(P)(\Omega^{\rm j}(P))^\dagger \mathbb G_n^{\rm j}(P) = \mathbb G_n^{\rm j}(P)$ for $\rm j \in \{\rm e,\rm i\}$, $\mathcal E^{\rm e}(P) \subset \Omega^{\rm e}(P)\mathcal V^{\rm e}(P)$, and $\mathcal E^{{\rm i},\tau_n}(P) \subseteq (AA^\prime)^\dagger \mathcal V^{\rm i}(P)$ imply that
\begin{align}\label{th:size20}
\text{med}\{& \max_{({\rm j},s)\in \mathcal A_n(P)} \mathbb N(({\rm j},s),P)\} \notag \\
\leq &\frac{1}{\underline{\sigma}(P)}\{\text{med}\{\sup_{s\in \mathcal V^{\rm e}(P)} \langle s,\mathbb G_n^{\rm e}(P)\rangle \vee\sup_{s\in \mathcal V^{\rm i}(P)} \langle A^\dagger s, A^\dagger \mathbb G_n^{\rm i}(P)\rangle\}  + c_n^{(1)}(\tau_n,P) + |c_n^{(2)}(\eta_n,P)|\} \notag \\
 = & \frac{\text{m(P)}}{\underline{\sigma}(P)} + \frac{c_n^{(1)}(\tau_n,P) + |c_n^{(2)}(\eta_n,P)|}{\underline{\sigma}(P)}
\end{align}
for all $P\in \mathbf P_0$ and $n$.
Furthermore, by Borell's inequality (see, for example, the corollary in pg. 82 of \cite{davydov:lifshits:smorodina:1998}) we also have the bound
\begin{equation}\label{th:size21}
c_n^{(1)}(\tau_n,P) \leq \text{m}(P) + z_{\tau_n}\bar\sigma(P)
\end{equation}
for all $P\in \mathbf P$ and $n$ sufficiently large due to $\tau_n \uparrow 1$.
Since $P\in \mathbf P_0$ implies $\langle s, \beta(P)\rangle \leq 0$ for any $s\in \mathcal E^{{\rm i},\tau_n}(P) \subset (AA^\prime)^\dagger \mathcal V^{\rm i}(P)$ by Theorem \ref{th:ineq}, we can obtain from Borell's inequality, $\eta_n \uparrow \eta > 1/2$, and definition of $\text{m}(P)$ that
\begin{equation}\label{th:size22}
c_n^{(2)}(\eta_n,P) \leq \text{m}(P) + \bar \sigma(P) z_{\eta_n}
\end{equation}
for $n$ sufficiently large.
Also, $\eta_n > 1/2$ for $n$ sufficiently large and $0 \geq \langle s, \sqrt n\beta(P)\rangle  \geq -c_n^{(1)}(\tau_n,P)$ for all $s\in \mathcal E^{{\rm i},\tau_n}(P)$ by result \eqref{th:size6} imply that
\begin{align}\label{th:size23}
c_n^{(2)} (\eta_n,P) & \geq\text{med}\{\max_{s\in \mathcal E^{\rm e}(P):\sigma^{\rm e}(s,P) >0} \mathbb A_n^{\rm e}(s,P) \vee \max_{s\in \mathcal E^{{\rm i},\tau_n}(P):\sigma^{\rm i}(s,P)>0} \langle s,\mathbb G_n^{\rm i}(P)\rangle\} - c_n^{(1)}(\tau_n,P) \notag \\
& \geq  - c_{n}^{(1)}(\tau_n,P) ,
\end{align}
where in the last inequality we employed that $E[\mathbb A_n^{\rm e}(s,P)] = 0$ for all $s \in \mathcal E^{\rm e}(P)$ and $E[\langle s, \mathbb G_n^{\rm i}(P)\rangle ] = 0$ for all $s\in \mathcal E^{\rm i}(P)$ imply $\text{med}\{\mathbb A_n^{\rm e}(s,P)\} \geq 0$ for any $s\in \mathcal E^{\rm e}(P)$ and $\text{med}\{\langle s,\mathbb G_n^{\rm i}(P)\rangle\} \geq 0$ for any $s\in \mathcal E^{\rm i}(P)$.
Therefore, results \eqref{th:size19}, \eqref{th:size20}, \eqref{th:size21}, \eqref{th:size22}, \eqref{th:size23}, $\tau_n \uparrow 1$ implying $z_{\tau_n} \uparrow \infty$, and $\ell_n$ satisfying \eqref{th:size8} yield
\begin{multline}\label{th:size24}
\limsup_{n\rightarrow \infty} \sup_{P\in \mathbf P_0} P(|\max_{({\rm j},s) \in \mathcal A_n(P)}  \mathbb A_n^{\rm j}(s,P) - c^{(2)}_n(\eta_n,P)|\leq 3\ell_n) \\
\lesssim \limsup_{n\rightarrow \infty} \sup_{P\in \mathbf P_0}  \frac{\ell_n(\text{m}(P) + z_{\tau_n}\bar \sigma(P))}{\underline{\sigma}^2(P)} = 0.
\end{multline}
Thus, \eqref{th:size17} and \eqref{th:size24} together with the definition of $c^{(2)}_n(\eta_n,P)$ and $\eta_n \uparrow \eta$ imply
\begin{equation*}
 \limsup_{n\rightarrow \infty} \sup_{P\in \mathbf P_0} P(\phi_n = 1)  \leq \limsup_{n\rightarrow \infty} \sup_{P\in \mathbf P_0} P(\max_{({\rm j},s) \in \mathcal A_n(P)}  \mathbb A_n^{\rm j}(s,P)  > c^{(2)}_n(\eta_n,P))
  \leq 1-\eta .
\end{equation*}
Since $\eta = 1-\alpha$, the claim of the theorem therefore follows. \qed

\begin{lemma}\label{lm:first}
Let Assumptions \ref{ass:weights}, \ref{ass:beta}, \ref{ass:moments}, \ref{ass:techrange}(i) hold, $\lambda_n \in [0,1]$, and $r_n = o(1)$.
Then, for any sequence $\ell_n$ satisfying $\lambda_n \sqrt{\log(1+p)} = o(\ell_n)$ it follows that
$$\liminf_{n\rightarrow \infty} \inf_{P\in \mathbf P_0} P(\sup_{s\in \hat{\mathcal V}_n^{\rm i}} \{\sqrt n \langle A^\dagger s , A^\dagger \beta(P)\rangle - \hat{\mathbb U}_n(s)\} \leq \ell_n) = 1.$$
\end{lemma}

\noindent \emph{Proof:} First note Theorem \ref{th:ineq} implies  $\langle A^\dagger s, A^\dagger \beta(P)\rangle \leq 0$ for all $s\in \hat {\mathcal V}_n^{\rm i}$ and $P\in \mathbf P_0$.
Therefore, the definitions of $\hat{\mathbb U}_n(s)$ and $\lambda_n\in[0,1]$ imply
\begin{multline}\label{lm:first1}
\sup_{s\in \hat{\mathcal V}_n^{\rm i}} \sqrt n \langle A^\dagger s, A^\dagger \beta(P)\rangle - \hat {\mathbb U}_n(s) \leq \sup_{s\in \hat{\mathcal V}_n^{\rm i}} \lambda_n \sqrt n \langle A^\dagger s, A^\dagger \{\beta(P) - \hat \beta_n^{\rm r}\}\rangle \\
\leq \sup_{s\in \hat{\mathcal V}_n^{\rm i}} \lambda_n \sqrt n |\langle A^\dagger s, \hat x_n^\star - A^\dagger \hat \beta_n^{\rm r}\rangle| + \sup_{s\in \hat{\mathcal V}_n^{\rm i}} \lambda_n \sqrt n |\langle A^\dagger s, A^\dagger \beta(P) - \hat x_n^\star\rangle|.
\end{multline}
Moreover, the definition of $\hat \beta_n^{\rm r}$ in \eqref{step1:eq2}, $\hat x_n^\star \equiv A^\dagger \hat C_n \hat \beta_n$ with $\hat C_n\beta(P) = \beta(P)$ for any $P\in \mathbf P_0$ by Assumption \ref{ass:beta}(ii), $\beta(P) \in R$ for any $P\in \mathbf P_0$, and \eqref{lm:first1} yield
\begin{multline}\label{lm:first2}
\sup_{s\in \hat{\mathcal V}_n^{\rm i}} \sqrt n \langle A^\dagger s, A^\dagger \beta(P)\rangle - \hat {\mathbb U}_n(s)  \leq \sup_{s\in \hat{\mathcal V}_n^{\rm i}} 2 \lambda_n |\langle A^\dagger s, \sqrt n \{\hat x_n^\star - A^\dagger \beta(P)\}\rangle|  \\
 = \sup_{s\in \hat{\mathcal V}_n^{\rm i}} 2 \lambda_n |\langle A^\dagger s, A^\dagger AA^\dagger \hat C_n \sqrt n \{\hat \beta_n - \beta(P)\}\rangle|.
\end{multline}
By applying Theorem \ref{th:master} twice, once with $\hat {\mathbb W}_n^{\rm i}(P)= AA^\dagger \hat C_n\sqrt n \{\hat \beta_n - \beta(P)\}$  and $\hat{\mathbb W}_n^{\rm e}(P) = \mathbb G_n^{\rm e}(P)$, and once with $\hat {\mathbb W}_n^{\rm i}(P)=  AA^\dagger \hat C_n\sqrt n \{\beta(P) - \hat \beta_n\}$ and $\hat{\mathbb W}_n^{\rm e}(P) = -\mathbb G_n^{\rm e}(P)$, and in both cases setting $\hat f_n(s,P) = 0$ for all $s \in \mathbf R^p$, we obtain from Lemma \ref{lm:betacoup} and $(-\mathbb G_n^{\rm e}(P)^\prime, -\mathbb G_n^{\rm i}(P)^\prime)^\prime\sim N(0,\Sigma(P))$ that uniformly in $P\in \mathbf P_0$
\begin{align}\label{lm:first3}
\sup_{s\in \hat {\mathcal V}_n^{\rm i}} \langle A^\dagger s, A^\dagger AA^\dagger \hat C_n\sqrt n \{\beta(P) - \hat \beta_n\}\rangle & = \sup_{s\in \mathcal V^{\rm i}(P)} \langle A^\dagger s, A^\dagger (-\mathbb G_n^{\rm i}(P))\rangle + O_P(r_n) \notag \\
\sup_{s\in \hat {\mathcal V}_n^{\rm i}} \langle A^\dagger s, A^\dagger AA^\dagger \hat C_n\sqrt n \{\hat \beta_n - \beta(P)\}\rangle & = \sup_{s\in \mathcal V^{\rm i}(P)} \langle A^\dagger s, A^\dagger \mathbb G_n^{\rm i}(P)\rangle + O_P(r_n).
\end{align}
Since $\Omega^{\rm i}(P)(\Omega^{\rm i}(P))^\dagger \mathbb G_n^{\rm i}(P) = \mathbb G_n^{\rm i}(P)$ almost surely due to $\mathbb G_n^{\rm i}(P) \in \text{range}\{\Sigma^{\rm i}(P)\}\subseteq \text{range}\{\Omega^{\rm i}(P)\}$ almost surely by Theorem 3.6.1 in \cite{bogachev1998gaussian} and Assumption \ref{ass:techrange}(i), we obtain from results \eqref{lm:first2}, \eqref{lm:first3}, and H\"older's inequality that
\begin{multline}\label{lm:first4}
\sup_{s\in \hat{\mathcal V}_n^{\rm i}} 2 \lambda_n |\langle A^\dagger s, A^\dagger AA^\dagger \hat C_n \sqrt n \{\hat \beta_n - \beta(P)\}\rangle| =
\sup_{s\in \mathcal V^{\rm i}(P)} 2 \lambda_n |\langle A^\dagger s, A^\dagger \mathbb G_n^{\rm i}(P)\rangle| + O_P(\lambda_n r_n) \\
\leq 2\lambda_n \|(\Omega^{\rm i}(P))^\dagger \mathbb G_n^{\rm i}(P)\|_\infty + O_P(\lambda_n r_n) = O_P(\lambda_n \sqrt{\log(1+p)})
\end{multline}
 uniformly in $P\in \mathbf P_0$, and where the final equality follows from $r_n = o(1)$, Markov's inequality, and $\sup_{P\in \mathbf P} E_P[\|(\Omega^{\rm i}(P))^\dagger \mathbb G_n^{\rm i}(P)\|_\infty] \lesssim \sqrt{\log(1+p)}$ by Lemma \ref{lm:obvious} and Assumption \ref{ass:moments}(ii).
The claim of the Lemma then follows from results \eqref{lm:first2}, \eqref{lm:first4}, and $\lambda_n\sqrt{\log(1+p)} = o(\ell_n)$ by hypothesis. \qed

\begin{theorem}\label{th:master}
Let Assumptions \ref{ass:weights}, \ref{ass:moments}(ii), \ref{ass:techrange}(i) hold, $a_n = o(1)$, set $\Sigma(P) \equiv E_P[\psi(X,P)\psi(X,P)^\prime]$, and suppose $(\hat {\mathbb W}_n^{\text{\rm e}}(P)^\prime,\hat {\mathbb W}_n^{\text{\rm i}}(P)^\prime)^\prime \equiv \hat {\mathbb W}_n(P)$ satisfies
\begin{equation}\label{th:masterdisp1}
\|(\Omega^{\rm e}(P))^\dagger\{\hat {\mathbb W}_n^{\rm e}(P) - \mathbb W_n^{\rm e}(P)\}\|_\infty \vee \|(\Omega^{\rm i}(P))^\dagger\{\hat {\mathbb W}_n^{\rm i}(P) - \mathbb W_n^{\rm i}(P)\}\|_\infty = O_P(\omega_n)
\end{equation}
for $\omega_n > 0$, $\mathbb W_n(P) \equiv (\mathbb W_n^{\text{\rm e}}(P)^\prime,\mathbb W_n^{\text{\rm i}}(P)^\prime)^\prime \sim N(0,\Sigma(P))$, and, for ${\rm j}\in \{\rm e, \rm i\}$, $\hat{\mathbb W}_n^{\rm j}(P) \in \text{\rm range}\{\Omega^{\rm j}(P)\}$ with probability tending to one uniformly in $P\in \mathbf P$.
Then, for any $\mathbf Q \subseteq \mathbf P$ and possibly random function $\hat f_n(\cdot,P) :\mathbf R^p \to \mathbf R$ satisfying
\begin{equation}\label{th:masterdisp2}
\gamma \hat f_n( s,P) \leq\hat f_n( \gamma s,P)\leq 0
\end{equation}
for all $s$ with $A^\dagger s \leq 0$, $\gamma \in [0,1]$, and $P\in \mathbf Q$, it follows uniformly in $P\in \mathbf Q$ that
\begin{align*}
\sup_{s\in \hat{\mathcal V}^{\rm e}_n} \langle s, \hat{\mathbb W}_n^{\rm e}(P)\rangle & = \sup_{s\in {\mathcal V}^{\rm e}(P)} \langle s, {\mathbb W}_n^{\rm e}(P)\rangle  + O_P(\omega_n + a_n) \\
\sup_{s\in \hat{\mathcal V}^{\rm i}_n} \langle A^\dagger s, A^\dagger \hat{\mathbb W}_n^{\rm i}(P)\rangle +\hat f_n(s,P) & = \sup_{s\in {\mathcal V}^{\rm i}(P)} \langle A^\dagger s, A^\dagger {\mathbb W}_n^{\rm i}(P)\rangle +\hat f_n(s,P) + O_P(\omega_n + a_n).
\end{align*}
\end{theorem}

\noindent \emph{Proof:}
We establish only the second claim of the theorem, noting that the first claim follows from slightly simpler but largely identical arguments.
First note that since $\Omega^{\rm i}(P)(\Omega^{\rm i}(P))^\dagger \hat{\mathbb W}_n^{\rm i}(P) = \hat {\mathbb W}_n^{\rm i}(P)$ whenever $\hat {\mathbb W}_n^{\rm i}(P) \in \text{range}\{\Omega^{\rm i}(P)\}$, it follows
\begin{equation}\label{th:master1}
\sup_{s\in \hat{\mathcal V}^{\rm i}_n} \langle A^\dagger s, A^\dagger \hat{\mathbb W}_n^{\rm i}(P)\rangle +\hat f_n(s,P) \\ = \sup_{s\in \hat{\mathcal V}^{\rm i}_n} \langle A^\dagger s, A^\dagger \Omega^{\rm i}(P)(\Omega^{\rm i}(P))^\dagger\hat {\mathbb W}_n^{\rm i}(P)\rangle +\hat f_n(s,P)
\end{equation}
with probability tending to one uniformly in $P\in \mathbf P$.
Further note that Lemma \ref{lm:auxpseudo} and Assumption \ref{ass:weights}(iii) imply $\hat \Omega_n^{\rm i}(\hat \Omega_n^{\rm i})^\dagger \Omega^{\rm i}(P) = \Omega^{\rm i}(P)$ with probability tending to one uniformly in $P\in \mathbf P$. Thus, since $\hat \Omega_n^{\rm i}$ and $\Omega^{\rm i}(P)$ are symmetric by Assumption \ref{ass:weights}(i)(ii), it follows that $\Omega^{\rm i}(P) = \Omega^{\rm i}(P)(\hat \Omega_n^{\rm i})^\dagger \hat \Omega_n^{\rm i}$ with probability tending to one uniformly in $P\in \mathbf P$.
The triangle inequality, the definition of $\hat {\mathcal V}_n^{\rm i}$, and $\hat \Omega_n^{\rm i} (\hat \Omega_n^{\rm i})^\dagger \hat \Omega_n^{\rm i} = \hat \Omega_n^{\rm i}$ by Proposition 6.11.1(6) in \cite{luenberger:1969} then yield
\begin{multline}\label{th:master2}
\sup_{s\in \hat{\mathcal V}_n^{\rm i}} \|\Omega^{\rm i}(P)(AA^\prime)^\dagger s\|_1 \leq 1+ \sup_{s\in \hat{\mathcal V}_n^{\rm i}} \|(\hat \Omega^{\rm i}_n - \Omega^{\rm i}(P))(AA^\prime)^\dagger s\|_1  \\ = 1+ \sup_{s\in \hat{\mathcal V}_n^{\rm i}} \|(\hat \Omega^{\rm i}_n - \Omega^{\rm i}(P))(\hat \Omega_n^{\rm i})^\dagger \hat \Omega^{\rm i}_n(AA^\prime)^\dagger s\|_1 \leq 1 + \|(\hat \Omega_n^{\rm i} - \Omega^{\rm i}(P))(\hat \Omega_n^{\rm i})^\dagger\|_{o,1}
\end{multline}
with probability tending to one uniformly in $P\in \mathbf P$.
Further note that Theorem 6.5.1 in \cite{luenberger:1969}, symmetry of $\hat \Omega_n^{\rm i}$ and $\Omega^{\rm i}(P)$, and Lemma \ref{lm:obviousmatrix} imply
\begin{equation}\label{th:master3}
\|(\hat \Omega_n^{\rm i} - \Omega^{\rm i}(P))(\hat \Omega_n^{\rm i})^\dagger\|_{o,1} = \|(\hat \Omega_n^{\rm i})^\dagger(\hat \Omega_n^{\rm i} - \Omega^{\rm i}(P))\|_{o,\infty} = O_P(\frac{a_n}{\sqrt{\log(1+p)}})
\end{equation}
uniformly in $P\in \mathbf P$.
Next, note that since $\Omega^{\rm i}(P)(A^\dagger)^\prime A^\dagger = \Omega^{\rm i}(P) (AA^\prime)^\dagger$ (see, e.g.,  \cite{seber2008matrix} pg. 139), H\"older's inequality, and results \eqref{th:master2} and \eqref{th:master3} yield
\begin{multline}\label{th:master4}
\sup_{s\in \hat{\mathcal V}_n^{\rm i}} |\langle A^\dagger s, A^\dagger \Omega^{\rm i}(P)(\Omega^{\rm i}(P))^\dagger(\hat {\mathbb W}_n^{\rm i}(P) - \mathbb W_n^{\rm i}(P))\rangle|
\\ \leq (1 + O_P(\frac{a_n}{\sqrt{\log(1+p)}}))\|(\Omega^{\rm i}(P))^\dagger(\hat {\mathbb W}_n^{\rm i}(P) - \mathbb W_n^{\rm i}(P))\|_\infty = O_P(\omega_n)
\end{multline}
uniformly in $P\in \mathbf P$, and where the final equality follows from $a_n = o(1)$.
Therefore, combining \eqref{th:master1} and \eqref{th:master4} we obtain uniformly in $P\in \mathbf P$ that
\begin{multline}\label{th:master5}
\sup_{s\in \hat{\mathcal V}^{\rm i}_n} \langle A^\dagger s, A^\dagger \hat{\mathbb W}_n^{\rm i}(P)\rangle +\hat f_n(s,P) \\
 = \sup_{s\in \hat{\mathcal V}^{\rm i}_n} \langle A^\dagger s, A^\dagger \Omega^{\rm i}(P)(\Omega^{\rm i}(P))^\dagger \mathbb W_n^{\rm i}(P)\rangle +\hat f_n(s,P) + O_P(\omega_n).
\end{multline}

We next replace $\hat {\mathcal V}_n^{\rm i}$ with $\mathcal V^{\rm i}(P)$ in \eqref{th:master5}.
To this end, let $\hat s_n \in \hat {\mathcal V}_n^{\rm i}$ satisfy
\begin{multline}\label{th:master6}
\langle A^\dagger \hat s_n ,A^\dagger \Omega^{\rm i}(P)(\Omega^{\text{i}}(P))^\dagger \mathbb W_n^{\text{i}}(P)\rangle + \hat f_n(\hat s_n,P)\\
 = \sup_{s\in \hat{\mathcal V}_n^{\rm i}} \langle A^\dagger s,A^\dagger \Omega^{\rm i}(P)(\Omega^{\text{i}}(P))^\dagger \mathbb W_n^{\text{i}}(P)\rangle + \hat f_n(s,P) + O(\omega_n ),
\end{multline}
where note $\hat s_n$ is random and \eqref{th:master6} is meant to hold surely.
Set $\bar s_n \equiv \gamma_n \hat s_n$ with
\begin{equation}\label{th:master7}
\gamma_n \equiv (\|\Omega^{\rm i}(P)(AA^\prime)^\dagger \hat s_n\|_{1} \vee 1)^{-1} \in [0,1],
\end{equation}
and note that since $\gamma_n \leq 1$, result \eqref{th:master2} and $\hat s_n \in \hat {\mathcal V}_n^{\rm i}$ allow us to conclude that
\begin{equation}\label{th:master8}
0 \leq 1- \gamma_n \leq 1 - (1+ \|(\hat \Omega_n^{\rm i} - \Omega^{\rm i}(P))(\hat \Omega_n^{\rm i})^\dagger\|_{o,1})^{-1}
\end{equation}
with probability tending to one uniformly in $P\in \mathbf P$.
Hence,  \eqref{th:master3}, \eqref{th:master8} yield
\begin{equation}\label{th:master9}
0 \leq 1 -\gamma_n \leq  O_P(\frac{a_n}{\sqrt{\log(1+p)}})
\end{equation}
uniformly $P\in \mathbf P$ due to $a_n = o(1)$.
Next, we note $A^\dagger \hat s_n \leq 0$ since $\hat s_n \in \hat {\mathcal V}^{\rm i}_n$ and therefore $A^\dagger \bar s_n = \gamma_nA^\dagger \hat s_n \leq 0$ because $\gamma_n \geq 0$.
Since $\bar s_n = \gamma_n \hat s_n$ and \eqref{th:master7} imply
\begin{equation}\label{th:master10}
\|\Omega^{\rm i}(P)(AA^\prime)^\dagger \bar s_n\|_1 = (\|\Omega^{\rm i}(P)(AA^\prime)^\dagger \hat s_n\|_{1} \vee 1)^{-1}\|\Omega^{\rm i}(P)(AA^\prime)^\dagger \hat s_n\|_1 \leq 1,
\end{equation}
it follows  $\bar s_n \in \mathcal V^{\rm i}(P)$.
Moreover, $\hat s_n - \bar s_n = (1-\gamma_n)\hat s_n$, $\gamma_n \hat f_n( \hat s_n,P) \leq \hat f_n( \gamma_n \hat s_n,P)$ and $\hat f_n(\hat s_n,P) \leq 0$ for all $P\in \mathbf Q$ by \eqref{th:masterdisp2}, and H\"older's inequality yield
\begin{align*}
\langle A^\dagger(\hat s_n - \bar s_n),& A^\dagger \Omega^{\rm i}(P)(\Omega^{\rm i}(P))^\dagger \mathbb W_n^{\text{i}}(P)\rangle + \hat f_n(\hat s_n,P) - \hat f_n(\bar s_n,P) \notag \\
& \leq (1-\gamma_n)\{\langle A^\dagger \hat s_n,A^\dagger \Omega^{\rm i}(P)(\Omega^{\rm i}(P))^\dagger  \mathbb W_n^{\text{i}}(P)\rangle + \hat f_n(\hat s_n,P)\} \notag \\
& \leq (1 -\gamma_n)\{\sup_{s\in \hat {\mathcal V}_n^{\rm i}} \|\Omega^{\rm i}(P)(AA^\prime)^\dagger s\|_1\}\|(\Omega^{\rm i}(P))^\dagger \mathbb W_n^{\text{i}}(P)\|_\infty .
\end{align*}
In particular, since $\sup_{P\in \mathbf P} E_P[\|(\Omega^{\rm i}(P))^\dagger \mathbb W_n^{\text{i}}(P)\|_\infty]\lesssim \sqrt{\log(1+p)}$ by Lemma \ref{lm:obvious} and Assumption \ref{ass:moments}(ii), Markov's inequality, results \eqref{th:master2}, \eqref{th:master3}, \eqref{th:master6}, and \eqref{th:master9}, and $\bar s_n \in \mathcal V^{\rm i}(P)$ allow us to conclude that uniformly in $P\in \mathbf Q$ we have
\begin{multline}\label{th:master11}
\sup_{s\in \hat{\mathcal V}^{\rm i}_n} \langle A^\dagger s,A^\dagger \Omega^{\rm i}(P)(\Omega^{\rm i}(P))^\dagger \mathbb W_n^{\text{i}}(P)\rangle + \hat f_n(s,P) \\
\leq \sup_{s\in {\mathcal V}^{\rm i}(P)} \langle A^\dagger s,A^\dagger \Omega^{\rm i}(P)(\Omega^{\rm i}(P))^\dagger \mathbb W_n^{\text{i}}(P)\rangle + \hat f_n(s,P) +O_P(\omega_n + a_n)
\end{multline}
uniformly in $P\in \mathbf Q$.
The reverse inequality to \eqref{th:master11} can be established by similar arguments, and therefore we can conclude that uniformly in $P\in \mathbf Q$ we have
\begin{multline}\label{th:master12}
\sup_{s\in \hat{\mathcal V}_n^{\rm i}} \langle A^\dagger s,A^\dagger \Omega^{\rm i}(P)(\Omega^{\rm i}(P))^\dagger \mathbb W_n^{\text{i}}(P)\rangle + \hat f_n(s,P) \\
= \sup_{s\in \mathcal V^{\rm i}(P)} \langle A^\dagger s,A^\dagger \Omega^{\rm i}(P)(\Omega^{\rm i}(P))^\dagger \mathbb W_n^{\text{i}}(P)\rangle + \hat f_n(s,P) + O_P(\omega_n + a_n).
\end{multline}
Finally, note $\mathbb W_n^{\rm i}(P)$ almost surely belongs to the range of $\Sigma^{\rm i}(P):\mathbf R^p \to \mathbf R^p$ by Theorem 3.6.1 in \cite{bogachev1998gaussian}.
Hence, since Assumption \ref{ass:techrange}(i) implies $\Omega^{\rm i}(P)(\Omega^{\rm i}(P))^\dagger \Sigma^{\rm i}(P)$ it follows that $\mathbb W_n^{\rm i}(P) = \Omega^{\rm i}(P)(\Omega^{\rm i}(P))^\dagger \mathbb W_n^{\rm i}(P)$ $P$-almost surely. The second claim of the theorem thus follows from \eqref{th:master5}, and \eqref{th:master12}. \qed

\begin{lemma}\label{aux:critbound}
Let Assumptions \ref{ass:weights}, \ref{ass:beta}(i)(ii), \ref{ass:moments}, \ref{ass:techrange}, \ref{ass:4bootnew} hold, $\eta \in (0.5,1)$, $\epsilon \in (0, \eta -0.5)$, $z_\eta$ be the $\eta$ quantile of $N(0,1)$, and $r_n \vee b_n = o(1)$.
If $\sup_{P\in \mathbf P} ({\rm m}(P) + \bar \sigma(P))/\underline{\sigma}^2(P) = o(r_n^{-1}\wedge b_n^{-1})$, then for each $P\in \mathbf P_0$ there are $\{E_n(P)\}$ with
\begin{equation}\label{aux:critbounddis}
\liminf_{n\rightarrow \infty} \inf_{P\in \mathbf P_0} P(\{Z_i\}_{i=1}^n \in E_n(P)) = 1
\end{equation}
and on $E_n(P)$ it holds that $\hat c_n(\eta) \geq (\underline{\sigma}(P)z_{\eta-\epsilon})/2$ whenever $T_n > 0$.
\end{lemma}

\noindent \emph{Proof:} First note that by Lemma \ref{lm:finitebootnew} there is a $(\mathbb G_n^{\text{e}\star}(P)^\prime, \mathbb G_n^{\text{i}\star}(P)^\prime)^\prime \equiv \mathbb G_n^\star(P) \sim N(0,\Sigma(P))$ that is independent of $\{Z_i\}_{i=1}^n$ and satisfies
\begin{equation*}
\|(\Omega^{\rm e}(P))^\dagger\{\hat {\mathbb G}^{\rm e}_n - \mathbb G_n^{\rm e \star}(P)\}\|_\infty \vee \|(\Omega^{\rm i}(P))^\dagger\{\hat {\mathbb G}^{\rm i}_n - \mathbb G_n^{\rm i \star}(P)\}\|_\infty = O_P(b_n)
\end{equation*}
uniformly in $P\in \mathbf P$.
Further define $\hat {\mathbb L}_n\in \mathbf R$ and $\mathbb L_n^\star(P) \in \mathbf R$ to be given by
\begin{align}
\hat{\mathbb L}_n & \equiv \max\{\sup_{s\in \hat{\mathcal V}_n^{\rm e}} \langle s, \hat{\mathbb G}_n^{\rm e}\rangle, \sup_{s\in \hat{\mathcal V}_n^{\rm i}} \langle A^\dagger s, A^\dagger \hat {\mathbb G}_n^{\rm i}\rangle  + \hat {\mathbb U}_n(s) \}\label{aux:critbound2}\\
\mathbb L_n^\star(P) & \equiv \max\{\sup_{s\in {\mathcal V}^{\rm e}(P)} \langle s, {\mathbb G}_n^{\rm e \star}(P)\rangle, \sup_{s\in \mathcal V^{\rm i}(P)} \langle A^\dagger s,A^\dagger  \mathbb G_n^{\text{i}\star}(P)\rangle + \hat {\mathbb U}_n(s)\} \label{aux:critbound3},
\end{align}
and note that since $\langle A^\dagger s, A^\dagger \hat \beta_n^{\rm r}\rangle \leq 0$ for all $s\in \mathbf R^p$ such that $A^\dagger s \leq 0$ by Theorem \ref{th:ineq}, it follows from Lemma \ref{lm:finitebootnew}, Assumptions \ref{ass:techrange}(i) and \ref{ass:4bootnew}(v), and Theorem \ref{th:master} applied with $\mathbb W_n(P) = \mathbb G_n^\star(P)$, $\hat {\mathbb W}_n = \hat {\mathbb G}_n$, and $\hat f_n(\cdot,P) = \hat {\mathbb U}_n(\cdot)$ that
\begin{align}
\sup_{s\in \hat{\mathcal V}_n^{\rm e}} \langle s,\hat{\mathbb G}_n^{\rm e\star}\rangle  & = \sup_{s\in {\mathcal V}^{\rm e}(P)} \langle s,{\mathbb G}_n^{\rm e\star}(P)\rangle + O_P(b_n) \label{aux:critbound4p1}\\
\sup_{s\in \hat{\mathcal V}_n^{\rm i}} \langle A^\dagger s,A^\dagger \hat{\mathbb G}_n^{\rm i\star}\rangle + \hat{\mathbb U}_n(s) & = \sup_{s\in {\mathcal V}^{\rm i}(P)} \langle A^\dagger s,A^\dagger {\mathbb G}_n^{\rm i\star}(P)\rangle + \hat{\mathbb U}_n(s)+ O_P(b_n) \label{aux:critbound4p2}
\end{align}
uniformly in $P\in \mathbf P$. We establish the lemma by studying three separate cases.

\noindent \underline{Case I:} Suppose $P\in \mathbf P_0^{\rm e} \equiv \{P\in \mathbf P_0 : \sigma^{\rm e}(s,P) > 0 \text{ for some } s \in \mathcal E^{\rm e}(P)\}$.
First set
\begin{equation*}
E_{n}(P)  \equiv \{P(|\sup_{s\in \hat{\mathcal V}_n^{\rm e}} \langle s, \hat{\mathbb G}_n^{\rm e}\rangle - \sup_{s\in {\mathcal V}^{\rm e}(P)} \langle s, {\mathbb G}_n^{\rm e\star}(P)\rangle| > (\underline{\sigma}(P)z_{\eta-\epsilon})/2 |\{Z_i\}_{i=1}^n) \leq \epsilon \}
\end{equation*}
and note that $z_{\eta - \epsilon} > 0$ due to $\eta - \epsilon > 0.5$, and therefore result \eqref{aux:critbound4p1}, Markov's inequality, and $b_n \times \sup_{P\in \mathbf P} 1/\underline{\sigma}(P)=o(1)$ by hypothesis, imply that
\begin{equation*}
\liminf_{n\rightarrow \infty}\inf_{P\in \mathbf P_0^{\rm e}} P(\{Z_i\}_{i=1}^n \in E_n(P)) = 1.
\end{equation*}
Then note that whenever $\{Z_i\}_{i=1}^n \in E_{n}(P)$ the triangle inequality yields
\begin{multline}\label{aux:critbound7}
P(\sup_{s\in {\mathcal V}^{\rm e}(P)} \langle s, \mathbb G_n^{\rm e\star}(P) \rangle \leq \hat c_n(\eta) + \frac{\underline{\sigma}(P)z_{\eta-\epsilon}}{2}|\{Z_i\}_{i=1}^n) \\
\geq P(\sup_{s\in \hat{\mathcal V}^{\rm e}_n} \langle s, \hat{\mathbb G}_n^{\rm e\star} \rangle \leq \hat c_n(\eta)|\{Z_i\}_{i=1}^n) - \epsilon
\geq P( \hat{\mathbb L}_n \leq \hat c_n(\eta)|\{Z_i\}_{i=1}^n) - \epsilon \geq \eta - \epsilon
\end{multline}
where the second inequality follows from \eqref{aux:critbound2}, while the final inequality holds by definition of $\hat c_n(\eta)$.
Also note that $\mathbb G_n^{\rm e\star}(P) \sim N(0,\Sigma^{\rm e}(P))$, Theorem 3.6.1 in \cite{bogachev1998gaussian}, and Assumption \ref{ass:techrange}(i) imply $\mathbb G_n^{\rm e \star}(P) = \Omega^{\rm e}(P)(\Omega^{\rm e}(P))^\dagger \mathbb G_n^{\rm e\star }(P)$ almost surely.
Therefore, by symmetry of $\Omega^{\rm e}(P)$ we obtain that almost surely
\begin{equation*}
\sup_{s\in \mathcal V^{\rm e}(P)} \langle s, \mathbb G_n^{\rm e \star}(P)\rangle  = \sup_{s\in \mathcal V^{\rm e}(P)} \langle \Omega^{\rm e}(P)s, (\Omega^{\rm e}(P))^\dagger \mathbb G_n^{\rm e \star}(P)\rangle = \max_{s\in \mathcal E^{\rm e}(P)} \langle s, (\Omega^{\rm e}(P))^\dagger \mathbb G_n^{\rm e \star}(P)\rangle,
\end{equation*}
where the second equality holds by Lemma \ref{lm:auxlp} and the supremum being finite by H\"older's inequality.
Hence,  the distribution of $\sup_{s\in \mathcal V^{\rm e}(P)} \langle s, \mathbb G_n^{\rm e \star}(P)\rangle$ first order stochastically dominates the distribution $N(0,\underline{\sigma}^2(P))$ whenever $P\in \mathbf P_0^{\text{e}}$ by definition of $\underline{\sigma}(P)$.
In particular, $\mathbb G_n^{\text{e}\star}(P)$ being independent of $\{Z_i\}_{i=1}^n$ and result \eqref{aux:critbound7} imply that whenever $\{Z_i\}_{i=1}^n \in E_n(P)$ and $P\in \mathbf P_0^{\text{e}}$ we must have
\begin{equation*}
\hat c_n(\eta) + \frac{\underline{\sigma}(P)z_{\eta-\epsilon}}{2} \geq \underline{\sigma}(P)z_{\eta-\epsilon},
\end{equation*}
which establishes the claim of the lemma for the subset $\mathbf P_0^{\text{e}} \subseteq \mathbf P_0$.

\noindent \underline{Case II:} Suppose $P\in \mathbf P_0^{\rm i} \equiv \{P\in \mathbf P_0 : \sigma^{\rm i}(s,P) > 0 \text{ for some } s\in \mathcal E^{\rm i}(P) \text{ and } \sigma^{\rm e}(s,P) = 0 \text{ for all } s\in \mathcal E^{\rm e}(P)\}$, and define the event $E_n(P) \equiv \bigcap_{j=1}^4 E_{j,n}(P)$, where
\begin{align*}
E_{1n}(P) & \equiv \{\hat {\mathcal V}_n^{\rm i} \subseteq 2\mathcal V^{\rm i}(P)\} \notag \\
E_{2n}(P) & \equiv \{AA^\dagger \hat C_n\{\hat \beta_n - \beta(P)\} \in \text{range}\{\Sigma^{\rm i}(P)\}\} \\
E_{3n}(P) & \equiv \{P(|\hat {\mathbb L}_n - \mathbb L_n^\star(P)| > (\underline{\sigma}(P)z_{\eta-\epsilon})/2 |\{Z_i\}_{i=1}^n) \leq \epsilon \} \\
E_{4n}(P) & \equiv \{T_n = \sup_{s\in \hat{\mathcal V}_n^{\rm i}} \langle A^\dagger s, \hat x_n^\star \rangle \}.
\end{align*}
Next note that $\Omega^{\rm i}(P)(\hat \Omega_n^{\rm i})^\dagger  \hat \Omega_n^{\rm i}= \Omega^{\rm i}(P)$ with probability tending to one uniformly in $P \in \mathbf P$ by Assumption \ref{ass:weights}(iii), Lemma \ref{lm:auxpseudo}, and symmetry of $\hat \Omega_n^{\rm i}$ and $\Omega^{\rm i}(P)$.
Since $\hat \Omega^{\rm i}_n (\hat \Omega^{\rm i}_n)^\dagger \hat \Omega_n^{\rm i} = \hat \Omega_n^{\rm i}$ by Proposition 6.11.1(6) in \cite{luenberger:1969}, we obtain from the definition of $\hat {\mathcal V}_n^{\rm i}$ that with probability tending to one uniformly in $P\in \mathbf P$
\begin{multline}\label{aux:critbound11}
\sup_{s\in \hat{\mathcal V}_n^{\rm i}} \|\Omega^{\rm i}(P)(AA^\prime)^\dagger s\|_1 \leq 1 + \sup_{s\in \hat{\mathcal V}_n^{\rm i}}\|(\hat \Omega_n^{\rm i}- \Omega^{\rm i}(P))(\hat \Omega_n^{\rm i})^\dagger \hat \Omega_n^{\rm i}(AA^\prime)^\dagger s\|_1 \\ \leq 1 + \|(\hat \Omega_n^{\rm i}- \Omega^{\rm i}(P))(\hat \Omega_n^{\rm i})^\dagger \|_{o,1} = 1+\|(\hat \Omega_n^{\rm i})^\dagger(\hat \Omega_n^{\rm i} - \Omega^{\rm i}(P))\|_{o,\infty} ,
\end{multline}
where the final equality follows from Assumptions \ref{ass:weights}(i)(ii) and Theorem 6.5.1 in \cite{luenberger:1969}.
Hence, \eqref{aux:critbound11} and Lemma \ref{lm:obviousmatrix} (for $E_{1n}(P)$), Assumption \ref{ass:techrange}(ii) (for $E_{2n}(P)$), \eqref{aux:critbound4p1} and \eqref{aux:critbound4p2} together with $\eta -\epsilon > 0.5$, Markov's inequality and $ \sup_{P\in \mathbf P} 1/\underline{\sigma}(P) = o(b_n^{-1})$ (for $E_{3n}(P)$), and Lemma \ref{lm:deg} (for $E_{4n}(P)$), yield
\begin{equation*}
\liminf_{n\rightarrow \infty}\inf_{P\in \mathbf P_0^{\rm i}} P(\{Z_i\}_{i=1}^n \in E_n(P)) = 1.
\end{equation*}

Next note that if $\{Z_i\}_{i=1}^n \in E_{n}(P)$ then the event $E_{1n}(P)$ allows us to conclude
\begin{equation}\label{aux:critbound13}
T_n = \sup_{s\in \hat{\mathcal V}_n^{\rm i}} \sqrt n \langle A^\dagger s, \hat x_n^\star \rangle \leq \sup_{s\in \mathcal V^{\rm i}(P)} 2 \sqrt n \langle A^\dagger s, \hat x_n^\star \rangle .
\end{equation}
Since $A^\dagger AA^\dagger = A^\dagger$ by Proposition 6.11.1(5) in \cite{luenberger:1969}, Assumption \ref{ass:beta}(ii), $AA^\dagger\beta(P) = \beta(P)$ whenever $P\in \mathbf P_0$ due to $\beta(P) \in R$, symmetry of $\Omega^{\rm i}(P)$, and $AA^\dagger \hat C_n\sqrt n\{\hat \beta_n - \beta(P)\} \in \text{range}\{\Omega^{\rm i}(P)\}$ whenever $\{Z_i\}_{i=1}^n \in E_{2n}(P)$ due to $\text{range}\{\Sigma^{\rm i}(P)\} \subseteq \text{range}\{\Omega^{\rm i}(P)\}$ by Assumption \ref{ass:techrange}(i) imply
\begin{multline}\label{aux:critbound14}
 \sqrt n \langle A^\dagger s, \hat x_n^\star \rangle =   \langle A^\dagger s, A^\dagger AA^\dagger \hat C_n\sqrt n\{\hat \beta_n - \beta(P)\}\rangle + \sqrt n \langle A^\dagger s, A^\dagger \beta(P)\rangle\\
=  \langle \Omega^{\rm i}(P) (AA^\prime)^\dagger s, (\Omega^{\rm i}(P))^\dagger AA^\dagger \hat C_n\sqrt n\{\hat \beta_n - \beta(P)\}\rangle + \sqrt n \langle (AA^\prime)^\dagger s, \beta(P)\rangle
\end{multline}
for any $s\in \mathcal V^{\rm i}(P)$ whenever $\{Z_i\}_{i=1}^n \in E_n(P)$.
Since $\langle A^\dagger s, A^\dagger \beta(P)\rangle \leq 0$ whenever $P\in \mathbf P_0$ and $s\in \mathcal V^{\rm i}(P)$ by Theorem \ref{th:ineq}, H\"older's inequality implying \eqref{aux:critbound14} is bounded in $s\in \mathcal V^{\rm i}(P)$ together with Lemmas \ref{lm:auxlp} and \ref{lm:Vpoly} implies that
\begin{multline}\label{aux:critbound14p5}
\sup_{s\in (AA^\prime)^\dagger \mathcal V^{\rm i}(P)} \langle \Omega^{\rm i}(P) s, (\Omega^{\rm i}(P))^\dagger AA^\dagger \hat C_n\sqrt n\{\hat \beta_n - \beta(P)\}\rangle + \sqrt n \langle s, \beta(P)\rangle
\\ =\max_{s\in \mathcal E^{\rm i}(P)} \langle \Omega^{\rm i}(P) s, (\Omega^{\rm i}(P))^\dagger AA^\dagger \hat C_n\sqrt n\{\hat \beta_n - \beta(P)\}\rangle + \sqrt n \langle s, \beta(P)\rangle.
\end{multline}
Hence, \eqref{aux:critbound13}, \eqref{aux:critbound14}, and \eqref{aux:critbound14p5} together establish that the set $\mathcal S^{\rm i}(P)$ given by
\begin{equation*}
{\mathcal S}^{\rm i}(P) \equiv \{s \in \mathcal E^{\rm i}(P) : \langle \Omega^{\rm i}(P) s,(\Omega^{\rm i}(P))^\dagger AA^\dagger\hat C_n \sqrt n\{\hat \beta_n - \beta(P)\} + \sqrt n \langle s, \beta(P)\rangle > 0\}
\end{equation*}
is such that $\mathcal S^{\rm i}(P) \neq \emptyset$ whenever $T_n > 0$ and $\{Z_i\}_{i=1}^n \in E_n(P)$.
Moreover, since $\sqrt n \langle s, \beta(P)\rangle \leq 0$ for all $s\in \mathcal S^{\rm i}(P)$ due to  $\mathcal S^{\rm i}(P) \subseteq \mathcal E^{\rm i}(P) \subset (AA^\prime)^\dagger \mathcal V^{\rm i}(P)$, $P\in \mathbf P_0$, and Theorem \ref{th:ineq}, it follows that whenever $\mathcal S^{\rm i}(P) \neq \emptyset$ we must have
\begin{equation}\label{aux:critbound14p7}
\langle \Omega^{\rm i}(P) s, (\Omega^{\rm i}(P))^\dagger AA^\dagger \hat C_n\sqrt n\{\hat \beta_n - \beta(P)\}\rangle > 0
\end{equation}
for all $s\in \mathcal S^{\rm i}(P)$.
Also note that if $\{Z_i\}_{i=1}^n \in E_n(P) \subseteq E_{2n}(P)$, then $\text{range}\{\Sigma^{\rm i}(P)\}$ equaling the support of $\mathbb G_n^{\rm i}(P)$ by Theorem 3.6.1 in \cite{bogachev1998gaussian} implies  $\sigma^{\rm i}( s,P) > 0$ for any $s$ satisfying \eqref{aux:critbound14p7}.
Thus, we have shown that if $P\in \mathbf P_0^{\rm i}$, then
\begin{equation}\label{aux:critbound14p71}
\mathcal S^{\rm i}(P)\neq \emptyset \text{ and }\sigma^{\rm i}(s,P) > 0 \text{ for all } s\in \mathcal S^{\rm i}(P)
\end{equation}
whenever $\{Z_i\}_{i=1}^n \in E_n(P)$ and $T_n > 0$.
We next aim to show that in addition
\begin{equation}\label{aux:critbound14p75}
\max_{s\in \mathcal S^{\rm i}(P)} \langle s, AA^\dagger \hat \beta_n^{\rm r}\rangle = 0
\end{equation}
whenever $\{Z_i\}_{i=1}^n \in E_n(P)$ and $T_n > 0$.
To this end, note Theorem \ref{th:ineq} yields that
\begin{equation}\label{aux:critbound14p8}
0 \geq \sup_{s\in \mathcal V^{\rm i}(P)} \langle A^\dagger s, A^\dagger \hat\beta_n^{\rm r}\rangle  = \sup_{s\in (AA^\prime)^\dagger \mathcal V^{\rm i}(P)} \langle s, AA^\dagger \hat \beta_n^{\rm r}\rangle  = \max_{s \in \mathcal E^{\rm i}(P)} \langle s, AA^\dagger \hat \beta_n^{\rm r}\rangle ,
\end{equation}
where the first equality follows from $A^\dagger AA^\dagger = A^\dagger$ by Proposition 6.11.1(5) in \cite{luenberger:1969} and the second from Lemmas \ref{lm:auxlp} and \ref{lm:Vpoly}.
Since $AA^\dagger \hat C_n \beta(P) = \beta(P)$ due to $\hat C_n \beta(P) = \beta(P)$ by Assumption \ref{ass:beta}(ii) and $\beta(P) \in R$, the symmetry of $\Omega^{\rm i}(P)$ and $AA^\dagger \hat C_n \sqrt n \{\hat \beta_n - \beta(P)\}\in \text{range}\{\Omega^{\rm i}(P)\}$ whenever $\{Z_i\}_{i=1}^n \in E_{2n}(P)$ due to $\text{range}\{\Sigma^{\rm i}(P)\}\subseteq \text{range}\{\Omega^{\rm i}(P)\}$ by Assumption \ref{ass:techrange}(i) imply
\begin{multline}\label{aux:critbound14p85}
\max_{s\in \mathcal E^{\rm i}(P)\setminus \mathcal S^{\rm i}(P)}\langle s, AA^\dagger \hat C_n \hat \beta_n\rangle =
\max_{s\in \mathcal E^{\rm i}(P)\setminus \mathcal S^{\rm i}(P)}\langle s, AA^\dagger \hat C_n \sqrt n\{\hat \beta_n - \beta(P)\}\rangle + \sqrt n \langle s, \beta(P)\rangle \\
=\max_{s\in \mathcal E^{\rm i}(P)\setminus \mathcal S^{\rm i}(P)} \langle \Omega^{\rm i}(P) s, (\Omega^{\rm i}(P))^\dagger AA^\dagger \hat C_n \sqrt n\{\hat \beta_n - \beta(P)\}\rangle + \sqrt n \langle s, \beta(P)\rangle \leq 0,
\end{multline}
where the inequality follows by definition of $\mathcal S^{\rm i}(P)$.
Thus, if we suppose by way of contradiction that \eqref{aux:critbound14p75} fails to hold, then \eqref{aux:critbound14p8}, \eqref{aux:critbound14p85}, $\mathcal S^{\rm i}(P) \subseteq \mathcal E^{\rm i}(P)$, and $\mathcal E^{\rm i}(P)$ being finite, imply there exists a $\gamma^\star \in (0,1)$ (depending on $\hat \beta_n$ and $\hat \beta_n^{\rm r})$ with
\begin{multline}\label{aux:critbound14p9}
0 \geq \max_{s\in \mathcal E^{\rm i}(P)} \langle s, AA^\dagger\{(1-\gamma^\star) \hat \beta_n^{\rm r} + \gamma^\star \hat C_n \hat \beta_n\}\rangle = \sup_{s\in  (AA^\prime)^\dagger \mathcal V^{\rm i}(P)} \langle s, AA^\dagger\{(1-\gamma^\star)\hat \beta_n^{\rm r} + \gamma^\star \hat C_n \hat \beta_n\}\rangle \\
= \sup_{s\in \mathcal V^{\rm i}(P)} \langle A^\dagger s, A^\dagger\{(1-\gamma^\star)\hat \beta_n^{\rm r} + \gamma^\star AA^\dagger \hat C_n \hat \beta_n\}\rangle
\end{multline}
where the equalities follow from Lemmas \ref{lm:auxlp} and \ref{lm:Vpoly}, and employing $A^\dagger AA^\dagger = A^\dagger$ by Proposition 6.11.1(5) in \cite{luenberger:1969}.
However, by construction $\hat \beta_n^{\rm r}\in R$ and $AA^\dagger \hat C_n \hat \beta_n \in R$, and therefore \eqref{aux:critbound14p9} and Theorem \ref{th:ineq} imply
\begin{equation*}
(1-\gamma^\star)\hat \beta_n^{\rm r} + \gamma^\star AA^\dagger \hat C_n \hat \beta_n = Ax \text{ for some }x\geq 0.
\end{equation*}
Next, note that if $T_n > 0$, then $\sup_{s\in \hat {\mathcal V}_n^{\rm i}} \langle A^\dagger s, \hat x_n^\star\rangle > 0$ whenever $\{Z_i\}_{i=1}^n \in E_n(P)\subseteq E_{4n}(P)$ and hence $T_n > 0$ implies $\sup_{s\in \hat{\mathcal V}_n^{\rm i}} \langle A^\dagger s, \hat x_n^\star - A^\dagger \hat \beta_n^{\rm r}\rangle > 0$ due to $\langle A^\dagger s, A^\dagger \hat \beta_n^{\rm r}\rangle \leq 0$ for all $s\in \hat{\mathcal V}_n$ by Theorem \ref{th:ineq}.
In particular, if $T_n >0$, then $\sup_{s\in \hat{\mathcal V}_n^{\rm i}} |\langle A^\dagger s, \hat x_n^\star - A^\dagger \hat \beta_n^{\rm r}\rangle| > 0$, and hence $\hat x_n^\star = A^\dagger \hat C_n \hat \beta_n$, $A^\dagger A A^\dagger = A^\dagger$, and $\gamma^\star \in (0,1)$ yield
\begin{multline*}
\sup_{s\in \hat {\mathcal V}_n^{\rm i}} |\langle A^\dagger s, \hat x_n^\star - A^\dagger \{(1-\gamma^\star)\hat \beta_n^{\rm r} + \gamma^\star AA^\dagger \hat C_n \hat\beta_n\}\rangle| \\
= (1-\gamma^\star) \sup_{s\in \hat {\mathcal V}_n^{\rm i}} |\langle A^\dagger s, \hat x_n^\star - A^\dagger \hat \beta_n^{\rm r}\rangle| < \sup_{s\in \hat{\mathcal V}_n^{\rm i}} |\langle A^\dagger s, \hat x_n^\star - A^\dagger \hat \beta_n^{\rm r}\rangle| ,
\end{multline*}
which is impossible by definition of $\hat \beta_n^{\rm r}$.
We thus obtain that if $\{Z_i\}_{i=1}^n \in E_n(P)$ and $T_n > 0$, then result \eqref{aux:critbound14p75} must hold.

To conclude, note \eqref{aux:critbound14p71} and \eqref{aux:critbound14p75} imply there is a $\hat s_n \in {\mathcal V}^{\rm i}(P)$ depending only on $P$ and $\{Z_i\}_{i=1}^n$ such that $(AA^\prime)^\dagger \hat s_n \in \mathcal E^{\rm i}(P)$, $\sigma((AA^\prime)^\dagger \hat s_n, P) > 0$, and $0=\lambda_n \langle A^\dagger \hat s_n, A^\dagger \hat \beta_n^{\rm r}\rangle \equiv \hat{\mathbb U}_n(\hat s_n) = 0$ whenever $\{Z_i\}_{i=1}^n \in E_n(P)$ and $T_n > 0$.
Hence,  the definitions of  $\hat {\mathbb L}_n$, $\mathbb L_n^\star(P)$, $\hat c_n(\eta)$, together with $\{Z_i\}_{i=1}^n \in E_{n}(P) \subseteq E_{3n}(P)$ yield
\begin{align}\label{aux:critbound16}
P(\langle A^\dagger \hat s_n,A^\dagger & \mathbb G_n^{\text{i}\star}(P) \rangle  \leq \hat c_n(\eta) + \frac{\underline{\sigma}(P)z_{\eta-\epsilon}}{2}|\{Z_i\}_{i=1}^n) \notag \\
& \geq P(\mathbb L_n^\star(P)  \leq \hat c_n(\eta) + \frac{\underline{\sigma}(P)z_{\eta-\epsilon}}{2}|\{Z_i\}_{i=1}^n) \notag \\
& \geq P(\hat {\mathbb L}_n \leq \hat c_n(\eta)|\{X_i\}_{i=1}^n) - \epsilon \notag \\
& \geq \eta - \epsilon
\end{align}
whenever $P\in \mathbf P_0^{\rm i}$, $\{Z_i\}_{i=1}^n \in E_n(P)$, and $T_n>0$.
Since $\mathbb G_n^{\rm i\star}(P)\in \text{range}\{\Omega^{\rm i}(P)\}$ by Assumption \ref{ass:techrange}(i) and Theorem 3.6.1 in \cite{bogachev1998gaussian}, we have
\begin{equation}\label{aux:critbound17}
\langle A^\dagger \hat s_n, A^\dagger \mathbb G_n^{\rm i\star}(P)\rangle = \langle \Omega^{\rm i}(P)(AA^\prime)^\dagger \hat s_n,(\Omega^{\rm i}(P))^\dagger \mathbb G_n^{\rm i\star}(P)\rangle
\end{equation}
almost surely.
Hence,  $\mathbb G_n^{\text{i}\star}(P)$ being independent of $\{Z_i\}_{i=1}^n$ implies $\langle A^\dagger \hat s_n,A^\dagger  \mathbb G_n^{\text{i}\star}(P) \rangle \sim N(0,(\sigma^{\text{i}}((AA^\prime)^\dagger\hat s_n,P))^2)$ conditional on $\{Z_i\}_{i=1}^n$.
Since  \eqref{aux:critbound17} and $\sigma^{\text{i}}((AA^\prime)^\dagger\hat s_n,P) > 0$ imply that the distribution of $\langle A^\dagger \hat s_n,A^\dagger \mathbb G_n^{\text{i}\star}(P) \rangle $ conditional on $\{Z_i\}_{i=1}^n$ first order stochastically dominates $N(0,\underline{\sigma}(P))$ random variable, \eqref{aux:critbound16} yields
\begin{equation*}
\hat c_n(\eta) + \frac{\underline{\sigma}(P)z_{\eta-\epsilon}}{2} \geq \underline{\sigma}(P)z_{\eta-\epsilon},
\end{equation*}
which establishes the claim of the lemma for the subset $\mathbf P^{\rm i}_0$.

\noindent \underline{Case III:} For the final case, suppose $P\in \mathbf P^{\rm d}_0 \equiv \{P\in \mathbf P_0 : \sigma^{\rm j}(s,P) = 0 \text{ for all } s\in \mathcal E^{\rm j}(P) \text{ and } \rm j \in \{\rm e,\rm i\}\}$.
Then, by Lemma \ref{lm:deg} we may set $E_n(P) \equiv \{T_n = 0\}$ and the claim of the lemma for the subset $\mathbf P^{\rm d}_0$ follows. \qed

\begin{lemma}\label{lm:betacoup}
Set $\Sigma(P) \equiv E_P[\psi(X,P)\psi(X,P)^\prime]$ and $r_n \equiv a_n + M_{3,\Psi} p^{1/3}(\log(1+p))^{5/6}/n^{1/6}$.
If Assumptions \ref{ass:beta}(i)(iii), \ref{ass:moments}, \ref{ass:techrange}(i) hold, and $r_n = o(1)$, then there exists $(\mathbb G_n^{\text{\rm e}}(P)^\prime,\mathbb G_n^{\text{\rm i}}(P)^\prime)^\prime \equiv \mathbb G_n(P) \sim N(0, \Sigma(P))$ satisfying uniformly in $P\in \mathbf P$:
\begin{align*}
\|(\Omega^{\text{\rm e}}(P))^\dagger \{(I_p - AA^\dagger\hat C_n)\sqrt n\{\hat \beta_n - \beta(P)\} - \mathbb G_n^{\text{\rm e}}(P)\}\|_\infty = O_P(r_n) \\ 
\|(\Omega^{\text{\rm i}}(P))^\dagger \{AA^\dagger\hat C_n\sqrt n\{\hat \beta_n - \beta(P)\} - \mathbb G_n^{\text{\rm i}}(P)\}\|_\infty = O_P(r_n). 
\end{align*}
\end{lemma}

\noindent \emph{Proof:} Set $\tilde \psi(Z,P) \equiv (((\Omega^{\rm e}(P))^\dagger \psi^{\rm e}(Z,P))^\prime, ((\Omega^{\rm i}(P))^\dagger \psi^{\rm i}(Z,P))^\prime)^\prime \in \mathbf R^{2p}$, define
\begin{equation}\label{lm:betacoup0}
\tilde \Sigma(P) \equiv E_P[\tilde \psi(Z,P)\tilde \psi(Z,P)^\prime],
\end{equation}
and let $S_n(P) \in \mathbf R^{2p}$ be normally distributed with mean zero and variance $\tilde \Sigma(P)/n$.
Next observe that since $\|a\|_2^2 \leq 2p \|a\|_\infty^2$ for any $a \in \mathbf R^{2p}$ we can conclude that
\begin{equation}\label{lm:betacoup1}
E_P[\|S_{n}(P)\|_2^2 \|S_{n}(P)\|_\infty] \leq 2p E_P[\|S_{n}(P)\|^3_\infty] \lesssim p(\frac{\sqrt{\log(1+p)}}{\sqrt n} )^3 ,
\end{equation}
where the second inequality follows from Lemma \ref{lm:obvious} and Assumption \ref{ass:moments}(ii).
By similar arguments, Assumption \ref{ass:moments}(iii), and result \eqref{lm:betacoup1} we can conclude
\begin{align}\label{lm:betacoup2}
n\{E_P[\|\frac{\tilde \psi(Z,P)}{\sqrt n}\|_2^2 &  \|\frac{\tilde \psi(Z,P)}{\sqrt n}\|_\infty] + E_P[\|S_n(P)\|_2^2 \|S_n(P)\|_\infty]\} \notag \\
& \lesssim  n\{\frac{p}{n^{3/2}} E_P[\Psi^3(Z,P)] + E_P[\|S_{n}(P)\|_2^2 \|S_{n}(P)\|_\infty]\} \notag \\
& \lesssim \frac{p}{\sqrt n}\{M_{3,\Psi}^3 +  (\log(1+p))^{3/2}\}.
\end{align}
For $\mathbb Z \sim N(0,I_{2p})$, we obtain by Assumptions \ref{ass:beta}(i), \ref{ass:moments}(i), Lemma 39 in \citet{belloni2019conditional}, and \eqref{lm:betacoup2} that for any $\delta > 0$ there is a $\tilde{\mathbb G}_{n}(P) \sim N(0,\tilde \Sigma(P))$ with
\begin{align}\label{lm:betacoup3}
P(\|\frac{1}{\sqrt n} \sum_{i=1}^n & \tilde \psi(Z_i,P) - \tilde{\mathbb G}_n(P)\|_\infty > \delta) \notag \\
& \lesssim \min_{t\geq 0} \{P(\|\mathbb Z\|_\infty > t) + \frac{t^2}{\delta^3}  \frac{p}{\sqrt n}\{M_{3,\Psi}^3 + (\log(1+p))^{3/2}\} \notag \\
& \lesssim \min_{t\geq 0} \{\exp\{-\frac{t^2}{8\log(1 + p)}\} + \frac{t^2}{\delta^3}  \frac{pM_{3,\Psi}^3(\log(1+p))^{3/2}}{\sqrt n}  \},
\end{align}
where the final inequality follows from Proposition A.2.1 in \cite{vandervaart:wellner:1996}, $E[\|\mathbb Z\|_\infty^2] \lesssim \log(1+p)$ by Lemma \ref{lm:obvious}, and we employed that $M_{3,\Psi} \geq 1$ by Assumption \ref{ass:moments}(iii).
Thus, by setting $t = K \sqrt{\log(1+p)}$ and $\delta^3 = K^3 p M_{3,\Psi}^3(\log(1+p))^{5/2}/\sqrt n$ in \eqref{lm:betacoup3} for any $K > 0$ we obtain
\begin{multline}\label{lm:betacoup4}
\lim_{K \uparrow \infty} \limsup_{n\rightarrow \infty} \sup_{P\in \mathbf P} P(\|\frac{1}{\sqrt n}\sum_{i=1}^n \tilde \psi(Z_i,P) - \tilde{\mathbb G}_n(P)\|_\infty > K \frac{M_{3,\Psi} p^{1/3}(\log(1+p))^{5/6}}{n^{1/6}}) \\
\lesssim \lim_{K \uparrow \infty}  \{\exp\{-\frac{K^2}{8}\} + \frac{1}{K}\} = 0.
\end{multline}
Since $r_n \equiv M_{3,\Psi} p^{1/3}(\log(1+p))^{5/6}/n^{1/6} + a_n$,  result \eqref{lm:betacoup4}, Assumption \ref{ass:beta}(iii), writing $\tilde{\mathbb G}_n(P) \equiv(\tilde {\mathbb G}_n^{\rm e}(P)^\prime,  \tilde {\mathbb G}_n^{\rm i}(P)^\prime)^\prime$, and the triangle inequality imply  that
\begin{align} \label{lm:betacoup5}
\| (\Omega^{\text{\rm e}}(P))^\dagger (I_p - AA^\dagger)\sqrt n\{\hat \beta_n - \beta(P)\} &- \tilde{\mathbb G}_n^{\text{\rm e}}(P)\|_\infty  = O_P(r_n) \notag\\
\| (\Omega^{\text{\rm i}}(P))^\dagger AA^\dagger\sqrt n\{\hat \beta_n - \beta(P)\} &- \tilde{\mathbb G}_n^{\text{\rm i}}(P)\|_\infty  = O_P(r_n)
\end{align}
uniformly in $P\in \mathbf P$.
To conclude, note $\tilde {\mathbb G}^{\rm j}_n(P) \sim N(0,(\Omega^{\rm j}(P))^\dagger \Sigma^{\rm j}(P)(\Omega^{\rm j}(P))^\dagger)$ for $\rm j \in \{\rm e,\rm i\}$, and therefore Theorem 3.6.1 in \cite{bogachev1998gaussian} implies that $\tilde {\mathbb G}^{\rm j}_n(P)$ belongs to the range of $(\Omega^{\rm j}(P))^\dagger \Sigma^{\rm j}(P)(\Omega^{\rm j}(P))^\dagger$ almost surely.
Thus, since for $\rm j \in \{\rm e, \rm i\}$ we have $(\Omega^{\rm j}(P))^\dagger \Omega^{\rm j}(P)(\Omega^{\rm j}(P))^\dagger = (\Omega^{\rm j}(P))^\dagger$ it follows $(\Omega^{\rm j}(P))^\dagger \Omega^{\rm j}(P)\tilde {\mathbb G}_n^{\rm j}(P) = \tilde{\mathbb G}_n^{\rm j}(P)$ almost surely.
Hence, setting $\mathbb G_n^{\rm j}(P) = \Omega^{\rm j}(P) \tilde {\mathbb G}_n^{\rm j}(P)$ for $\rm j \in \{\rm e, \rm i\}$ we obtain
\begin{align} \label{lm:betacoup6}
\| (\Omega^{\text{\rm e}}(P))^\dagger \{(I_p - AA^\dagger)\sqrt n\{\hat \beta_n - \beta(P)\} &- {\mathbb G}_n^{\text{\rm e}}(P)\}\|_\infty  = O_P(r_n) \notag\\
\| (\Omega^{\text{\rm i}}(P))^\dagger\{ AA^\dagger\sqrt n\{\hat \beta_n - \beta(P)\} &- {\mathbb G}_n^{\text{\rm i}}(P)\}\|_\infty  = O_P(r_n)
\end{align}
uniformly in $P\in \mathbf P$ by \eqref{lm:betacoup5}.
Since $\tilde {\mathbb G}_n(P) \sim N(0,\tilde \Sigma(P))$, and Assumption \ref{ass:techrange}(i) implies $\Omega^{\rm j}(P)(\Omega^{\rm j}(P))^\dagger \psi^{\rm j}(Z,P) = \psi^{\rm j}(Z,P)$ $P$-almost surely, we can conclude from the definition of $\tilde \Sigma(P)$ in \eqref{lm:betacoup0} and $\mathbb G_n(P) = ((\Omega^{\rm e}(P) \tilde{\mathbb G}_n^{\rm e}(P))^\prime, (\Omega^{\rm i}(P) \tilde{\mathbb G}_n^{\rm i}(P))^\prime)^\prime$ that $\mathbb G_n(P)\sim N(0,\Sigma(P))$ and thus the claim of the lemma follows. \qed

\begin{lemma}\label{lm:finitebootnew}
Let Assumptions \ref{ass:beta}(i), \ref{ass:moments}, \ref{ass:techrange}(i), \ref{ass:4bootnew}(i)-(iv) hold, and define
$$b_n \equiv \frac{\sqrt{p\log(1+n)}M_{3,\Psi}}{n^{1/4}} + (\frac{p\log^{5/2}(1+p)M_{3,\Psi}}{\sqrt n})^{1/3} + (\frac{p\log^3(1+p) n^{1/q} M_{q,\Psi^2}}{n})^{1/4}+ a_n.$$
If $b_n = o(1)$, then there is a Gaussian vector $(\mathbb G_n^{\text{\rm e}\star}(P)^\prime,\mathbb G_n^{\text{\rm i}\star}(P)^\prime) \equiv \mathbb G_n^\star(P) \sim N(0,\Sigma(P))$ independent of $\{Z_i\}_{i=1}^n$ and satisfying uniformly in $P\in \mathbf P$:
$$\|(\Omega^{\rm e}(P))^\dagger\{\hat {\mathbb G}^{\rm e}_n - \mathbb G^{\rm e\star}_n(P)\}\|_\infty \vee \|(\Omega^{\rm i}(P))^\dagger\{\hat {\mathbb G}^{\rm i}_n - \mathbb G^{\rm i\star }_n(P)\}\|_\infty= O_P(b_n).$$
\end{lemma}

\noindent \emph{Proof:}
For ease of exposition we divide the proof into multiple steps.
In the arguments that follow, we let $\varphi(Z,P) \equiv (\varphi^{\rm e}(Z,P)^\prime, \varphi^{\rm i}(Z,P)^\prime)^\prime \in \mathbf R^{2p}$, where
\begin{equation}\label{lm:finitebootnew0}
\varphi^{\rm e}(Z,P) \equiv (\Omega^{\rm e}(P))^\dagger \psi^{\rm e}(Z,P) \hspace{0.5 in} \varphi^{\rm i}(Z,P) \equiv (\Omega^{\rm i}(P))^\dagger \psi^{\rm i}(Z,P).
\end{equation}

\noindent \underline{Step 1:} (Distributional Representation).
Let $\{U_i\}_{i=1}^\infty$ be an i.i.d.\ sequence independent of $\{Z_i,W_{i,n}\}_{i=1}^n$ with $U_i$ uniformly distributed on $(0,1]$.
We further set $(U_{(1),n}, \ldots U_{(n),n})$ to denote the order statistics of $\{U_i\}_{i=1}^n$ and $R_{i,n}$ to denote the rank of each $U_i$ (i.e.,\ $U_i = U_{(R_{i,n}),n}$).
By Lemma 13.1(iv) in \cite{vandervaart:1999}, the vector $R_n \equiv (R_{1,n},\ldots, R_{n,n})$ is uniformly distributed on the set of all $n!$ permutations of $\{1,\ldots, n\}$ and hence by Assumption \ref{ass:4bootnew}(i) we can conclude that
\begin{equation*}
(\frac{1}{\sqrt n} \sum_{i=1}^n (W_{i,n} - \bar W_n)\varphi(Z_i,P),\{Z_i\}_{i=1}^n) \stackrel{d}{=} (\frac{1}{\sqrt n} \sum_{i=1}^n (W_{R_{i},n} - \bar W_n)\varphi(Z_i,P),\{Z_i\}_{i=1}^n),
\end{equation*}
where $\stackrel{d}{=}$ denotes equality in distribution and $\bar W_n \equiv \sum_{i=1}^n W_{i,n}/n$.

\noindent \underline{Step 2:} (Couple to i.i.d.).
We next define $\tau_n : [0,1] \to \{W_{i,n} - \bar W_n\}_{i=1}^n$ to be equal
\begin{equation*}
\tau_n(u) \equiv \inf \{ c: \frac{1}{n} \sum_{i=1}^n 1\{W_{i,n} - \bar W_n \leq c\} \geq u\};
\end{equation*}
i.e., $\tau_n$ is the empirical quantile function of the sample $\{W_{i,n} - \bar W_n\}_{i=1}^n$.
Also set
\begin{align*}
S_n(P) & \equiv  \frac{1}{\sqrt n} \sum_{i=1}^n (W_{R_{i},n} - \bar W_n)\varphi(Z_i,P) \\ 
L_n(P) & \equiv  \frac{1}{\sqrt n} \sum_{i=1}^n (\varphi(Z_i,P)- \bar \varphi_n(P))\tau_n(U_i) 
\end{align*}
where $\bar \varphi_n(P) \equiv \sum_{i=1}^n \varphi(Z_i,P)/n$.
Letting $S_{j,n}(P)$ and $L_{j,n}(P)$ denote the $j^{th}$ coordinates of $S_n(P)$ and $L_n(P)$ respectively, we then observe that Theorem 3.1 in \cite{Hajek1961RankCLT} (see in particular equation (3.11) in page 512) yields
\begin{multline}\label{lm:finitebootnew3}
E[(S_{j,n}(P) - L_{j,n}(P))^2|\{Z_i,W_{i,n}\}_{i=1}^n] \\
\lesssim \text{Var}\{L_{j,n}(P)|\{Z_i,W_{i,n}\}_{i=1}^n\} \frac{\max_{1\leq i \leq n} |W_{i,n} - \bar W_n|}{(\sum_{i=1}^n (W_{i,n} - \bar W_n)^2)^{1/2}}.
\end{multline}
In order to study the properties of $L_n(P)$ it is convenient to define $\xi_{i,n}(P)$ to equal
\begin{equation}\label{lm:finitebootnew4}
\xi_{i,n}(P) \equiv (\varphi(Z_i,P) - \bar \varphi_n(P))\frac{\tau_n(U_i)}{\sqrt n}.
\end{equation}
Moreover, since $\{U_i\}_{i=1}^n$ are i.i.d.\ uniform on $(0,1]$ and independent of $\{Z_i,W_{i,n}\}_{i=1}^n$, and $\tau_n$ is the empirical quantile function of $\{W_{i,n} - \bar W_n\}_{i=1}^n$ it follows that
\begin{align}\label{lm:finitebootnew5}
& E[\xi_{i,n}(P) \xi_{i,n}(P)^\prime|\{Z_i,W_{i,n}\}_{i=1}^n]  = \frac{\hat \sigma_n^2}{n} (\varphi(Z_i,P) - \bar \varphi_{n}(P))(\varphi(Z_i,P) - \bar \varphi_{n}(P))^\prime \notag \\
& E[\xi_{i,n}(P)|\{Z_i,W_{i,n}\}_{i=1}^n]  = \frac{1}{\sqrt n}(\varphi(Z_i,P) - \bar \varphi_{n}(P))(\frac{1}{n}\sum_{i=1}^n W_{i,n} - \bar W_n) = 0,
\end{align}
where $\hat \sigma_n^2 \equiv \sum_{i=1}^n (W_{i,n} - \bar W_n)^2/n$.
Hence, since $\{\xi_{i,n}(P)\}_{i=1}^n$ are independent conditional on $\{Z_i,W_{i,n}\}_{i=1}^n$ it follows from $L_n(P) = \sum_{i=1}^n \xi_{i,n}(P)$ that
\begin{equation}\label{lm:finitebootnew6}
\text{Var}\{L_{j,n}(P)|\{Z_i,W_{i,n}\}_{i=1}^n\} = \frac{\hat \sigma_n^2}{n} \sum_{i=1}^n (\varphi_j(Z_i,P) - \bar \varphi_{j,n}(P))^2,
\end{equation}
where $\varphi_j(Z_i,P)$ and $\bar \varphi_{j,n}(P)$ denote the $j^{th}$ coordinates of $\varphi(Z_i,P)$ and $\bar \varphi_n(P)$ respectively.
Thus, since for any random variable $(V_1,\ldots, V_{2p}) \equiv V\in \mathbf R^{2p}$ Jensen's inequality implies $E[\|V\|_\infty] \leq \sqrt {2p} \max_{1\leq j \leq 2p} (E[V_j^2])^{1/2}$,  \eqref{lm:finitebootnew3} and \eqref{lm:finitebootnew6} yield
\begin{multline}\label{lm:finitebootnew7}
E[\|S_n(P) - L_n(P)\|_\infty|\{Z_i,W_{i,n}\}_{i=1}^n] \\
\lesssim \sqrt{p}  \max_{1\leq j \leq 2p} (\frac{\hat \sigma_n}{n^{3/2}}\sum_{i=1}^n (\varphi_j(Z_i,P) - \bar \varphi_{j,n}(P))^2)^{1/2} (\max_{1\leq i \leq n} |W_{i,n} - \bar W_n|)^{1/2}.
\end{multline}
Next, we note that the definition of $\varphi(z,P)$ implies that $\Psi(z,P)$, as introduced in Assumption \ref{ass:moments}(iii), satisfies $\Psi(z,P) = \|\varphi(z,P)\|_\infty$.
Hence, for $M_{3,\Psi}$ as introduced in Assumption \ref{ass:moments}(iii), Markov and Jensen's inequalities imply for any $C > 0$ that
\begin{multline*}
\sup_{P\in \mathbf P} P( |\frac{1}{n}\sum_{i=1}^n \Psi^2(Z_i,P)| > CM_{3,\Psi}^{2}) \\ \leq \frac{1}{CM_{3,\Psi}^{2}} \sup_{P\in \mathbf P} E_P[|\frac{1}{n}\sum_{i=1}^n \Psi^2(Z_i,P)|] \leq \frac{1}{CM_{3,\Psi}^{2}} \sup_{P\in \mathbf P} \|\Psi(\cdot,P)\|_{P,2}^2 \leq \frac{1}{C}.
\end{multline*}
Thus, using that $\Psi(z,P) = \|\varphi(z,P)\|_\infty$ we conclude uniformly in $P\in \mathbf P$ that
\begin{equation}\label{lm:finitebootnew9}
\max_{1\leq j \leq 2p} \frac{1}{n}\sum_{i=1}^n (\varphi_j(Z_i,P) - \bar \varphi_{j,n}(P))^2  \leq \frac{1}{n}\sum_{i=1}^n \Psi^2(Z_i,P) = O_P(M_{3,\Psi}^{2}).
\end{equation}
Moreover, by the triangle inequality, Assumption \ref{ass:4bootnew}(ii), Lemma 2.2.10 in \cite{vandervaart:wellner:1996}, and $E[|V|] \leq \|V\|_{\psi_1}$ for any random variable $V$ and $\|\cdot\|_{\psi_1}$ the Orlicz norm based on $\psi_1 = e^x -1$,  we can conclude that
\begin{align}\label{lm:finitebootnew10}
E[\max_{1\leq i \leq n} |W_{i,n} - \bar W_n|] & \leq E[\max_{1\leq i \leq n} |W_{i,n} - E[W_{1,n}]|] + E[|\bar W_n - E[W_{1,n}]|] \notag \\
& \lesssim \log(1+n)  + E[|W_{1,n}|].
\end{align}
Thus, $\hat \sigma_n^2 \stackrel{P}{\rightarrow}1$ by Assumption \ref{ass:4bootnew}(iii), $E[|W_{1,n}|]$ being uniformly bounded in $n$ by Jensen's inequality and Assumption \ref{ass:4bootnew}(iii), and  \eqref{lm:finitebootnew7}, \eqref{lm:finitebootnew9}, \eqref{lm:finitebootnew10} yield
\begin{equation*}
E[\|S_n(P) - L_n(P)\|_\infty | \{Z_i,W_{i,n}\}_{i=1}^n] = O_P(\frac{\sqrt{p\log(1+n)}M_{3,\Psi}}{n^{1/4}})
\end{equation*}
uniformly in $P\in \mathbf P$.
By Fubini's theorem and Markov's inequality we may therefore conclude that unconditionally (on $\{Z_i,W_{i,n}\}_{i=1}^n$) and uniformly in $P\in \mathbf P$
\begin{equation*}
\|S_n(P) - L_n(P)\|_\infty = O_P(\frac{\sqrt{p\log(1+n)}M_{3,\Psi}}{n^{1/4}}).
\end{equation*}

\noindent \underline{Step 3:} (Couple to Gaussian).
We next couple $L_n(P)$ to a (conditionally) Gaussian vector.
To this end, recall the definition of $\xi_{i,n}(P)$ in \eqref{lm:finitebootnew4} and let
\begin{equation*}
\bar G_{i,n}(P) \sim N(0, \text{Var}\{\xi_{i,n}(P)|\{Z_i,W_{i,n}\}_{i=1}^n\})
\end{equation*}
and $\{\bar G_{i,n}(P)\}_{i=1}^n$ be mutually independent conditional on $\{Z_i,W_{i,n}\}_{i=1}^n$.
Then note that $\|a\|_2^2 \leq 2p\|a\|_\infty^2$ for any $a\in \mathbf R^{2p}$, Lemma \ref{lm:obvious}, and result \eqref{lm:finitebootnew5} imply
\begin{multline}\label{lm:finitebootnew14}
\sum_{i=1}^n E[\|\bar G_{i,n}(P)\|_2^2 \|\bar G_{i,n}(P)\|_\infty | \{Z_i,W_{i,n}\}_{i=1}^n] \leq 2p \sum_{i=1}^n E[\|\bar G_{i,n}(P)\|_\infty^3|\{Z_i,W_{i,n}\}_{i=1}^n] \\
\lesssim p \log^{3/2}(1+p) \frac{\hat \sigma^{3}_n}{n^{3/2}}\sum_{i=1}^n \|\varphi(Z_i,P) - \bar \varphi_{n}(P)\|_\infty^{3/2}.
\end{multline}
Similarly, the definition of $\xi_{i,n}(P)$, $\{U_i\}_{i=1}^n$ being independent of $\{Z_i,W_{i,n}\}_{i=1}^n$, and $\tau_n$ being the empirical quantile function of $\{W_{i,n} - \bar W_n\}_{i=1}^n$, yield
\begin{multline}\label{lm:finitebootnew15}
\sum_{i=1}^n E[\|\xi_{i,n}(P)\|_2^2 \|\xi_{i,n}(P)\|_\infty | \{Z_i,W_{i,n}\}_{i=1}^n] 
 \\ \leq \frac{2p}{\sqrt n} (\frac{1}{n}\sum_{i=1}^n \|\varphi(Z_i,P) - \bar \varphi_n(P)\|_\infty^3)( \frac{1}{n}\sum_{i=1}^n |W_{i,n} - \bar W_n|^3).
\end{multline}
Therefore, results \eqref{lm:finitebootnew14} and \eqref{lm:finitebootnew15}, $\Psi(Z_i,P) = \|\varphi(Z_i,P)\|_\infty$, and multiple applications of the triangle and Jensen's inequalities yield the upper bound
\begin{multline}\label{lm:finitebootnew16}
\sum_{i=1}^n E[\|\bar G_{i,n}(P)\|_2^2 \|\bar G_{i,n}(P)\|_\infty + \|\xi_{i,n}(P)\|_2^2 \|\xi_{i,n}(P)\|_\infty | \{Z_i,W_{i,n}\}_{i=1}^n] \\
\lesssim \frac{p\log^{\frac{3}{2}}(1+p)}{\sqrt n}(\frac{1}{n}\sum_{i=1}^n |W_{i,n}|^3)(\frac{1}{n}\sum_{i=1}^n \{\Psi^3(Z_i,P)+\Psi^{\frac{3}{2}}(Z_i,P)\}) \equiv B_n(P),
\end{multline}
where the final equality is definitional.
Next, let $\mathcal B$ denote the Borel $\sigma$-field on $\mathbf R^{2p}$ and for any $A\in \mathcal B$ and $\epsilon > 0$ set $A^{\epsilon} \equiv \{a \in \mathbf R^{2p} : \inf_{\tilde a \in A} \|a - \tilde a\|_\infty \leq \epsilon\}$.
Strassen's Theorem (see Theorem 10.3.1 in \cite{pollard2002user}), Lemma 39 in \cite{belloni2019conditional}, and result \eqref{lm:finitebootnew16} then establish for any $\delta > 0$ that
\begin{multline}\label{lm:finitebootnew17}
\sup_{A \in \mathcal B} \{P(L_n(P) \in A |\{Z_i,W_{i,n}\}_{i=1}^n) - P(\frac{1}{\sqrt n} \sum_{i=1}^n \bar G_{i,n}(P) \in A^{3\delta}| \{Z_i,W_{i,n}\}_{i=1}^n)\} \\
\lesssim \min_{t \geq 0} (2P(\|\mathbb Z\|_\infty > t) + \frac{B_n(P)}{\delta^3}t^2)
\end{multline}
where $\mathbb Z \in \mathbf R^{2p}$ satisfies $\mathbb Z \sim N(0, I_{2p})$.
Furthermore, Proposition A.2.1 in \cite{vandervaart:wellner:1996} and Lemma \ref{lm:obvious} imply for some $C<\infty $
\begin{align}\label{lm:finitebootnew18}
\sup_{P\in \mathbf P} E_P[\min_{t \geq 0} & (2P(\|\mathbb Z\|_\infty > t) + \frac{B_n(P)}{\delta^3}t^2)] \notag \\
& \lesssim \min_{t\geq 0} (\exp\{-\frac{t^2}{C \log(1+p)}\} + \sup_{P \in \mathbf P} E_P[B_n(P)]\frac{t^2}{\delta^3})\notag  \\
& \lesssim \min_{t\geq 0} ( \exp\{-\frac{t^2}{C \log(1+p)}\} + \frac{p\log^{3/2}(1+p)M_{3,\Psi}^3}{\sqrt n}\frac{t^2}{\delta^3} ),
\end{align}
where the final inequality follows from  \eqref{lm:finitebootnew16}, $\sup_nE[|W_{i,n}|^3] < \infty$ by Assumption \ref{ass:4bootnew}(iii), Jensen's inequality, $\sup_{P\in \mathbf P}\|\Psi\|_{P,3} \leq M_{3,\Psi}$ with $M_{3,\Psi} \geq 1$ by Assumption \ref{ass:moments}(iii), and $\{W_{i,n}\}_{i=1}^n$ being independent of $\{Z_i\}_{i=1}^n$ by Assumption \ref{ass:4bootnew}(i).
Hence, for any $K > 0$, $p\log^{5/2}(1+p)M_{3,\Psi}^3/\sqrt n \leq b_n^3$, \eqref{lm:finitebootnew17}, and \eqref{lm:finitebootnew18} imply
\begin{multline}\label{lm:finitebootnew19}
\sup_{P\in \mathbf P} E_P[\sup_{A \in \mathcal B} E_P[1\{L_n(P) \in A\} - 1\{\frac{1}{\sqrt n} \sum_{i=1}^n \bar G_{i,n}(P) \in A^{3Kb_n}\}| \{Z_i,W_{i,n}\}_{i=1}^n]] \\
\lesssim  \min_{t\geq 0} ( \exp\{-\frac{t^2}{C\log(1+p)}\} + \frac{t^2}{K^3 \log(1+p)}) \leq \exp\{-\frac{K^2}{C}\} + \frac{1}{K},
\end{multline}
where in the final inequality we set $t = K\sqrt{\log(1+p)}$.
Theorem 4 in \cite{monrad1991nearby} and result \eqref{lm:finitebootnew19} then imply that there exists a $\bar {\mathbb G}_n(P)$ such that
\begin{equation*}
\|L_n(P) - \bar {\mathbb G}_n(P)\|_\infty = O_P(b_n)
\end{equation*}
uniformly in $P\in \mathbf P$, and its distribution conditional on $\{Z_i,W_{i,n}\}_{i=1}^n$ is given by
\begin{equation}\label{lm:finitebootnew21}
\bar {\mathbb G}_n(P) \sim N(0, \sum_{i=1}^n \text{Var}\{\xi_{i,n}(P)|\{Z_i,W_{i,n}\}_{i=1}^n\}).
\end{equation}

\noindent \underline{Step 4:} (Remove Dependence). We next couple $\bar {\mathbb G}_n(P)$ to a Gaussian vector $\tilde {\mathbb G}_n^\star(P)$ that is independent of $\{Z_i,W_{i,n}\}_{i=1}^n$.
To this end, we first note result \eqref{lm:finitebootnew5} implies
\begin{equation*}
\hat \Lambda_n(P) \equiv \sum_{i=1}^n \text{Var}\{\xi_{i,n}(P)|\{Z_i,W_{i,n}\}_{i=1}^n\}  = \frac{\hat \sigma_n^2}{n} \sum_{i=1}^n (\varphi(Z_i,P)\varphi(Z_i,P)^\prime - \bar \varphi_n(P)\bar \varphi_n(P)^\prime).
\end{equation*}
Moreover, $E_P[\varphi(Z,P)] = 0$ and $\sup_{P\in \mathbf P} \max_{1\leq j \leq 2p} \|\varphi_j(\cdot,P)\|_{P,2}$ being bounded in $n$ by Assumptions \ref{ass:moments}(i)(ii), and $\|aa^\prime\|_{o,2} = \|a\|_2^2$ for any $a\in \mathbf R^{2p}$ imply
\begin{multline}\label{lm:finitebootnew23}
\sup_{P\in \mathbf P} E_P[\|\bar \varphi_n(P)\bar \varphi_n(P)^\prime\|_{o,2}] = \sup_{P\in \mathbf P} E_P[\|\bar \varphi_n(P)\|_2^2] \\
= \sup_{P\in \mathbf P} \sum_{j=1}^{2p} E_P[(\frac{1}{n}\sum_{i=1}^n \varphi_{j}(Z_i,P))^2] \lesssim \frac{p}{n}.
\end{multline}
Jensen's inequality, $\|\varphi(Z_i,P)\|_2^2 \leq 2p \Psi^2(Z_i,P)$, and Assumption \ref{ass:4bootnew}(iv) imply
\begin{multline*}
\sup_{P\in \mathbf P} E_P[\max_{1\leq i \leq n} \|\varphi(Z_i,P)\|_2^2] \lesssim \sup_{P\in \mathbf P} pE_P[\max_{1\leq i \leq n}\Psi^2(Z_i,P)] \\
\leq \sup_{P\in \mathbf P} p( E_P[\max_{1\leq i \leq n} \Psi^{2q}(Z_i,P)])^{1/q} \leq \sup_{P\in \mathbf P} p(n E_P[\Psi^{2q}(Z_i,P)])^{1/q} \leq pn^{1/q}M_{q,\Psi^2}.
\end{multline*}
Setting $\Lambda(P) \equiv E_P[\varphi(Z,P)\varphi(Z,P)^\prime]$, we then note that  $b_n = o(1)$, Lemma \ref{lm:matrixlln}, $\|\Lambda(P)\|_{o,2}$ being uniformly bounded in $n$ and $P\in \mathbf P$ by Assumption \ref{ass:moments}(ii) and definition of $\varphi(Z,P)$, and Markov's inequality allow us to conclude that
\begin{equation}\label{lm:finitebootnew25}
\|\frac{1}{n}\sum_{i=1}^n \varphi(Z_i,P)\varphi(Z_i,P)^\prime - \Lambda(P)\|_{o,2} = O_P(\{\frac{p\log(1+p)n^{1/q} M_{q,\Psi^2}}{n}\}^{1/2})
\end{equation}
uniformly in $P\in \mathbf P$.
Therefore, the triangle inequality, \eqref{lm:finitebootnew23}, \eqref{lm:finitebootnew25}, $\|\Lambda(P)\|_{o,2}$ being bounded in $n$ and $P\in \mathbf P$ and Assumption \ref{ass:4bootnew}(iii) yield
\begin{align*}
\|\hat\Lambda_n(P) - \Lambda(P)\|_{o,2} & \leq |\hat \sigma_n^2 -1| \|\Lambda(P)\|_{o,2} + O_P(\{\frac{p\log(1+p)n^{1/q} M_{q,\Psi^2}}{n}\}^{1/2}) \notag \\
& = O_P(\{\frac{p\log(1+p)n^{1/q} M_{q,\Psi^2}}{n}\}^{1/2})
\end{align*}
uniformly in $P\in \mathbf P$.
Since the distribution of $\bar {\mathbb G}_n(P)$ conditional on $\{Z_i,W_{i,n}\}_{i=1}^n$ equals \eqref{lm:finitebootnew21}, applying Lemma \ref{aux:finiteboot} with $V_n = \{Z_i,W_{i,n}\}_{i=1}^n$ implies there is a $\tilde {\mathbb G}_n^\star(P) \sim N(0,\Lambda(P))$ independent of $\{Z_i,W_{i,n}\}_{i=1}^n$ satisfying uniformly in $P\in \mathbf P$
\begin{equation*}
\|\bar {\mathbb G}_n(P) - \tilde {\mathbb G}_n^\star(P)\|_\infty = O_P((\frac{p\log^3(1+p) n^{1/q} M_{q,\Psi^2}}{n})^{1/4}).
\end{equation*}

\noindent \underline{Step 5:} (Couple $\hat {\mathbb G}_n$).
Combining Steps 2, 3, and 4, we obtain that there exists a Gaussian vector $\tilde {\mathbb G}_n^\star(P)$ that is independent of $\{Z_i,W_{i,n}\}_{i=1}^n$ and satisfies
\begin{equation*}
\|S_n(P) - \tilde {\mathbb G}_n^\star(P)\|_\infty = O_P(b_n)
\end{equation*}
uniformly in $P\in \mathbf P$.
Since $\tilde {\mathbb G}_n^\star(P)$ is independent of $\{Z_i\}_{i=1}^n$, the representation in Step 1 and Lemma 2.11 in \cite{DudleyPhilipp1983Invariance} imply that there exists a $(\breve{\mathbb G}_n^{\rm e\star}(P)^\prime, \breve{\mathbb G}_n^{\rm i\star}(P)^\prime)^\prime \equiv \breve{\mathbb G}_n^\star(P)\sim N(0,\Lambda(P))$ independent of $\{Z_i\}_{i=1}^n$ and such that
\begin{equation}\label{lm:finitebootnew29}
\|\frac{1}{\sqrt n} \sum_{i=1}^n (W_{i,n} - \bar W_n)\varphi(Z_i,P) - \breve{\mathbb G}_n^\star(P)\|_\infty = O_P(b_n)
\end{equation}
uniformly in $P\in \mathbf P$.
To conclude, set $\mathbb G_n^{\rm j \star}(P) \equiv \Omega^{\rm j}(P) \breve{\mathbb G}_n^{\rm j\star}(P)$ for $\rm j \in \{\rm e,\rm i\}$ and $\mathbb G_n^\star(P) \equiv(\mathbb G_n^{\rm e \star}(P)^\prime, \mathbb G_n^{\rm i \star}(P)^\prime)^\prime$.
Then note that since $\Omega^{\rm j}(P)(\Omega^{\rm j}(P))^\dagger \psi^{\rm j}(Z,P) = \psi^{\rm j}(Z,P)$ $P$-almost surely for $\rm j \in \{\rm e,\rm i\}$ by Assumption \ref{ass:techrange}(i), it follows from $\Lambda(P) \equiv E_P[\varphi(Z,P)\varphi(Z,P)^\prime]$ and the definition of $\varphi$ that $\mathbb G_n^\star(P) \sim N(0,\Sigma(P))$.
Furthermore, since $\breve{\mathbb G}_n^\star(P)$ belongs to the range of $\Lambda(P)$ almost surely by Theorem 3.6.1 in \cite{bogachev1998gaussian}, it follows that $\breve{\mathbb G}_n^{\rm j\star}(P) = (\Omega^{\rm j}(P))^\dagger \Omega^{\rm j}(P) \breve{\mathbb G}_n^{\rm j \star}(P) = (\Omega^{\rm j}(P))^\dagger \mathbb G_n^{\rm j \star}(P)$ for $\rm j \in \{\rm e, \rm i\}$.
The lemma thus follows from \eqref{lm:finitebootnew29}, the definition of $\varphi(Z,P)$, and Assumption \ref{ass:4bootnew}(i). \qed

\begin{lemma}\label{lm:deg}
Let Assumptions \ref{ass:weights}, \ref{ass:beta}(ii), \ref{ass:techrange}, \ref{ass:4bootnew}(v) hold, $a_n = o(1)$, and for $\rm j \in \{\rm e, \rm i\}$ set $\mathbf D_0^{\rm j} \equiv \{P\in \mathbf P_0 : \sigma^{\rm j}(s,P) = 0 \text{ for all } s\in \mathcal E^{\rm j}(P)\}$. Then:
$$\liminf_{n\rightarrow \infty} \inf_{P \in \mathbf D_0^{\rm e}} P(\sup_{s\in \hat{\mathcal V}_n^{\rm e}} |\sqrt n \langle s,\hat \beta_n -A\hat x_n^\star \rangle| = \sup_{s\in \hat{\mathcal V}_n^{\rm e}} |\langle s,\hat{\mathbb G}_n^{\rm e}\rangle| = 0) = 1$$
$$\liminf_{n\rightarrow \infty} \inf_{P \in \mathbf D_0^{\rm i}} P(\sup_{s\in \hat{\mathcal V}_n^{\rm i}} |\langle A^\dagger s, A^\dagger \hat{\mathbb G}_n^{\rm i}\rangle |= \sup_{s\in \hat{\mathcal V}_n^{\rm i}} |\langle A^\dagger s, A^\dagger \beta(P) - \hat x_n^\star\rangle | = \sup_{s\in \hat{\mathcal V}_n^{\rm i}} \langle A^\dagger s,\hat x_n^\star\rangle = 0) = 1.$$
\end{lemma}

\noindent \emph{Proof:} See Supplemental Appendix II. \qed

\begin{lemma}\label{lm:obviousmatrix}
If Assumption \ref{ass:weights} holds and $a_n = o(\sqrt{\log(1+p)})$, then $\|(\hat \Omega_n^{\text{\rm e}})^\dagger(\hat \Omega_n^{\text{\rm e}} - \Omega^{\rm e}(P))\|_{o,\infty}\vee\|(\hat \Omega_n^{\text{\rm i}})^\dagger(\hat \Omega_n^{\text{\rm i}} - \Omega^{\rm i}(P))\|_{o,\infty} = O_P(a_n/\sqrt{\log(1+p)})$ uniformly in $P\in \mathbf P$.
\end{lemma}

\noindent \emph{Proof:} See Supplemental Appendix II. \qed

\begin{lemma}\label{aux:finiteboot}
Let $\{V_n\}_{n=1}^\infty$ be random variables with distribution parametrized by $P \in \mathbf P$ and $\bar {\mathbb G}_n(P) \in \mathbf R^{d_n}$ be such that $\bar {\mathbb G}_n(P) \sim N(0,\hat \Sigma_n(P))$ conditionally on $V_n$.
If there exist non-random matrices $\Sigma_n(P)$ such that $\|\hat \Sigma_n(P) - \Sigma_n(P)\|_{o,2} = O_P(\delta_n)$ uniformly in $P \in \mathbf P$, then there exists a $\mathbb G_n^\star(P) \sim N(0,\Sigma_n(P))$ independent of $V_n$ and satisfying $\|\bar {\mathbb G}_n(P) - \mathbb G_n^\star(P)\|_\infty = O_P(\sqrt{\log(1+d_n)\delta_n})$ uniformly in $P\in \mathbf P$.
\end{lemma}

\noindent \emph{Proof:} See Supplemental Appendix II. \qed

\begin{lemma}\label{lm:obvious}
Let $Z = (Z_1,\ldots, Z_p)\in \mathbf R^p$ be jointly Gaussian with $E[Z_j] = 0$ and $E[Z_j^2] \leq \sigma^2$ for all $1\leq j \leq p$. Then, there is a universal $K < \infty$ such that for any $q \geq 1$ we have $E[\|Z\|_\infty^q] \leq (q!\sqrt{\log(1+p)} \sigma K/\sqrt{\log(2)})^q.$
\end{lemma}

\noindent \emph{Proof:} See Supplemental Appendix II. \qed

\begin{lemma}\label{lm:matrixlln}
Let $\{V_i\}_{i=1}^n$ be i.i.d.\ with $V_i \in \mathbf R^k$ and $\Sigma\equiv E[VV']$. Then:
$$E[\|\frac{1}{n}\sum_{i=1}^{n}V_iV_i'-\Sigma\|_{o,2}]\leq \max\{\|\Sigma\|_{o,2}^{1/2}\delta,\delta^2\},$$
where $\delta\equiv D\sqrt{E[\max_{1\leq i \leq n}\|V_i\|_2^2]\log(1+k)/n}$ for some universal constant $D$.
\end{lemma}

\noindent \emph{Proof:} See Supplemental Appendix II. \qed

\begin{lemma}\label{lm:auxpseudo}
Let $\Omega_1$ and $\Omega_2$ be $k\times k$ symmetric matrices such that $\text{\rm range}\{\Omega_1\} = \text{\rm range}\{\Omega_2\}$.
It then follows that $\Omega_2 \Omega_2^\dagger \Omega_1 = \Omega_1$ and $\Omega_2^\dagger \Omega_2 \Omega_1^\dagger = \Omega_1^\dagger$.
\end{lemma}

\noindent \emph{Proof:} See Supplemental Appendix II. \qed

\begin{lemma}\label{lm:auxdensity}
Let $(\mathbb Z_1,\ldots, \mathbb Z_d)^\prime \equiv \mathbb Z \in \mathbf R^d$ be Gaussian with $E[\mathbb Z_j] \geq 0$, $\text{\rm Var}\{\mathbb Z_j\} = \sigma^2 > 0$ for all $1\leq j \leq d$, and define $\mathbb S \equiv \max_{1\leq j \leq d} \mathbb Z_j$ and $\text{\rm m} \equiv \text{\rm med}\{\mathbb S\}$. Then, the distribution of $\mathbb S$ is absolutely continuous and its density is bounded on $\mathbf R$ by $(2/\sigma) \max\{{\rm m}/\sigma,1\}.$
\end{lemma}

\noindent \emph{Proof:} See Supplemental Appendix II. \qed

\begin{lemma}\label{lm:auxlp}
Let $C\subseteq \mathbf R^k$ be a nonempty, closed, polyhedral set containing no lines, and $\mathcal E$ denote its extreme points. Then: $\mathcal E\neq \emptyset$ and for any $y \in \mathbf R^k$ such that $\sup_{c \in C} \langle c,y\rangle < \infty$, it follows that $\sup_{c \in C} \langle c,y\rangle= \max_{c\in \mathcal E} \langle c,y\rangle$.
\end{lemma}

\noindent \emph{Proof:} See Supplemental Appendix II. \qed

\begin{lemma}\label{lm:Vpoly}
Let $\mathcal V^{\rm i}(P)$ be as defined in \eqref{eq:Vdef}. Then, $(AA^\prime)^\dagger \mathcal V^{\rm i}(P)$ is nonempty, closed, polyehdral, contains no lines, and zero is one of its extreme points.
\end{lemma}

\noindent \emph{Proof:} See Supplemental Appendix II. \qed



\phantomsection
\addcontentsline{toc}{section}{References}

{\small
\setstretch{0.99}

\putbib }

\end{bibunit}

\newpage


\begin{bibunit}

\renewcommand{\thesection}{M.\arabic{section}}
\renewcommand{\theequation}{M.\arabic{equation}}
\renewcommand{\thelemma}{M.\arabic{section}.\arabic{lemma}}
\renewcommand{\thecorollary}{M.\arabic{section}.\arabic{corollary}}
\renewcommand{\thetheorem}{M.\arabic{section}.\arabic{theorem}}
\renewcommand{\theassumption}{M.\arabic{section}.\arabic{assumption}}
\setcounter{lemma}{0}
\setcounter{theorem}{0}
\setcounter{corollary}{0}
\setcounter{equation}{0}
\setcounter{remark}{0}
\setcounter{section}{0}
\setcounter{assumption}{0}

\setcounter{page}{1}

\begin{center}
    {\Large {\sc Supplemental Appendix II}} \vspace{0.1 in}\\
    {\Large {\sc Not Intended for Publication}}
\end{center}

\onehalfspacing

This Supplemental Appendix is organized as follows.
Section \ref{sec:computational-details} contains computational details on the implementation of our test.
Section \ref{sec:auxproofs} contains the proofs of auxiliary results that where stated in Supplemental Appendix I.

\section{Computational Details} \label{sec:computational-details}

In this appendix, we provide details on how we compute our test statistic, $T_n$, defined in \eqref{eq:Tndef}, the restricted estimator $\hat \beta_n^{\rm r}$, defined in  \eqref{step1:eq2}, and obtain a critical value.
One computational theme that we found important in our simulations is that the pseudoinverse $A^{\dagger}$ can be poorly conditioned.
As we show below, however, it is possible to implement our procedure without computing $A^{\dagger}$ explicitly.

First, we need to select an estimator $\hat{x}_{n}^{\star}$.
In the mixed logit simulation in Section \ref{sec:simulations}, the parameter $\beta(P)$ can be decomposed into $\beta(P) = (\beta_{\rm u}(P)^\prime,\beta_{\rm k}^\prime)^\prime$, where $\beta_{\rm u}(P)\in \mathbf R^{p_{\rm u}}$ and $ \beta_{\rm k}\in \mathbf R^{p_{\rm k}}$ is a known constant for all $P\in \mathbf P_0$.
Similarly, we decompose any $b\in \mathbf R^p$ into $b = (b_{\rm u}^\prime, b_{\rm k}^\prime)^\prime$ with $b_{\rm u}\in \mathbf R^{p_{\rm u}}$ and $b_{\rm k}\in \mathbf R^{p_{\rm k}}$, and partition the matrix $A$ into the corresponding submatrices $A_{\rm u}$ (dimension $p_{\rm u}\times d$) and $A_{\rm k}$ (dimension $p_{\rm k}\times d$).
In our simulations, we then set $\hat{x}_{n}^{\star}$ to be a solution to
\begin{align}
    \label{eq:range-problem}
    \min_{x \in \re^{d}}
    \left(
        \hat \beta_{{\rm u},n}
        -
        A_{\rm u} x
    \right)'
    \hat{\Xi}_{n}^{-1}
    \left(
        \hat \beta_{{\rm u},n}
        -
        A_{\rm u} x
    \right)
~
    \text{s.t.}
    ~
    A_{\rm k}
    x
    =
    \beta_{\rm k},
\end{align}
where $\hat \beta_n = (\hat \beta_{{\rm u},n}^\prime, \beta_{\rm k}^\prime)^\prime$ and $\hat{\Xi}_{n}$ is an estimate of the asymptotic variance matrix of $\hat{\beta}_{{\rm u}, n}$.
While the solution to \eqref{eq:range-problem} may not be unique, any two minimizers $x_1$ and $x_2$ of \eqref{eq:range-problem} must satisfy $A x_1 = Ax_2$.
Since in our reformulations below $\hat x_n^\star$ only enters through $A\hat x_n^\star$, the specific choice of minimizer in \eqref{eq:range-problem} is immaterial.

Throughout, we let $\hat{\Omega}_{n}^{\rm e}$ be the sample standard deviation matrix of $\hat{\beta}_{n}$.
Note that, since $\hat \beta_n = (\hat \beta_{{\rm u},n}^\prime, \beta_{\rm k}^\prime)^\prime$ and $\beta_{\rm k}$ is non-stochastic, $\hat{\Omega}_{n}^{\rm e}$ has the form
\begin{align}
    \label{eq:omegahate-block-form}
    \hat{\Omega}_{n}^{\rm e}
    =
    \left[\begin{array}{cc}
        \hat{\Xi}_{n}^{1/2} & 0 \\
        0 & 0
    \end{array}\right].
\end{align}
We further let $\hat{\Omega}_{n}^{\rm i}$ be the sample standard deviation of $A\hat{x}_{n}^{\star}$, although this choice of studentization plays no special computational role in what follows.

Consider the first component of $T_n$ (see \eqref{eq:Tndef}), which we reproduce here as
\begin{align}
    \label{eq:tnrange}
    T_{n}^{\rm e}
    \equiv
    \sup_{s\in \hat{\mathcal V}_n^{\rm e}} \sqrt n \langle s, \hat \beta_n - A\hat x_n^\star\rangle
    ~
    \text{where}
    ~
    \hat {\mathcal V}_n^{\rm e}  \equiv \{s \in \mathbf R^p : \|\hat \Omega_n^{\rm e} s\|_1 \leq 1\}.
\end{align}
As in the main text, the superscript ``e'' alludes to the relation to the ``equality" condition in Theorem \ref{th:ineq}. 
As noted in the main text, $\hat \beta_n = A \hat x_n^\star$ and hence $T_{n}^{\rm e} = 0$ whenever $A$ is full rank and $d \geq p$.
In other cases, we use the fact that $\hat{x}_{n}^{\star}$, as the solution to \eqref{eq:range-problem}, must satisfy $A_{\rm k} \hat{x}_{n}^{\star} = \beta_{\rm k}$, and that our choice of $\hat{\Omega}_{n}^{\rm e}$ in \eqref{eq:omegahate-block-form} has $\hat{\Xi}_{n}^{1/2}$ as its upper left block.
From these observations, we deduce that
\begin{align}\label{eq:compTnefinal}
    T_{n}^{\rm e}
    &=
    \sup_{s_{\rm u} \in \re^{p_{\rm u}}}
    \sqrt n \langle s_{\rm u}, \hat \beta_{{\rm u}, n} - A_{\rm u}\hat x_n^\star\rangle
    ~
    \text{s.t.}
    ~
    \|\hat \Xi_n^{1/2} s_{\rm u}\|_1 \leq 1 \notag \\
    &=
    \Vert\sqrt{n}\hat{\Xi}_{n}^{-1/2}
        \left(
            \hat{\beta}_{{\rm u}, n}
            -
            A_{\rm u}
            \hat{x}_{n}^{\star}
        \right)
    \Vert_{\infty}.
\end{align}
Thus, $T_{n}^{\rm e}$ can be computed by taking the maximum of a vector of length $p_{\rm u}$.

The second component of $T_n$, defined in \eqref{eq:Tndef}, is reproduced here as
\begin{equation}
    \label{eq:tnpos-def}
    T_{n}^{\rm i}
    \equiv
    \sup_{s\in \hat {\mathcal V}_n^{\rm i}} \sqrt n \langle A^\dagger s, \hat x_n^\star \rangle
    ~
    \text{where}
    ~
    \hat {\mathcal V}_n^{\rm i}  \equiv \{s \in \mathbf R^p : A^\dagger s \leq 0 \text{ and } \|\hat \Omega_n^{\rm i}(AA^\prime)^\dagger s\|_1 \leq 1\},
\end{equation}
and the superscript ``i'' alludes to the relation to the ``inequality" condition in Theorem \ref{th:ineq}. 
To compute $T_n^{\rm i}$ without explicitly using $A^\dagger$, we first note
\begin{align}
    \label{eq:pseudoinverse-identity}
    A^{\dagger} = A'(AA')^{\dagger},
\end{align}
see, e.g., Proposition 6.11.1(9) in \cite{luenberger:1969}.
Then, we observe that
\begin{align}
    \label{eq:positive-part-simplification}
    \text{range}\{(AA')^{\dagger}\}
    =
    \text{null}\{AA'\}^{\perp}
    =
    \text{range}\{AA'\}
    =
    \text{range}\{A\},
\end{align}
where the first equality is a property of pseudoinverses, see \citet[][pg. 164]{luenberger:1969}.
The second equality is a standard result in linear algebra, see Theorem 6.6.1 in \cite{luenberger:1969}.
This result is also used in the third equality, which uses the following logic: if $t = As$ for some $s \in \mathbf R^{p}$, then also $t = As_{1}$, where $s_{1} \in \text{null}\{A\}^{\perp} = \text{range}\{A'\}$ is determined from the orthogonal decomposition $s = s_{0} + s_{1}$  with $s_{0} \in \text{null}\{A\}$, and hence $t \in \text{range}\{AA'\}$ implying $\text{range}\{A\} \subseteq \text{range}\{AA^\prime\}$. Since $\text{range}\{AA^\prime\} \subseteq \text{range}\{A\}$ the third equality follows.
Thus,
\begin{align} \label{eq:tnpos-intermediate}
    T_{n}^{\rm i} & = \sup_{s \in \mathbf R^{p}}
    \sqrt n \langle A'(AA^\prime)^\dagger s, \hat x_n^\star \rangle
    ~
    \text{s.t.}
    ~
    A'(AA^\prime)^\dagger s \leq 0
    ~
    \text{and}
    ~
    \|\hat \Omega_n^{\rm i} (AA^\prime)^\dagger s\|_1 \leq 1, \notag \\
    & = \sup_{x \in \mathbf R^{d}}
    \sqrt n \langle A'Ax, \hat x_n^\star \rangle
    ~
    \text{s.t.}
    ~
    A'Ax \leq 0
    ~
    \text{and}
    ~
    \|\hat \Omega_n^{\rm i}A x\|_1 \leq 1, \notag \\
    & = \sup_{x \in \mathbf R^{d}, s \in \mathbf{R}^{p}}
    \sqrt n \langle s, A\hat{x}_{n}^{\star} \rangle
    ~
    \text{s.t.}
    ~
    Ax = s,
    ~
    A's \leq 0
    ~
    \text{and}
    ~
    \|\hat \Omega_n^{\rm i}s\|_1 \leq 1,
\end{align}
where the first equality follows from \eqref{eq:pseudoinverse-identity}, the second from \eqref{eq:positive-part-simplification}, and in the third we substituted $s = Ax$.
The final program in \eqref{eq:tnpos-intermediate} can be written explicitly as a linear program by introducing non-negative slack variables, so that
\begin{multline}\label{eq:pos-program-final}
    T_{n}^{\rm i}
    =
    \sup_{x \in \mathbf R^{d}, s \in \mathbf{R}^{p}, \phi^{+} \in \mathbf{R}_{+}^{p}, \phi^{-} \in \mathbf{R}_{+}^{p}}
    \sqrt n \langle s, A\hat{x}_{n}^{\star} \rangle \\
    \text{s.t.} ~
    Ax = s,~
    A's \leq 0, ~
    \langle \mathbf{1}_{p}, \phi^{+}\rangle + \langle \mathbf{1}_{p},\phi^{-}\rangle  \leq 1, ~
    \phi^{+} - \phi^{-} = \hat{\Omega}_{n}^{\rm i}s,
\end{multline}
where $\mathbf 1_p \in \mathbf R^p$ is the vector with all coordinates equal to one.
Note that if $d \geq p$ and $A$ has full rank, then the constraint $Ax = s$ is redundant since $Ax$ ranges across all of $\mathbf{R}^{p}$ as $x$ varies across $\mathbf{R}^{d}$.
In these cases, the constraint $Ax = s$ together with the variable $x$ can be entirely removed from the linear program in \eqref{eq:pos-program-final}.
Taking the maximum of \eqref{eq:compTnefinal} and \eqref{eq:pos-program-final} yields our test statistic $T_n$.

Turning to our bootstrap procedure, we first show how to solve \eqref{step1:eq2} to find $\hat{\beta}_{n}^{\rm r}$.
The optimization problem to solve is here reproduced as:
\begin{equation}
\label{step1:eq2-appendix}
\min_{\tilde x \in \mathbf{R}^{d}_{+}, b=(b_{\rm u}^\prime, b_{\rm k}^\prime)^\prime} \left[\sup_{s\in \hat {\mathcal V}_n^{\rm i}}  |\langle A^\dagger s, \hat x_n^\star - A^\dagger b\rangle|\right] \text{ s.t. } b_{\rm k} = \beta_{\rm k}, ~A\tilde x = b.
\end{equation}
With probability tending to one, the inner problem is finite when evaluated at $b = \beta(P)$, and hence we may restrict attention to $b$ for which the inner problem is finite.
Moreover, the inner problem has the same structure as \eqref{eq:tnpos-def}, but with $\hat{x}_{n}^{\star}$ replaced by $\hat{x}_{n}^{\star} - A^{\dagger}b$. 
Hence, applying the same logic employed in \eqref{eq:tnpos-intermediate} allows us to rewrite the inner problem in \eqref{step1:eq2-appendix} as being equal to
\begin{align}
    \label{eq:betahatr-intermediate}
    \sup_{x \in \mathbf R^{d}}
    |\langle A'Ax, \hat x_n^\star - A^{\dagger}b \rangle|
    ~
    \text{s.t.}
    ~
    A'Ax \leq 0
    ~
    \text{and}
    ~
    \|\hat \Omega_n^{\rm i}A x\|_1 \leq 1.
\end{align}
It is in turn possible to establish that the optimization problem in \eqref{eq:betahatr-intermediate} equals
\begin{align}\label{eq:betahatr-intermediate1}
\sup_{x \in \mathbf R^{d}}
    \langle A'Ax, \hat x_n^\star - A^{\dagger}b \rangle
    ~
    \text{s.t.}
    ~
    x \in \text{co}\{v \in \mathbf R^d : A'Av \leq 0, ~  \|\hat \Omega_n^{\rm i}A v\|_1 \leq 1\},
\end{align}
where $\text{co}\{\cdot\}$ denotes the convex hull of a set.
By introducing slack variables as in \eqref{eq:pos-program-final}, we may rewrite \eqref{eq:betahatr-intermediate1} explicitly as a linear program
\begin{align}\label{eq:betahatr-intermediate2}
& \sup_{v_1,v_2\in \mathbf R^d, \phi_1^+, \phi_1^-,\phi_2^+,\phi_2^-\in \mathbf R^p_+, a_1,a_2 \in \mathbf R_+}
 \langle A'A(v_1 + v_2), \hat x_n^\star - A^{\dagger}b \rangle \notag \\
     & \text{s.t. }~  A^\prime A v_1 \leq 0, ~  -A^\prime A v_2 \leq 0, ~ \langle \mathbf 1_p,\phi_1^+\rangle +\langle \mathbf 1_p,\phi_1^-\rangle \leq a_1, ~ \langle \mathbf 1_p,\phi_2^+\rangle +\langle \mathbf 1_p,\phi_2^-\rangle \leq a_2, \notag \\
     &\phi_1^+ - \phi_1^- = \hat \Omega_n^{\rm i} A v_1, ~  \phi_2^+ - \phi_2^- = \hat \Omega_n^{\rm i} A v_2, ~ a_1 + a_2 = 1.
\end{align}
In turn, the dual of the linear program in \eqref{eq:betahatr-intermediate2} can be shown to be equal to
\begin{align}
    \label{justlp:eq1}
    & \inf_{\phi_0\in \mathbf R, \phi_1^s,\phi_2^s\in \mathbf R^d_+, \phi_1^n,\phi_2^n\in \mathbf R_+, \phi_1^e,\phi_2^e\in \mathbf R^p}\phi_0
    ~ \text{s.t.}
    ~ A^\prime A \phi_1^s - A^\prime \hat \Omega_n^{\rm i} \phi_1^e = A^\prime A(\hat x_n^\star - A^\dagger b), \notag  \\
         & -A^\prime A \phi_2^s - A^\prime \hat \Omega_n^{\rm i} \phi_2^e = A^\prime A(\hat x_n^\star - A^\dagger b), ~ \phi_1^n \mathbf 1_p + \phi_1^e \geq 0, ~
         \phi_1^n \mathbf 1_p - \phi_1^e \geq 0, \notag \\ & \phi_2^n \mathbf 1_p + \phi_2^e \geq 0, ~ \phi_2^n \mathbf 1_p - \phi_2^e \geq 0, ~ -\phi_1^n + \phi_0 \geq 0, ~ -\phi_2^n + \phi_0 \geq 0.
\end{align}
Let $V \equiv \text{range}\{AA^\prime\}$ and note $A^\dagger = A^\prime(AA^\prime)^\dagger$ (see Proposition 6.11.1(8) in \cite{luenberger:1969}) implies $A^\prime AA^\dagger b = A^\prime AA^\prime (AA^\prime)^\dagger b = A^\prime \Pi_V b$. However, by \eqref{eq:positive-part-simplification}, $V \equiv \text{range}\{AA^\prime\} = \text{range}\{A\} = \text{null}\{A^\prime\}^\perp$, where the final equality follows by Theorem 6.6.1 in \cite{luenberger:1969}. Hence, $A^\prime \Pi_V b = A^\prime b$ and \eqref{justlp:eq1} equals
\begin{align}
    \label{justlp:eq2}
        & \inf_{\phi_0\in \mathbf R, \phi_1^s,\phi_2^s\in \mathbf R^d_+, \phi_1^n,\phi_2^n\in \mathbf R_+, \phi_1^e,\phi_2^e\in \mathbf R^p}\phi_0
    ~ \text{s.t.}
    ~ A^\prime A \phi_1^s - A^\prime \hat \Omega_n^{\rm i} \phi_1^e = A^\prime( A\hat x_n^\star - b), \notag  \\
         & -A^\prime A \phi_2^s - A^\prime \hat \Omega_n^{\rm i} \phi_2^e = A^\prime( A\hat x_n^\star - b), ~ \phi_1^n \mathbf 1_p + \phi_1^e \geq 0, ~
         \phi_1^n \mathbf 1_p - \phi_1^e \geq 0, \notag \\ & \phi_2^n \mathbf 1_p + \phi_2^e \geq 0, ~ \phi_2^n \mathbf 1_p - \phi_2^e \geq 0, ~ -\phi_1^n + \phi_0 \geq 0, ~ -\phi_2^n + \phi_0 \geq 0.
\end{align}
Substituting \eqref{justlp:eq2} back into the inner problem in \eqref{step1:eq2-appendix} then yields a single linear program that determines $\hat \beta_n^{\rm r}$.
Given $\hat \beta_n^{\rm r}$ it is then straightforward to compute our bootstrap statistic.
For instance, in the simulations in Section \ref{sec:simulations}, we let
$$\hat {\mathbb G}_n^{\rm e}  = \sqrt n\{(\hat \beta_{b,n} - A\hat x_{b,n}^\star) - (\hat \beta_n - A\hat x_n^\star)\} \hspace{0.3 in} \hat {\mathbb G}_n^{\rm i} = \sqrt n A(\hat x^\star_{b,n} - \hat x_n^\star)$$
where $\hat \beta_{b,n}$ and $\hat x_{b,n}^\star$ are nonparametric bootstrap analogues to $\hat \beta_n$ and $\hat x_n^\star$.
Arguing as in result \eqref{eq:compTnefinal} it is then straightforward to show that
\begin{equation}\label{eq:suppappboot1}
\sup_{s \in \hat{\mathcal V}_n^{\rm e}}\langle s,\hat {\mathbb G}_n^{\rm e}\rangle = \Vert\sqrt{n}\hat{\Xi}_{n}^{-1/2}\hat{\mathbb G}_n^{\rm e}    \Vert_{\infty}.
\end{equation}
In analogy to \eqref{eq:compTnefinal}, \eqref{eq:suppappboot1} equals zero whenever $A$ is full rank and $d\geq p$.
Next, we may employ the same arguments as in \eqref{eq:tnpos-intermediate} and \eqref{eq:pos-program-final} and note $AA^\dagger \hat {\mathbb G}_n^{\rm i} = \hat {\mathbb G}_n^{\rm i}$ and $AA^\dagger \hat \beta_n^{\rm r} = \hat \beta_n^{\rm r}$ because $\hat {\mathbb G}_n^{\rm i}$ and $\hat \beta_n^{\rm r}$ are on the range of $A$ to obtain
\begin{align}\label{eq:suppappboot2}
\sup_{s\in \hat {\mathcal V}_n^{\rm i}} \langle A^\dagger s, & A^\dagger(\hat {\mathbb G}_n^{\rm i} + \sqrt n \lambda_n \hat \beta_n^{\rm r})\rangle \notag \\
 =  & \sup_{x \in \mathbf R^{d}, s \in \mathbf{R}^{p}, \phi^{+} \in \mathbf{R}_{+}^{p}, \phi^{-} \in \mathbf{R}_{+}^{p}}    \langle s, \hat {\mathbb G}_n^{\rm i} + \sqrt n \lambda_n \hat \beta_n^{\rm r} \rangle  \notag\\ & \text{s.t.} ~     Ax = s,~     A's \leq 0, ~     \langle \mathbf{1}_{p}, \phi^{+}\rangle + \langle \mathbf{1}_{p},\phi^{-}\rangle  \leq 1, ~
    \phi^{+} - \phi^{-} = \hat{\Omega}_{n}^{\rm i}s.
\end{align}
As in \eqref{eq:pos-program-final}, we note that if $A$ is full rank and $d\geq p$, then the constraint $Ax = s$ and the variable $x$ may be dropped from \eqref{eq:suppappboot2}.
The critical value is then obtained by computing the $1-\alpha$ quantile of the maximum of \eqref{eq:suppappboot1} and \eqref{eq:suppappboot2} across bootstrap iterations.
Finally, we note that the problem \eqref{eq:lambda2} used to determine $\lambda_{n}^{\rm b}$ is equivalent to \eqref{eq:pos-program-final} with $A\hat{x}_{n}^{\star}$ replaced by $\hat{\mathbb{G}}_{n}^{\rm i}$.

\section{Additional Proofs} \label{sec:auxproofs}

\noindent \emph{Proof of Lemma \ref{lm:deg}:} Theorem 3.6.1 in \cite{bogachev1998gaussian} and Assumption \ref{ass:techrange}(i) imply $\mathbb G_n^{\rm e}(P) \in \text{range}\{\Sigma^{\rm e}(P)\} \subseteq \text{range}\{\Omega^{\rm e}(P)\}$ almost surely.
Hence, symmetry of $\Omega^{\rm e}(P)$ and $\Omega^{\rm e}(P)(\Omega^{\rm e}(P))^\dagger \mathbb G_n^{\rm e}(P) = \mathbb G_n^{\rm e}(P)$ almost surely imply for any $P\in \mathbf D_0^{\rm e}$
\begin{multline}\label{lm:deg1}
\sup_{s\in \mathcal V^{\rm e}(P)} |\langle s, \mathbb G_n^{\rm e}(P)\rangle| = \sup_{s\in \mathcal V^{\rm e}(P)} |\langle \Omega^{\rm e}(P) s, (\Omega^{\rm e}(P))^\dagger \mathbb G_n^{\rm e}(P)\rangle| \\
= \max_{s\in \mathcal E^{\rm e}(P)} |\langle s, (\Omega^{\rm e}(P))^\dagger \mathbb G_n^{\rm e}(P)\rangle| = 0,
\end{multline}
where the second equality follows from H\"older's inequality implying the supremum is finite and Lemma \ref{lm:auxlp}.
Also note $\hat \Omega_n^{\rm e}(\hat \Omega_n^{\rm e})^\dagger \Omega^{\rm e}(P) = \Omega^{\rm e}(P)$ with probability tending to one uniformly in $P\in \mathbf P$ by Assumption \ref{ass:weights}(iii) and Lemma \ref{lm:auxpseudo}.
Thus, the symmetry of $\hat \Omega_n^{\rm e}$ and $\Omega^{\rm e}(P)$ imply $\Omega^{\rm e}(P) = \Omega^{\rm e}(P)(\hat \Omega_n^{\rm e})^\dagger \hat \Omega_n^{\rm e}$, which together with the definition of $\hat {\mathcal V}_n^{\rm e}$ and $\hat \Omega^{\rm e}_n(\hat \Omega_n^{\rm e})^\dagger \hat \Omega_n^{\rm e} = \hat \Omega_n^{\rm e}$ by Proposition 6.11.1(6) in \cite{luenberger:1969} imply with probability tending to one uniformly in $P\in \mathbf P$ that
\begin{multline*}
\sup_{s \in \hat {\mathcal V}_n^{\rm e}} \|\Omega^{\rm e}(P)s\|_1 \leq 1+ \sup_{s\in \hat{\mathcal V}_n^{\rm e}} \|(\hat \Omega_n^{\rm e} - \Omega^{\rm e}(P))s\|_1 \\ = 1+ \sup_{s\in \hat{\mathcal V}_n^{\rm e}}\|(\hat \Omega_n^{\rm e} - \Omega^{\rm e}(P)) (\hat\Omega_n^{\rm e})^\dagger \hat \Omega_n^{\rm e} s\|_1 \leq 1 + \|(\hat \Omega_n^{\rm e})^\dagger(\hat \Omega_n^{\rm e} - \Omega^{\rm e}(P))\|_{o,\infty},
\end{multline*}
where the final inequality follows from Theorem 6.5.1 in \cite{luenberger:1969}.
Therefore, Lemma \ref{lm:obviousmatrix} and $a_n = o(1)$ imply that $\hat {\mathcal V}_n^{\rm e} \subseteq 2\mathcal V^{\rm e}(P)$ with probability tending to one uniformly in $P\in \mathbf P$.
We can thus conclude from $0\in \hat{\mathcal V}_n^{\rm e}$,  \eqref{lm:deg1}, Assumption \ref{ass:techrange}(ii), and the support of $\mathbb G_n^{\rm e}(P)$ being equal to the range of $\Sigma^{\rm e}(P)$ by Theorem 3.6.1 in \cite{bogachev1998gaussian} that with probability tending to one uniformly in $P\in \mathbf D_0^{\rm e}$
\begin{equation}\label{lm:deg3}
0 \leq \sup_{s\in \hat{\mathcal V}_n^{\rm e}} |\sqrt n \langle s,\hat \beta_n -A\hat x_n^\star \rangle| \leq \sup_{s \in 2\mathcal V^{\rm e}(P)} |\langle s, (I_p - AA^\dagger\hat C_n)\sqrt n \{\hat \beta_n - \beta(P)\}\rangle| = 0.
\end{equation}
Identical arguments but relying on Assumption \ref{ass:4bootnew}(v) instead of \ref{ass:techrange}(i) also yield
\begin{equation}\label{lm:deg4}
0 \leq \sup_{s\in \hat{\mathcal V}_n^{\rm e}} |\langle s, \hat{\mathbb G}_n^{\rm e}\rangle |\leq \sup_{s\in 2 {\mathcal V}^{\rm e}(P)} |\langle s, \hat{\mathbb G}_n^{\rm e}\rangle| = 0
\end{equation}
with probability tending to one uniformly in $P\in \mathbf D_0^{\rm e}$. The first claim of the lemma therefore follows from results \eqref{lm:deg3} and \eqref{lm:deg4}.

For the second claim of the lemma, we note that identical arguments to those employed for the first claim readily establish that $\hat {\mathcal V}_n^{\rm i} \subseteq 2\mathcal V^{\rm i}(P)$ and
\begin{equation}\label{lm:deg5}
\sup_{s\in \hat{\mathcal V}_n^{\rm i}} |\langle A^\dagger s, A^\dagger AA^\dagger\hat C_n\sqrt n\{\hat \beta_n - \beta(P)\}\rangle | = \sup_{s\in \hat{\mathcal V}_n^{\rm i}} |\langle A^\dagger s, A^\dagger \hat{\mathbb G}_n^{\rm i}\rangle| = 0
\end{equation}
with probability tending to one uniformly in $P\in \mathbf D_0^{\rm i}$.
Since $A^\dagger A A^\dagger = A^\dagger$ by Proposition 6.11.1(5) in \cite{luenberger:1969}, it follows that $A^\dagger A A^\dagger \hat C_n \hat \beta_n = \hat x_n^\star$ due to $\hat x_n^\star \equiv A^\dagger \hat C_n \hat \beta_n$ by Assumption \ref{ass:beta}(ii) and therefore \eqref{lm:deg5} yields
\begin{equation}\label{lm:deg6}
\sup_{s\in \hat{\mathcal V}_n^{\rm i}} |\langle A^\dagger s,\hat x_n^\star  - A^\dagger \beta(P)\}\rangle | =0
\end{equation}
with probability tending to one uniformly in $P\in \mathbf D_0^{\rm i}$.
Since $\langle A^\dagger s, A^\dagger \beta(P)\rangle \leq 0$ for any $P\in \mathbf P_0$ and $s \in \mathcal V^{\rm i}(P)$ by Theorem \ref{th:ineq}, we obtain from $0\in \hat{\mathcal V}_n^{\rm i}$ and \eqref{lm:deg6}
\begin{equation*}
0 \leq \sup_{s \in \hat {\mathcal V}_n^{\rm i}} \sqrt n \langle A^\dagger s, \hat x_n^\star\rangle \\ \leq \sup_{s\in  \hat {\mathcal V}_n^{\rm i}} |\langle A^\dagger s,\hat x_n^\star  - A^\dagger \beta(P)\}\rangle | + \sup_{s\in \hat {\mathcal V}_n^{\rm i}} \langle A^\dagger s, A^\dagger \beta(P)\rangle = 0
\end{equation*}
with probability tending to one uniformly in $P\in \mathbf D_0^{\rm i}$. \qed

\noindent \emph{Proof of Lemma \ref{lm:obviousmatrix}:} Assumption \ref{ass:weights} and Lemma \ref{lm:auxpseudo} imply $\Omega^{\rm e}(P)(\Omega^{\rm e}(P))^\dagger\hat \Omega_n^{\rm e} = \hat \Omega_n^{\rm e}$ and $(\hat \Omega_n^{\rm e})^\dagger \hat \Omega_n^{\rm e}(\Omega^{\rm e}(P))^\dagger = (\Omega^{\rm e}(P))^\dagger$ with probability tending to one uniformly in $P\in \mathbf P$.
Since $\Omega^{\rm e}(P)(\Omega^{\rm e}(P))^\dagger \Omega^{\rm e}(P) = \Omega^{\rm e}(P)$ by Proposition 6.11.1(6) in \cite{luenberger:1969}, we obtain, with probability tending to one uniformly in $P\in \mathbf P$:
\begin{align}\label{lm:obviousmatrix1}
\|& (\hat \Omega_n^{\text{\rm e}})^\dagger (\hat \Omega_n^{\text{\rm e}}- \Omega^{\rm e}(P))\|_{o,\infty} \notag \\
& = \|(\hat \Omega_n^{\text{\rm e}})^\dagger\Omega^{\rm e}(P)(\Omega^{\rm e}(P))^\dagger (\hat \Omega_n^{\text{\rm e}} - \Omega^{\rm e}(P))\|_{o,\infty} \notag \\
& \leq \|(\hat \Omega_n^{\text{\rm e}})^\dagger(\Omega^{\rm e}(P)-\hat \Omega_n^{\rm e})\|_{o,\infty} \times o_P(1) +\|(\hat \Omega_n^{\rm e})^\dagger \hat \Omega_n^{\rm e}(\Omega^{\rm e}(P))^\dagger (\hat \Omega_n^{\text{\rm e}} - \Omega^{\rm e}(P))\|_{o,\infty} \notag \\
& = \|(\hat \Omega_n^{\text{\rm e}})^\dagger(\Omega^{\rm e}(P)-\hat \Omega_n^{\rm e})\|_{o,\infty} \times o_P(1) + \|(\Omega^{\rm e}(P))^\dagger (\hat \Omega_n^{\text{\rm e}} - \Omega^{\rm e}(P))\|_{o,\infty},
\end{align}
where the inequality follows from $a_n/\sqrt{1+\log(p)} = o(1)$ and $\|(\Omega^{\rm e}(P))^\dagger (\hat \Omega_n^{\rm e} - \Omega^{\rm e}(P))\|_{o,\infty} = O_P(a_n/\sqrt{1+\log(p)})$ uniformly in $P\in \mathbf P$ by Assumption \ref{ass:weights}(ii).
Since $\|(\Omega^{\rm e}(P))^\dagger (\hat \Omega_n^{\rm e} - \Omega^{\rm e}(P))\|_{o,\infty} = O_P(a_n/\sqrt{1+\log(p)})$ uniformly in $P\in \mathbf P$ by Assumption \ref{ass:weights}(ii),  \eqref{lm:obviousmatrix1} implies $\|(\hat \Omega_n^{\text{\rm e}})^\dagger(\hat \Omega_n^{\text{\rm e}} - \Omega^{\rm e}(P))\|_{o,\infty} = O_P(a_n/\sqrt{\log(1+p)}$ uniformly in $P\in \mathbf P$.
The claim $\|(\hat \Omega_n^{\text{\rm i}})^\dagger(\hat \Omega_n^{\text{\rm i}} - \Omega^{\rm i}(P))\|_{o,\infty} = O_P(a_n/\sqrt{\log(1+p)}$ uniformly in $P\in \mathbf P$ can be established by identical arguments. \qed

\noindent \emph{Proof of Lemma \ref{aux:finiteboot}:}  Let $\{\hat \nu_j(P)\}_{j=1}^{d_n}$ and $\{\hat \lambda_j(P)\}_{j=1}^{d_n}$ denote the unit length eigenvectors and corresponding eigenvalues of $\hat \Sigma_n(P)$.
Further letting $\mathcal N_{d_n}$ be independent of $(V_n, \bar {\mathbb G}_n(P))$ and distributed according to $\mathcal N_{d_n} \sim N(0,I_{d_n})$, we then define
\begin{equation*}
\mathbb Z_n(P) \equiv \sum_{j : \hat \lambda_j(P) \neq 0} \hat \nu_j(P) \frac{\hat \nu_j(P)^\prime \bar {\mathbb G}_n(P)}{\hat \lambda_j^{1/2}(P)} + \sum_{j : \hat \lambda_j(P) = 0} \hat \nu_j(P)(\hat \nu_j(P)^\prime \mathcal N_{d_n}).
\end{equation*}
Since $\mathcal N_{d_n}$ is independent of $V_n$ and $\bar {\mathbb G}_n(P)$ is Gaussian conditional on $V_n$ it follows that $\mathbb Z_{n}(P)$ is Gaussian conditional on $V_n$ as well.
Moreover, we have
\begin{equation*}
E[\mathbb Z_n(P)\mathbb Z_n(P)^\prime|V_n] = \sum_{j=1}^{d_n} \hat \nu_j(P)\hat \nu_j(P)^\prime = I_{d_n},
\end{equation*}
by direct calculation, and hence we conclude that $\mathbb Z_n(P) \sim N(0,I_{d_n})$ and is independent of $V_n$.
Moreover, Theorem 3.6.1 in \cite{bogachev1998gaussian} implies that $\bar {\mathbb G}_n(P)$ belongs to the range of $\hat \Sigma_n(P)$ almost surely.
Thus, since $\{\hat \nu_j(P) : \hat \lambda_j(P) \neq 0 \}$ is an orthonormal basis for the range of $\hat \Sigma_n(P)$, we obtain that almost surely
\begin{equation}\label{aux:finiteboot3}
\hat \Sigma_n^{1/2}(P)\mathbb Z_n(P) = \sum_{j : \hat \lambda_j(P)\neq 0} \hat \nu_j(P)(\hat \nu_j(P)^\prime \bar{\mathbb G}_n(P)) = \bar {\mathbb G}_n(P).
\end{equation}
Employing that $\mathbb Z_n(P)$ is independent of $V_n$, we then define the desired $\mathbb G_n^\star(P)$ by
\begin{equation}\label{aux:finiteboot4}
\mathbb G_n^\star(P) \equiv \Sigma_n^{1/2}(P) \mathbb Z_n(P).
\end{equation}

Next, set $\hat \Delta_n(P) \equiv \hat \Sigma_n^{1/2}(P) - \Sigma^{1/2}_n(P)$ and let $\hat \Delta^{(j,k)}_n(P)$ denote its $(j,k)$ entry.
Note \eqref{aux:finiteboot3}, \eqref{aux:finiteboot4}, Lemma \ref{lm:obvious}, and $\sup_{\|v\|_2 = 1} \langle v,a\rangle = \|a\|_2$ for any $a$ yield
\begin{multline}\label{aux:finiteboot5}
E[\|\bar {\mathbb G}_n(P) - \mathbb G_n^\star(P)\|_\infty|V_n] \lesssim \sqrt{\log(1+d_n)} \max_{1 \leq j \leq d_n} (\sum_{k=1}^{d_n} (\hat \Delta_n^{(j,k)}(P))^2)^{1/2} \\
= \sqrt{\log(1+d_n)} \sup_{\|v\|_2 = 1} \|\hat \Delta_n(P) v\|_\infty \leq \sqrt{\log(1+d_n)} \|\hat \Delta_n(P)\|_{o,2},
\end{multline}
where $\|\hat \Delta_n(P)\|_{o,2}$ denotes the operator norm of $\hat \Delta_n(P) : \mathbf R^{d_n} \to \mathbf R^{d_n}$ when $\mathbf R^{d_n}$ is endowed with the norm $\|\cdot\|_2$, and the final inequality follows from $\|\cdot\|_\infty \leq \|\cdot\|_2$.
Moreover, Theorem X.1.1 in \cite{bhatia:1997} further implies that
\begin{equation}\label{aux:finiteboot6}
\|\hat \Delta_n(P)\|_{o,2}^2 \leq \|\hat \Sigma_n(P) - \Sigma(P)\|_{o,2} = O_P(\delta_n),
\end{equation}
where the equality holds uniformly in $P \in \mathbf P$ by hypothesis.
Therefore, Fubini's theorem, Markov's inequality, and \eqref{aux:finiteboot5} imply for any $C>0$ that
\begin{multline}\label{aux:finiteboot7}
\sup_{P\in \mathbf P} P( \|\bar {\mathbb G}_n(P) - {\mathbb G}_n^\star(P)\|_\infty > C^2 \sqrt{\log(1+d_n)\delta_n} \text{ and } \|\hat \Delta_n(P)\|_{o,2} \leq C\sqrt{\delta_n}) \\
\leq \sup_{P\in \mathbf P}  E_P[\frac{ \|\hat\Delta_n(P)\|_{o,2}}{C^2\sqrt{\delta_n}}1\{\|\hat \Delta_n(P)\|_{o,2} \leq C\sqrt {\delta_n}\}] \leq \frac{1}{C}.
\end{multline}
The claim of the lemma then follows from results \eqref{aux:finiteboot6} and \eqref{aux:finiteboot7}. \qed

\noindent \emph{Proof of Lemma \ref{lm:obvious}:} The result is well known and stated here for ease of reference.
Define $\psi_2 : \mathbf R \to \mathbf R$ to equal $\psi_2(u) = \exp\{u^2\} - 1$ for any $u \in \mathbf R$ and recall that for any random variable $V\in \mathbf R$ its Orlicz norm $\|V\|_{\psi_2}$ is given by $\|V\|_{\psi_2} \equiv \inf \{ C > 0 : E[\psi(|V|/C)] \leq 1\}.$
Further note that for any $q \geq 1$ and random variable $V$ we have $(E[|V|^q])^{1/q} \leq q!\|V\|_{\psi_2}/\sqrt{\log(2)}$; see \cite{vandervaart:wellner:1996} pg. 95.
Hence, Lemmas 2.2.1 and 2.2.2 in \cite{vandervaart:wellner:1996} imply that there exist finite $K_0$ and $K_1$ such that for all $q\geq 1$ we have
\begin{multline*}
E[\|Z\|_\infty^q] \leq (\frac{q!}{\sqrt{\log(2)}})^q \|\max_{1 \leq j \leq p} |Z_j|\|_{\psi_2}^q \\ \leq (\frac{ q!}{\sqrt{\log(2)}})^q\{K_0\sqrt{\log(1 + p)} \max_{1 \leq j \leq p} \| Z_j \|_{\psi_2}\}^q
\leq (\frac{q!\sqrt{\log(1+p)} \sigma K_1}{\sqrt{\log(2)}})^q,
\end{multline*}
for all $q\geq 1$. The claim of the lemma therefore follows. \qed

\noindent \emph{Proof of Lemma \ref{lm:matrixlln}:} This is essentially Theorem E.1 in \citet{Kato2013QuasiB} if $k\ge 2$.
Suppose $k=1$.
Then by Lemma 2.3.1 in \citet{vandervaart:wellner:1996}
\begin{equation}\label{lm:matrixlln1}
E[\|\frac{1}{n}\sum_{i=1}^{n}V_iV_i'-\Sigma\|_{o,2}]\le 2 E[|\frac{1}{n}\sum_{i=1}^{n}\epsilon_iV_i^2|],
\end{equation}
where $\{\epsilon_i\}_{i=1}^n$ are i.i.d.\ Rademacher random variables independent of $\{V_i\}_{i=1}^n$.
For $\|\cdot\|_{\psi_2}$ the Orlicz norm induced by $\psi_2(u) = \exp\{u^2\} - 1$, it then follows from $E[|U|] \leq \|U\|_{\psi_2}/\sqrt{\log(2)}$ for any random variable $U\in \mathbf R$ (see \cite{vandervaart:wellner:1996} pg. 95) and Lemma 2.2.7 in  \citet{vandervaart:wellner:1996} that
\begin{multline}\label{lm:matrixlln2}
E[|\frac{1}{n}\sum_{i=1}^{n}\epsilon_i V_i^2|] = E[E[|\frac{1}{n}\sum_{i=1}^{n}\epsilon_iV_i^2||\{V_i\}_{i=1}^n]]\\
\leq \frac{\sqrt{6}}{\sqrt{\log(2)}}  E[\{\sum_{i=1}^{n}(\frac{V_i^2}{n})^2\}^{1/2}] \leq \frac{\sqrt{6}}{\sqrt{\log(2)}}  E[\max_{1\leq i \leq n} |V_i|\{\sum_{i=1}^{n}(\frac{V_i}{n})^2\}^{1/2}].
\end{multline}
Therefore, the Cauchy-Schwarz's inequality and result \eqref{lm:matrixlln2} imply
\begin{equation*}
E[|\frac{1}{n}\sum_{i=1}^{n}\epsilon_iV_i^2|]\leq \frac{\sqrt 6}{\sqrt{\log(2)}} \{E[\max_{1\leq i \leq n} |V_i|^2]\}^{1/2}\{\frac{E[V^2]}{n}\}^{1/2},
\end{equation*}
which together with \eqref{lm:matrixlln1} establishes the claim of the lemma. \qed

\noindent \emph{Proof of Lemma \ref{lm:auxpseudo}:} For any $k\times k$ matrix $M$, let $R(M) \subseteq \mathbf R^k$ and $N(M)\subseteq \mathbf R^k$ denote its range and null space.
Also recall that for any set $V\subseteq \mathbf R^k$ we let $V^\perp \equiv \{s \in \mathbf R^k : \langle s,v\rangle = 0 \text{ for all } v\in V\}$.
To establish the first claim of the lemma, let $s_1\in \mathbf R^k$ and note that $\Omega_1 s_1 = \Omega_2 s_2$ for some $s_2\in \mathbf R^k$ because $R(\Omega_1) = R(\Omega_2)$.
Therefore, Proposition 6.11.1(6) in \cite{luenberger:1969} yields $\Omega_2 \Omega_2^\dagger \Omega_1 s_1 = \Omega_2 \Omega_2^\dagger \Omega_2 s_2 = \Omega_2 s_2 = \Omega_1 s_1$. Hence, since $s_1 \in \mathbf R^k$ was arbitrary, it follows that $\Omega_2 \Omega_2^\dagger \Omega_1 = \Omega_1$.

In order to establish the second claim of the lemma, first note that $R(M^\dagger) = N(M)^\perp$ for any $k\times k$ matrix $M$.
Thus, since for $j\in \{1,2\}$ we have $N(\Omega_j)^\perp = R(\Omega_j)$ due to $\Omega_j^\prime = \Omega_j$ and Theorem 6.7.3(2) in \cite{luenberger:1969}, we can conclude
\begin{equation*}
R(\Omega_2^\dagger) = N(\Omega_2)^\perp = R(\Omega_2) = R(\Omega_1) = N^\perp(\Omega_1) = R(\Omega_1^\dagger),
\end{equation*}
where the third equality holds by assumption.
Letting $s_1 \in \mathbf R^k$ be arbitrary, it then follows that there exists an $s_2 \in \mathbf R^k$ for which $\Omega_1^\dagger s_1 = \Omega_2^\dagger s_2$, and thus
\begin{equation*}
\Omega_2^\dagger \Omega_2 \Omega_1^\dagger s_1 = \Omega_2^\dagger \Omega_2 \Omega_2^\dagger s_2 = \Omega_2^\dagger s_2 = \Omega_1^\dagger s_1,
\end{equation*}
where the second equality holds by Proposition 6.11.1(5) in \cite{luenberger:1969}. Since $s_1\in \mathbf R^k$ was arbitrary, it follows that $\Omega_2^\dagger \Omega_2 \Omega_1^\dagger = \Omega_1^\dagger$. \qed

\noindent \emph{Proof of Lemma \ref{lm:auxdensity}:} The result follows from results in Chapter 11 of \cite{davydov:lifshits:smorodina:1998}.
Let $F$ denote the c.d.f.\ of $\mathbb S$ and note that Theorem 11.2 in \cite{davydov:lifshits:smorodina:1998} implies that $F$ is absolutely continuous with density $F^\prime$ satisfying
\begin{equation}\label{lm:auxdensity1}
F^\prime(r) = q(r) \exp\{-\frac{r^2}{2\sigma^2}\},
\end{equation}
where $q : \mathbf R \to \mathbf R_+$ is a nondecreasing function. Moreover, we can conclude that
\begin{equation}\label{lm:auxdensity2}
q(r) \int_r^\infty \exp\{-\frac{u^2}{2\sigma^2}\}du \leq \int_r^\infty q(u)\exp\{-\frac{u^2}{2\sigma^2}\}du = P(\mathbb S \geq r) \leq 1,
\end{equation}
where the first inequality follows from $q:\mathbf R \to \mathbf R_+$ being nondecreasing and the equality follows from \eqref{lm:auxdensity1}.
Setting $\Phi$ and $\Phi^\prime$ to denote the c.d.f.\ and density of a standard normal random variable respectively, then note that we may write
\begin{equation}\label{lm:auxdensity3}
\int_r^\infty \exp\{-\frac{u^2}{2\sigma^2}\}du = \sqrt{2\pi} \int_r^\infty \Phi^\prime(u/\sigma)du = \sqrt{2\pi}\sigma(1-\Phi(r/\sigma)).
\end{equation}
Therefore, we can combine \eqref{lm:auxdensity1}, \eqref{lm:auxdensity2}, and \eqref{lm:auxdensity3} to obtain the bound
\begin{equation}\label{lm:auxdensity4}
F^\prime(r) \leq  \frac{ \exp\{- r^2/2\sigma^2\} }{\sqrt{2\pi}\sigma(1-\Phi(r/\sigma))} = \frac{\Phi^\prime(r/\sigma)}{\sigma (1-\Phi(r/\sigma))} \leq \frac{2}{\sigma} \max\{\frac{r}{\sigma},1\},
\end{equation}
where the final result follows from Mill's inequality implying $\Phi^\prime(r)/(1-\Phi^\prime(r)) \leq 2\max\{r,1\}$ for all $r\in \mathbf R$ (see, e.g., pg. 64 in \cite{chernozhukov2014comparison}).

Next note that for any $\eta > 0$, the definitions of $\mathbb S$ and $\text{m}$, and the distribution of $\mathbb S$ first order stochastically dominating that of $\mathbb Z_j$ for any $1\leq j \leq d$ imply that
\begin{equation*}
P(\mathbb S \leq \text{m} + \eta) \geq P(\mathbb S \leq \max_{1\leq j \leq p} \text{med}\{\mathbb Z_j\} + \eta) \geq P(\max_{1\leq j \leq d}(\mathbb Z_j - E[\mathbb Z_j]) \leq \eta) > 0,
\end{equation*}
where the final inequality follows from $E[\mathbb Z]$ belonging to the support of $\mathbb Z$.
Theorem 11.2 in \cite{davydov:lifshits:smorodina:1998} thus implies $q : \mathbf R \to \mathbf R_+$ is continuous at any $r > \text{m}$, which together with \eqref{lm:auxdensity1} and the first fundamental theorem of calculus establishes $F$ is in fact differentiable at any $r > \text{m}$ with derivative given by $F^\prime$. Setting $\Gamma \equiv \Phi^{-1} \circ F$, then observe $F = \Phi \circ \Gamma$ and hence at any $r > \text{m}$ we obtain
\begin{equation}\label{lm:auxdensity6}
F^\prime(r) = \Phi^\prime(\Gamma(r)) \Gamma^\prime(r)
\end{equation}
for $\Gamma^\prime$ the derivative of $\Gamma$.
However, $\Gamma^\prime$ is decreasing since $\Gamma$ is concave by Proposition 11.3 in \cite{davydov:lifshits:smorodina:1998}, while $\Phi^\prime(\Gamma(r))$ is decreasing on $[\text{m},+\infty)$ due to $\Phi^\prime$ being decreasing on $[0,\infty)$ and $\Gamma(r) \in [0,\infty)$ for any $r > \text{m}$.
In particular, \eqref{lm:auxdensity6} implies $F^\prime$ is decreasing on $(\text{m},+\infty)$ which together with \eqref{lm:auxdensity4} yields
\begin{equation}\label{lm:auxdensity7}
\sup_{r\in (\text{m},+\infty)} F^\prime(r) = \limsup_{r\downarrow \text{m}} F^\prime(r) \leq \limsup_{r\downarrow \text{m}} \frac{\Phi^\prime(r/\sigma)}{\sigma (1-\Phi(r/\sigma))} = \frac{\Phi^\prime(\text{m}/\sigma)}{\sigma (1-\Phi(\text{m}/\sigma))}.
\end{equation}
Since result \eqref{lm:auxdensity4} implies $F^\prime(r)$ is bounded by $2\max\{\text{m}/\sigma,1\}/\sigma$ on $(-\infty,\text{m}]$ and result \eqref{lm:auxdensity7} implies the same bound applies on $(\text{m},+\infty)$, the lemma follows. \qed

\noindent \emph{Proof of Lemma \ref{lm:auxlp}:} The claim that $\mathcal E \neq \emptyset$ follows from Corollary 18.5.3 in \cite{rockafellar1970convex}.
Moreover, for $\mathcal D$ the set of extreme directions of $C$, Corollary 19.1.1 in \cite{rockafellar1970convex} implies both $\mathcal E$ and $\mathcal D$ are finite.
Thus, writing $\mathcal E = \{a_j\}_{j=1}^{m}$ and $\mathcal D \equiv \{a_j\}_{j=m+1}^{n}$ (with $n =m$ when $\mathcal D = \emptyset$), Theorem 18.5 in \cite{rockafellar1970convex} yields the representation
\begin{equation}\label{lm:auxlp1}
C \equiv \{ c \in \mathbf R^k : c = \sum_{j = 1}^n a_j \lambda_j \text{ s.t. } \sum_{j=1}^m \lambda_j =1 \text{ and } \lambda_j \geq 0 \text{ for all } j\}.
\end{equation}
Next note that if $\sup_{c\in C} \langle c, y \rangle$ is finite, then Corollary 5.3.7 in \cite{borwein2010convex} implies that the supremum is attained.
Hence, by result \eqref{lm:auxlp1} we obtain
\begin{align}\label{lm:auxlp2}
\sup_{c \in C} \langle c, y\rangle & = \max_{\{\lambda_j\}_{j=1}^n} \langle y, \sum_{j=1}^n \lambda_j a_j \rangle \text{ s.t. } \sum_{j=1}^m \lambda_j =1, ~ \lambda_j \geq 0 \text{ for } 1\leq j \leq n  \notag \\
& = \max_{\{\lambda_j\}_{j=1}^m} \langle y, \sum_{j=1}^m \lambda_j a_j \rangle \text{ s.t. } \sum_{j=1}^m \lambda_j =1, ~ \lambda_j \geq 0 \text{ for } 1\leq j \leq m ,
\end{align}
where the second equality follows due to $\sup_{c\in C}\langle c,y\rangle$ being finite implying we must have $\langle y,a_j\rangle \leq 0$ for all $m+1 \leq j \leq n$. Since $\mathcal E = \{a_j\}_{j=1}^m$ and the maximization in \eqref{lm:auxlp2} is solved by setting $\lambda_{j^\star} = 1$ for some $1\leq j^\star \leq m$, the lemma follows. \qed

\noindent \emph{Proof of Lemma \ref{lm:Vpoly}:} First note that $0 \in (AA^\prime)^\dagger \mathcal V^{\rm i}(P)$ and therefore $(AA^\prime)^\dagger \mathcal V^{\rm i}(P)$ is non-empty.
To show $(AA^\prime)^\dagger \mathcal V^{\rm i}(P)$ is closed, suppose $\{v_j\}_{j=1}^\infty \in (AA^\prime)^\dagger \mathcal V^{\rm i}(P)$ and $\|v_j - v^\star\|_2 = o(1)$ for some $v^\star \in \mathbf R^p$.
Since $v_j \in (AA^\prime)^\dagger \mathcal V^{\rm i}(P)$ it follows that there is an $s_j \in \mathcal V^{\rm i}(P)$ such that $v_j = (AA^\prime)^\dagger s_j$.
Next, let $\tilde s_j \equiv A A^\dagger s_j$ and note that
\begin{equation}\label{lm:Vpoly1}
(AA^\prime)^\dagger \tilde s_j = (AA^\prime)^\dagger A A^\dagger  s_j = (A^\prime)^\dagger A^\dagger s_j = (AA^\prime)^\dagger s_j
\end{equation}
since $(AA^\prime)^\dagger A = (A^\prime)^\dagger$ by Proposition 6.11.1(8) in \cite{luenberger:1969} and $(A^\prime)^\dagger A^\dagger = (AA^\prime)^\dagger$ (see \cite{seber2008matrix} pg. 139).
Also note $A^\dagger \tilde s_j = A^\dagger A A^\dagger s_j = A^\dagger s_j$ by Proposition 6.11.1(5) in \cite{luenberger:1969}, while \eqref{lm:Vpoly1} implies $\|\Omega^{\rm i}(P)(AA^\prime)^\dagger \tilde s_j\|_1 = \|\Omega^{\rm i}(P)(AA^\prime)^\dagger s_j\|_1$.
Hence, if $s_j\in \mathcal V^{\rm i}(P)$, then $\tilde s_j \in \mathcal V^{\rm i}(P)$, and by \eqref{lm:Vpoly1} we have $(AA^\prime)^\dagger \tilde s_j = v_j$.
Furthermore, by construction $\tilde s_j \in R$ and hence $(AA^\prime)(AA^\prime)^\dagger \tilde s_j = \tilde s_j$, which together with $(AA^\prime)^\dagger \tilde s_j = v_j$ implies $\tilde s_j = AA^\prime v_j$.
By continuity, it then follows from $\|v_j - v^\star\|_2 = o(1)$ that $\|\tilde s_j - s^\star\|_2 = o(1)$ for $s^\star = AA^\prime v^\star$ and thus $s^\star \in \mathcal V^{\rm i}(P)$ due to $\mathcal V^{\rm i}(P)$ being closed.
Furthermore, $v_j = (AA^\prime)^\dagger \tilde s_j$ yields
\begin{equation}\label{lm:Vpoly2}
\|v^\star - (AA^\prime)^\dagger s^\star\|_2 \leq \lim_{n\rightarrow \infty} \|v_j - v^\star\|_2 + \|(AA^\prime)^\dagger (\tilde s_j - s^\star)\|_2 = 0
\end{equation}
due to $\|v_j - v^\star\|_2 = o(1)$ and $\|\tilde s_j - s^\star\|_2 = o(1)$. Since, as argued, $s^\star \in \mathcal V^{\rm i}(P)$, we can conclude that $v^\star \in (AA^\prime)^\dagger \mathcal V^{\rm i}(P)$ and hence that $(AA^\prime)^\dagger \mathcal V^{\rm i}(P)$ is closed.

The fact that $(AA^\prime)^\dagger \mathcal V^{\rm i}(P)$ is polyhedral is immediate from definition of $\mathcal V^{\rm i}(P)$, and thus we next show $(AA^\prime)^\dagger \mathcal V^{\rm i}(P)$ contains no lines.
To this end, suppose $v \in (AA^\prime)^\dagger \mathcal V^{\rm i}(P)$, which implies $v = (AA^\prime)^\dagger s$ for some $s \in \mathcal V^{\rm i}(P)$.
Since $A^\prime (AA^\prime)^\dagger = A^\dagger$ by Proposition 6.11.1(9) in \cite{luenberger:1969}, we are able to conclude that
\begin{equation}\label{lm:Vpoly3}
A^\prime v = A^\prime(AA^\prime)^\dagger s = A^\dagger s \leq 0
\end{equation}
due to $s\in \mathcal V^{\rm i}(P)$.
Similarly, if $-v \in (AA^\prime)^\dagger \mathcal V^{\rm i}(P)$, then $A^\prime(-v) \leq 0$ and thus $-v,v\in (AA^\prime)^\dagger \mathcal V^{\rm i}(P)$ imply that $A^\prime v = 0$.
However, for $N(A^\prime)^\perp$ the orthocomplement to the null space of $A^\prime$, note that $v = (AA^\prime)^\dagger s = (A^\prime)^\dagger A^\dagger s$ implies that
\begin{equation}\label{lm:Vpoly4}
v \in N(A^\prime)^\perp .
\end{equation}
Since $v\in N(A^\prime)^\perp$ and $A^\prime v =0$ imply $v = 0$, it follows that if $-v,v\in (AA^\prime)^\dagger \mathcal V^{\rm i}(P)$, then $v =0$ and hence $(AA^\prime)^\dagger \mathcal V^{\rm i}(P)$ contains no lines as claimed.

Finally, to see that zero is an extreme point of $(AA^\prime)^\dagger \mathcal V^{\rm i}(P)$ suppose that $0 = \lambda v_1 + (1-\lambda)v_2$ for some $v_1,v_2 \in (AA^\prime)^\dagger \mathcal V^{\rm i}(P)$ and $\lambda \in (0,1)$.
By result \eqref{lm:Vpoly3} holding for any $v\in (AA^\prime)^\dagger \mathcal V^{\rm i}(P)$, $\lambda \in (0,1)$, and $0 = A^\prime 0 = A^\prime (\lambda v_1 + (1-\lambda)v_2)$, it then follows that $A^\prime v_1 = A^\prime v_2 = 0$. Therefore, \eqref{lm:Vpoly4} holding for any $v\in (AA^\prime)^\dagger \mathcal V^{\rm i}(P)$ implies $v_1 = v_2 = 0$, which verifies that zero is an extreme point. \qed

\phantomsection
\addcontentsline{toc}{section}{References}

{\small
\setstretch{0.99}

\putbib }

\end{bibunit}

\end{document}